\documentclass[aps,twocolumn,prb,showpacs,floatfix]{revtex4}
\usepackage{graphics,amssymb,amsmath,epsfig}

\newcommand {\half} { \frac{1}{2} }

\newcommand {\avg} [1]    { \left\langle #1\right\rangle }
\newcommand {\ket} [1]    { \left|#1\right\rangle }

\newcommand {\nn}[1] { \left\langle #1\right\rangle }

\newcommand {\varH}  {{\mathcal{H}}}
\newcommand {\varO}     {{\mathcal{O}}}

\newcommand {\bfB}  {{\mathbf{B}}}

\newcommand {\bfr}  {{\mathbf{r}}}
\newcommand {\bfx}  {{\mathbf{x}}}
\newcommand {\bfy}  {{\mathbf{y}}}
\newcommand {\bfz}  {{\mathbf{z}}}

\newcommand{\ux}        {{\hat\bfx}}
\newcommand{\uy}        {{\hat\bfy}}
\newcommand{\uz}        {{\hat\bfz}}

\newcommand {\ic}    {\Delta_2}
\newcommand {\icavg} {\avg{\Delta_2}}
\newcommand {\icstd} {\delta\Delta_2}
\newcommand {\icres} {\icstd/\icavg}

\newcommand{\eqnref}[1]{Eq.~(\ref{#1})}

\newcommand{\figref}[1]{Fig.~\ref{#1}}
\newcommand{\figsref}[1]{Figs.~\ref{#1}}
\newcommand{\Figref}[1]{Figure~\ref{#1}}

\newcommand{\secref}[1]{Sec.~\ref{#1}}
\newcommand{\secsref}[1]{Secs.~\ref{#1}}
\newcommand{\Secref}[1]{Section~\ref{#1}}

\begin{document}
\title{Quantum and frustration effects on fluctuations of the inverse compressibility
in two-dimensional Coulomb glasses}
\author{Minchul Lee,$^1$ Gun Sang Jeon,$^2$ and M.Y. Choi$^{1,3}$}
\affiliation{$^1$Department of Physics,
  Seoul National University, Seoul 151-747, Korea\\
  $^2$Center for Strongly Correlated Materials Research,
  Seoul National University, Seoul 151-747, Korea\\
  $^3$Korea Institute for Advanced Study, Seoul 130-012, Korea}

\begin{abstract}
  We consider interacting electrons in a two-dimensional quantum Coulomb glass and
  investigate by means of the Hartree-Fock approximation the combined effects of the
  electron-electron interaction and the transverse magnetic field on fluctuations of the
  inverse compressibility.  Preceding systematic study of the system in the absence of the
  magnetic field identifies the source of the fluctuations, interplay of disorder and
  interaction, and effects of hopping.  Revealed in sufficiently clean samples with strong
  interactions is an unusual right-biased distribution of the inverse compressibility,
  which is neither of the Gaussian nor of the Wigner-Dyson type.  While in most cases weak
  magnetic fields tend to suppress fluctuations, in relatively clean samples with weak
  interactions fluctuations are found to grow with the magnetic field.  This is
  attributed to the localization properties of the electron states, which may be measured
  by the participation ratio and the inverse participation number.  It is also observed
  that at the frustration where the Fermi level is degenerate, localization or
  modulation of electrons is enhanced, raising fluctuations.  Strong frustration in
  general suppresses effects of the interaction on the inverse compressibility and on the
  configuration of electrons.
\end{abstract}


\maketitle

%
\section{Introduction}

Since the experimental observation of the crucial effects of the Coulomb
interaction,\cite{Sivan96} mesoscopic fluctuations of the inverse compressibility in
two-dimensional quantum dots have been investigated intensively, both theoretically and
experimentally.\cite{Sivan96,Simmel99,Patel98,Blanter97,Levit99,Koulakov97,Jeon99,Walker99b,Vallejos98,Alhassid00}
Recent experiments\cite{Sivan96,Simmel99,Patel98} on the distribution of the inverse
compressibility, which can be measured by conductance peak spacing, in the
Coulomb-blockade regime have shown that the standard random matrix theory fails to explain
the observed fluctuations; this implies that charging energy fluctuations may play a
dominant role in the inverse-compressibility fluctuations.  In particular, the
distribution was observed to take a Gaussian-like symmetric form with non-Gaussian
tails.\cite{Sivan96,Simmel99,Patel98} Subsequent analytical and numerical studies on
interacting electrons in disordered systems revealed the possibility of larger
inverse-compressibility fluctuations induced by Coulomb
interactions.\cite{Sivan96,Blanter97,Levit99,Koulakov97,Jeon99,Walker99b} It was also
proposed that the shape deformation of the dot due to the gate-voltage
sweeping\cite{Vallejos98} or the irregular shape of the dot\cite{Alhassid00} may
contribute to the deviation of the experimental results from the Wigner-Dyson (WD)
statistics predicted by the random matrix theory.  However, the recent experiment with a
stacked gate structure, which allows one to vary the electron density without significant
deformation of the dot, was supportive of the claim that the observed shape of
fluctuations originates from electron-electron interactions.\cite{Simmel99} Another recent
experiment reported the crossover behavior of the compressibility around the
metal-insulator transition in two dimensions.\cite{Ilani00}

The influence of the magnetic field, either perpendicular\cite{Sivan96,Patel98} or
parallel,\cite{Duncan00} on the distribution of the inverse compressibility was
also investigated experimentally.  In the presence of a perpendicular magnetic
field, the shape of the distribution is found qualitatively similar to that
in zero magnetic field, only with narrower width.\cite{Patel98}
There was an attempt to explain the qualitative behavior of
peak positions in the system under the magnetic field;\cite{Avishai99} the considered
classical model is based on the electrostatics of several electron islands with quantum
orbitals, the energies of which depend on the external magnetic field.
The possibility of nearly vanishing peak spacing was thus suggested,
which is in qualitative agreement with the bunching of the addition spectra.\cite{Zhitenev97}
However, it does not take into account the quantum interference effects between electron wave
functions, which can be crucial in mesoscopic systems.

In this paper, we study numerically fluctuations of the inverse compressibility in the
two-dimensional Coulomb glass.\cite{classical,quantum,jeon2} To treat the interaction
between electrons, we employ the Hartree-Fock (HF) approximation, which makes it possible
to investigate larger samples than those feasible in exact diagonalization.  For a
systematic study, we first examine the classical system, revealing the detailed dependence
of the inverse-compressibility fluctuations upon disorder and interaction.  Then the
effects of electron hopping on the fluctuations are investigated.  Observed is an unusual
right-biased distribution for very weak disorder and strong interactions as well as a
symmetric Gaussian-like distribution with non-Gaussian tails.  It is also found that
relative fluctuations decrease eventually for sufficiently strong interactions, which
appears to be inconsistent with the quadratic interaction dependence of the fluctuations
suggested in a previous numerical study.\cite{Walker99b}
We further examine the effects of magnetic fields, which introduce frustration to the
system, on the distributions and fluctuations of the inverse compressibility, and observe
that weak magnetic fields tend to suppress fluctuations.  Nevertheless the fluctuations
may be enhanced in some range of the frustration for weak disorder and interaction.  Such
opposite tendency of the inverse-compressibility fluctuations in response to the applied
magnetic fields is explained in terms of the localization properties of the system.
Finally, it is also found that strong frustration suppresses significantly the effects of
the interaction.

This paper is organized as follows: In \secref{sec:cs} we consider the classical Coulomb
glass and investigate in detail the effects of disorder and interaction on the
inverse-compressibility fluctuations.  The effects of hopping on the fluctuations in the
quantum Coulomb glass are examined in \secref{sec:qs}.  \Secref{sec:mf} is devoted to the
effects of magnetic fields.  We examine the system at both weak and strong frustration,
and describe the electron distributions and localization properties by means of the
participation ratio and the inverse participation number.
The main results are summarized in \secref{sec:s}.

%
%

\section{Classical Coulomb Glass\label{sec:cs}}

We address the Hamiltonian for the classical Coulomb glass model:\cite{classical}
\begin{equation}
  \label{eq:cH}
  \varH = \sum_i w_i \hat{n}_i + \half \sum_{i\ne j} (\hat{n}_i - K) U_{ij} (\hat{n}_j - K),
\end{equation}
where $i$ is the site index on an $L\times L$ square lattice, $\hat{n}_i$ is the number
operator of the electron at site $i$, and $w_i$ is the random on-site potential uniformly
distributed in the range $[-W,W]$.  We assume the Coulomb interaction potential between
electrons:
\begin{equation}
  U_{ij} = \frac{e^2}{\epsilon |\bfr_i-\bfr_j|} \equiv \frac{V}{|\bfr_i-\bfr_j|}
\end{equation}
with $\epsilon$ being the dielectric constant.  In Eq.~(\ref{eq:cH}) $K$ represents the
uniform neutralizing background charge per site in units of $+e$, which is demanded by the
charge neutrality condition over the whole system.  Throughout this paper we concentrate
on the half-filling case $K=1/2$ in the free boundary conditions, and write the distance
in units of the lattice constant $a$.

The ground state of the classical Hamiltonian in \eqnref{eq:cH} is obtained by means of
simulated annealing procedure.  The inverse compressibility $\ic$ for a given disorder
configuration is then calculated via the relation
\begin{equation}
  \label{eq:ic}
  \ic = E_{M+1} - 2E_M + E_{M-1},
\end{equation}
where $M \equiv L^2/2$ and $E_n$ denotes the $n$-electron ground-state energy.  We have
performed the calculation for size $L=4$, 6, and 8, and averaged the data over 2000 different
disorder configurations.

\begin{figure}
  \centerline{\epsfig{file=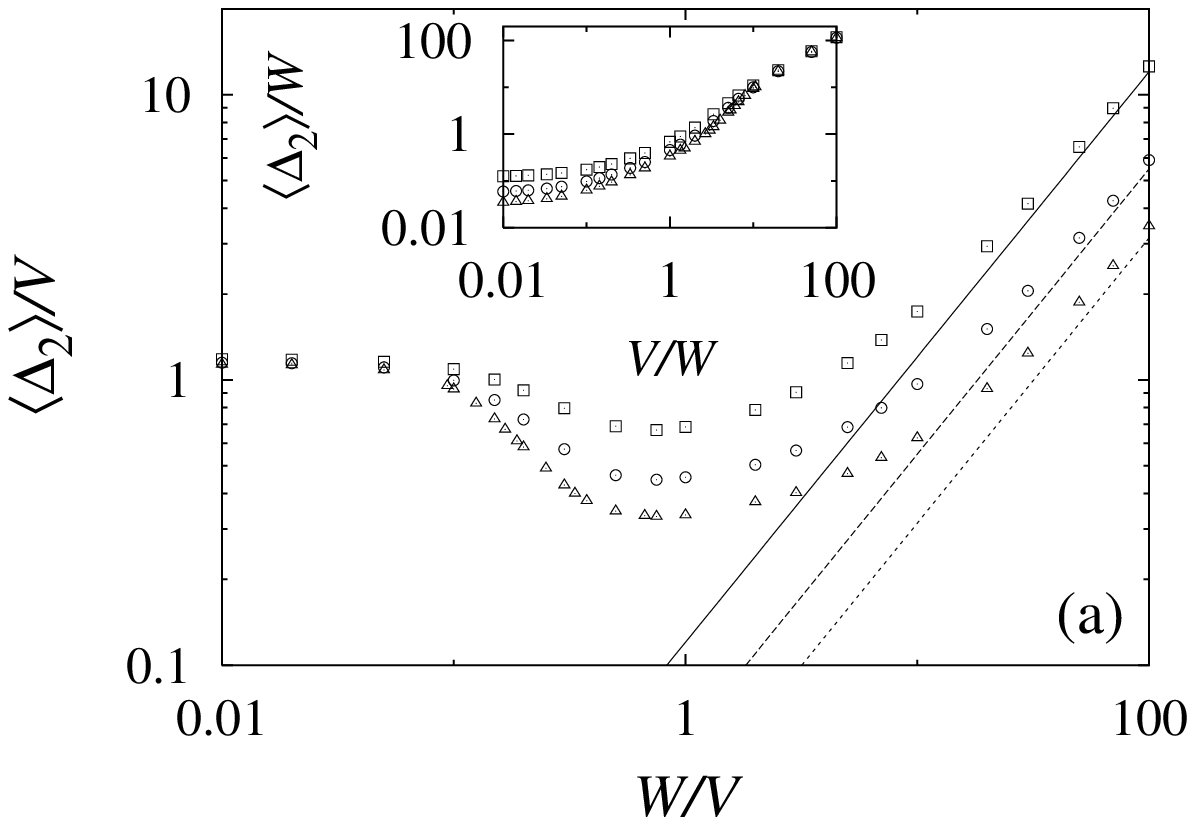,width=8cm}}
  \centerline{\epsfig{file=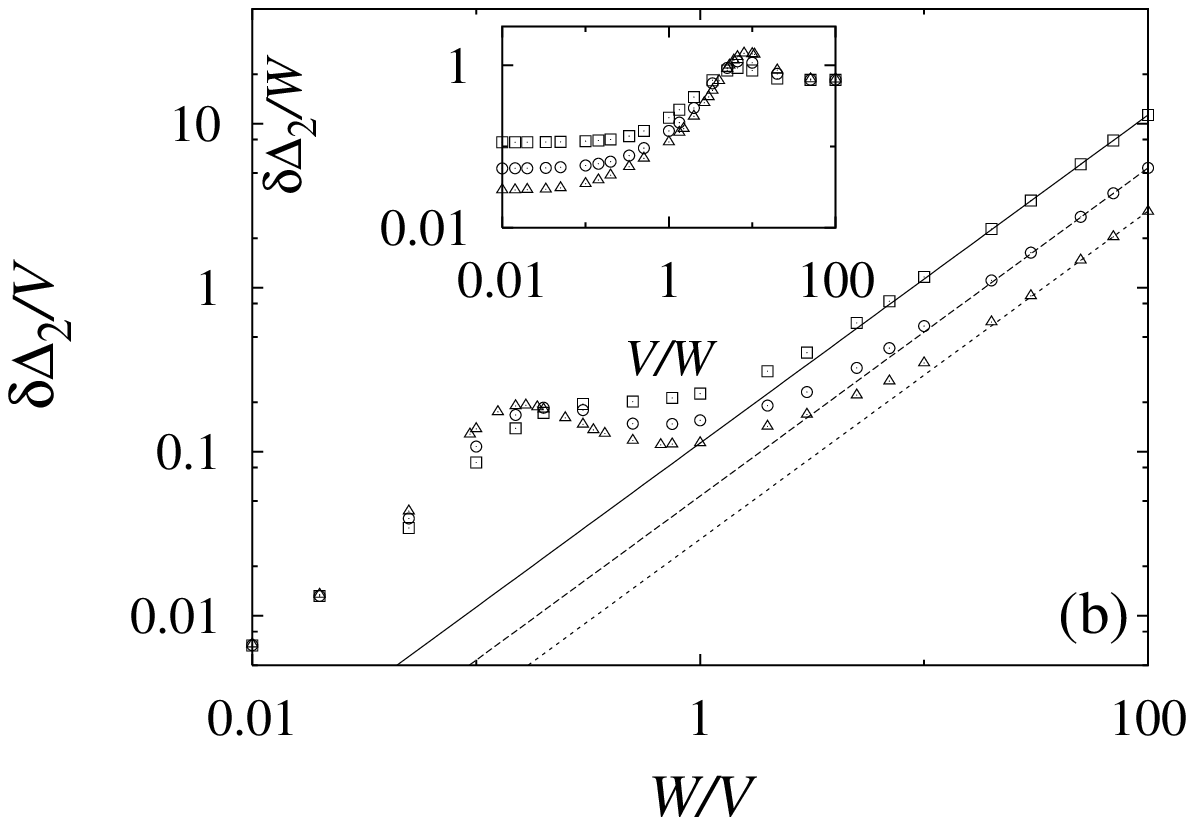,width=8cm}}
  \caption{(a) Averages and (b) fluctuations of the inverse compressibility in units
    of $V$ as functions of $W/V$ in the classical system of size $L=4(\square)$,
    $6(\bigcirc)$, and $8(\triangle)$.  Each straight line represents the behavior of the
    corresponding noninteracting system ($V=0$).  The insets display the averages and the
    fluctuations in units of $W$ as functions of $V/W$.}
  \label{fig:cicmndv}
\end{figure}
\Figref{fig:cicmndv}(a) shows the average inverse compressibility as a function of the
relative disorder strength $W/V$.  For weak disorder, $\icavg/V$ depends rather weakly on
both the disorder strength $W$ and the size $L$ since $\icavg$ is proportional to $V$ in
this regime as shown in the inset.  As the disorder is increased, $\icavg/V$ is suppressed
up to $W/V\approx 1$, then increases for $W/V > 1$, and approaches asymptotically the
straight line describing the behavior in the corresponding noninteracting system.  One can
also observe that $\icavg$ decreases with the system size, which is prominent for strong
disorder.  It is remarkable that the disorder-scaled average $\icavg/W$, which is plotted
in the inset of \figref{fig:cicmndv}(a), is monotonically increasing with $V/W$, thus
indicating that larger interactions lead to larger values of the inverse compressibility
for given disorder strength.

In contrast to the average $\icavg$, the fluctuations $\icstd$ of the inverse
compressibility, shown in \figref{fig:cicmndv}(b), scale with $W$ independently of $V$ for
weak disorder as well as for strong disorder.  In the crossover regime, on the other hand,
the fluctuations display non-monotonic behavior, with a broad peak at $W/V \approx 0.2$.
This becomes conspicuous as the system size grows: Whereas for $W/V > 0.2$, $\icstd$
decreases with the size similarly to $\icavg$, slight increase of $\icstd$ can be observed
for $W/V \lesssim 0.2$.  It is of interest to note that $\icstd$ approaches the
noninteracting limit rather fast as $W/V$ is increased, demonstrating that for strong
disorder ($W/V \gg 1$) the interaction merely shifts the average inverse compressibility
toward slightly larger values.

\begin{figure}
  \centerline{\epsfig{file=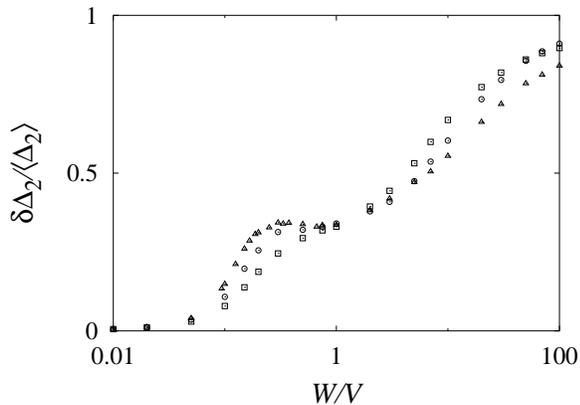,width=8cm}}
  \caption{Relative fluctuations versus $W/V$ in the classical system. The
    symbols are the same as those in \figref{fig:cicmndv}.}
  \label{fig:cicrdv}
\end{figure}
In \figref{fig:cicrdv} we plot the relative fluctuations $\icres$ of the inverse
compressibility versus the relative disorder strength $W/V$.  The relative fluctuations in
general grow with $W/V$, saturating to a finite value in the noninteracting limit.  At the
point $W/V=1$ the relative fluctuations are apparently independent of the system size,
which has been reported in previous numerical studies.\cite{Koulakov97,Jeon99} However, it
should be noted that such universal behavior shows up only at $W/V=1$: While the relative
fluctuations grow with the system size for $W/V \lesssim 1$, they are reduced by the
increase of the system size for $W/V \gtrsim 1$.  Again a broad peak, which becomes
prominent for larger systems, is observed for $W/V \approx 0.2$; this can be attributed to
the presence of the peak in $\icstd$ mentioned above.

In the limit of strong disorder ($W/V \gg 1$), the electrons are distributed spatially to
minimize the total random potential energy, and firmly pinned, rarely rearrange their
whole distribution in response to the addition or removal of an electron.  Consequently,
as in the noninteracting system, both $\icavg$ and $\icstd$ are expected to increase with
the disorder strength $W$, while the interaction simply renormalizes the values slightly
without qualitative change.  Note that such tendency begins at relatively small disorder
$W/V \gtrsim\varO(1)$, as shown in \figref{fig:cicmndv}.

In the opposite limit of strong interaction ($W/V \ll 1$), on the other hand, the half-filled
electron state $\ket{M}$ is expected to form a Wigner crystal (WC). In case that two WC
states are nearly degenerate even in the presence of the disorder, the addition or
removal of an electron can change the ground state into that similar to the other WC
state; however, the disorder configuration, which allows such phenomena, is very rare
and accordingly, the WC state in general does not experience significant change by the
insertion or extraction of an electron.  The inverse compressibility can then be
calculated as follows:
\begin{eqnarray}
  \nonumber
  \ic & = & (E_{M+1} - E_M) + (E_{M-1} - E_M) \\
  \nonumber
  & = &
  w_{i_+} + \sum_i (n_i-K) U_{ii_+} - w_{i_-} - \sum_i (n_i-K) U_{ii_-} \\
  \label{eq:siric}
  & = & w_{i_+} - w_{i_-} + \sum_i (n_i-K) (U_{ii_+} - U_{ii_-}),
\end{eqnarray}
where $n_i$ is 0/1 for site $i$ empty/occupied in the state $\ket{M}$, and $i_{\pm}$
represents the site to/from which an electron is inserted/extracted.  In this limit, the
interaction part plays a major role in choosing $i_\pm$, which is very likely to be one of
the corners.  Accordingly, the strength $w_{i_\pm}$ of the impurity at site $i_\pm$ is
random and uncorrelated, leading to the disorder-independent average given by the third
term on the right-hand side of \eqnref{eq:siric}; this quantity decreases slightly with
$L$.  In contrast, the fluctuations arise mainly from the first two terms which have their
origin in the random potential and thus increase with the disorder strength $W$.  The
numerical data presented above are indeed in good agreement with these conclusions based
on analytic argument.  \Figref{fig:cicmndv} demonstrates that this regime persists until
$W/V\approx0.02$.

\begin{figure*}
  \parbox{0.32\textwidth}{
    \centerline{\epsfig{width=0.33\textwidth,file=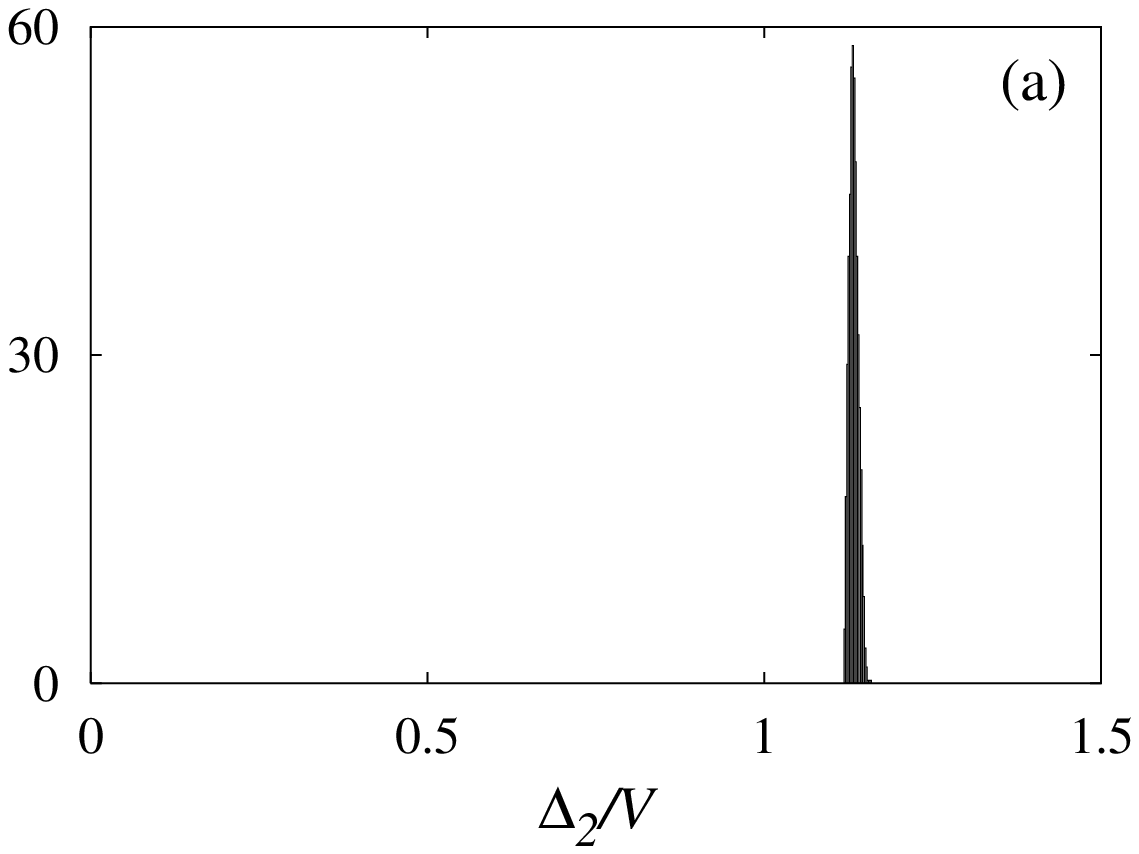}}
  }
  \parbox{0.32\textwidth}{
    \centerline{\epsfig{width=0.33\textwidth,file=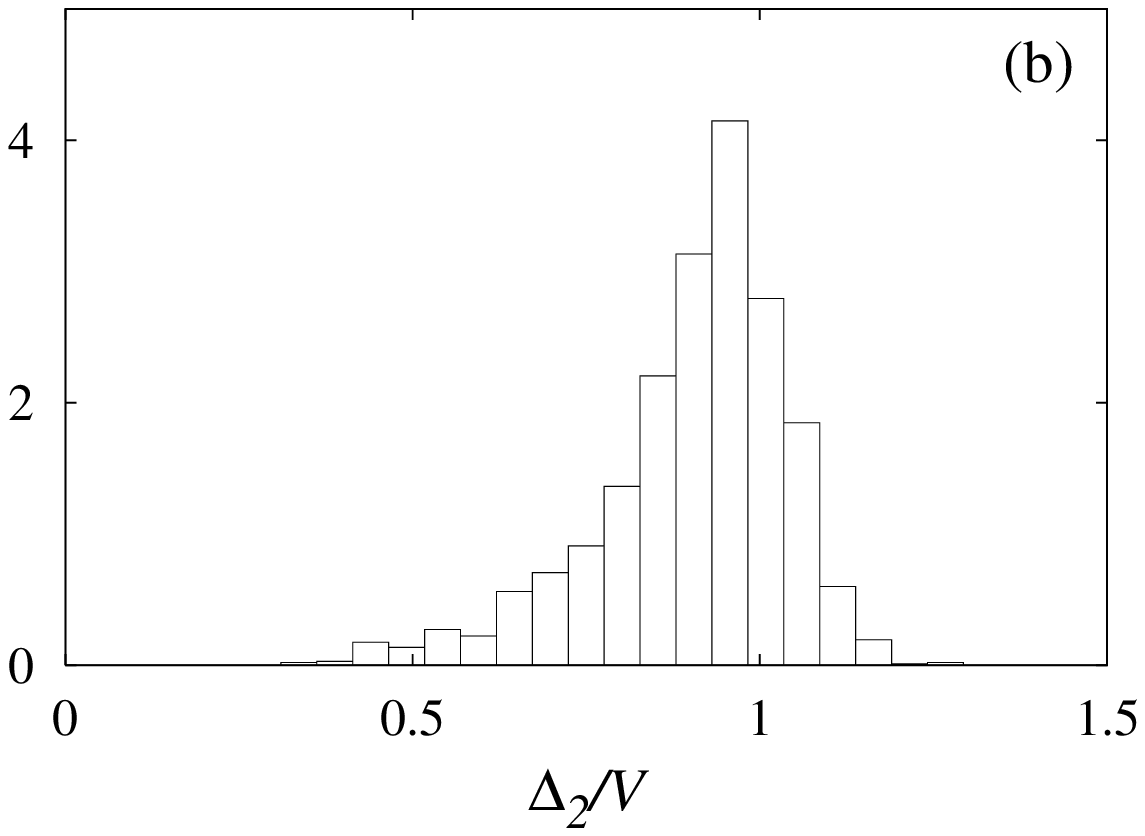}}
  }
  \parbox{0.32\textwidth}{
    \centerline{\epsfig{width=0.33\textwidth,file=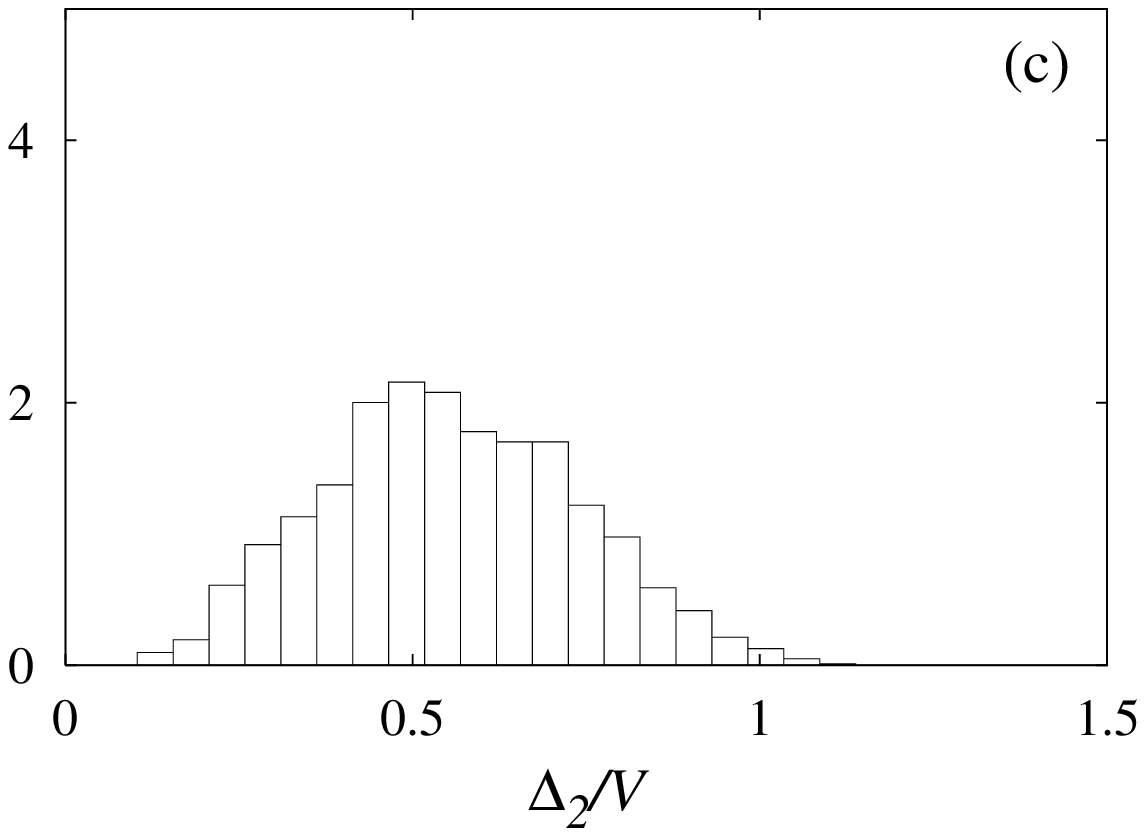}}
  }\\
  \parbox{0.32\textwidth}{
    \centerline{\epsfig{width=0.33\textwidth,file=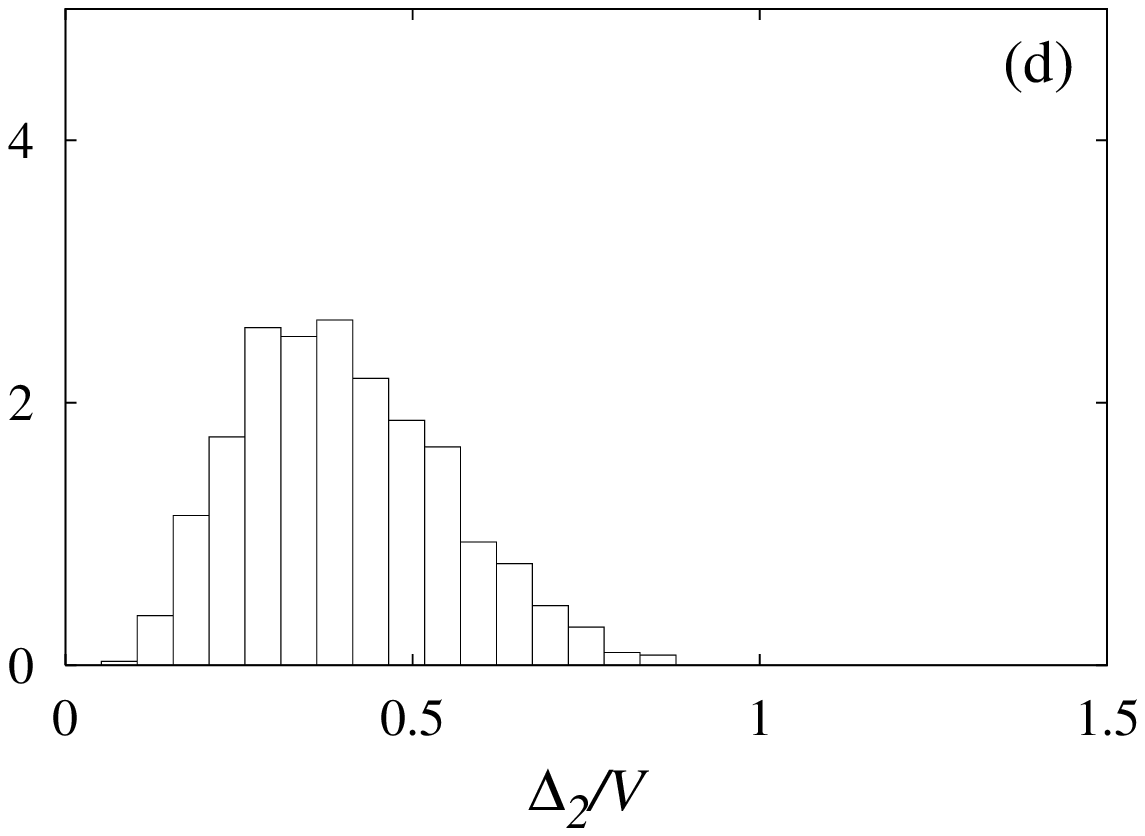}}
  }
  \parbox{0.32\textwidth}{
    \centerline{\epsfig{width=0.33\textwidth,file=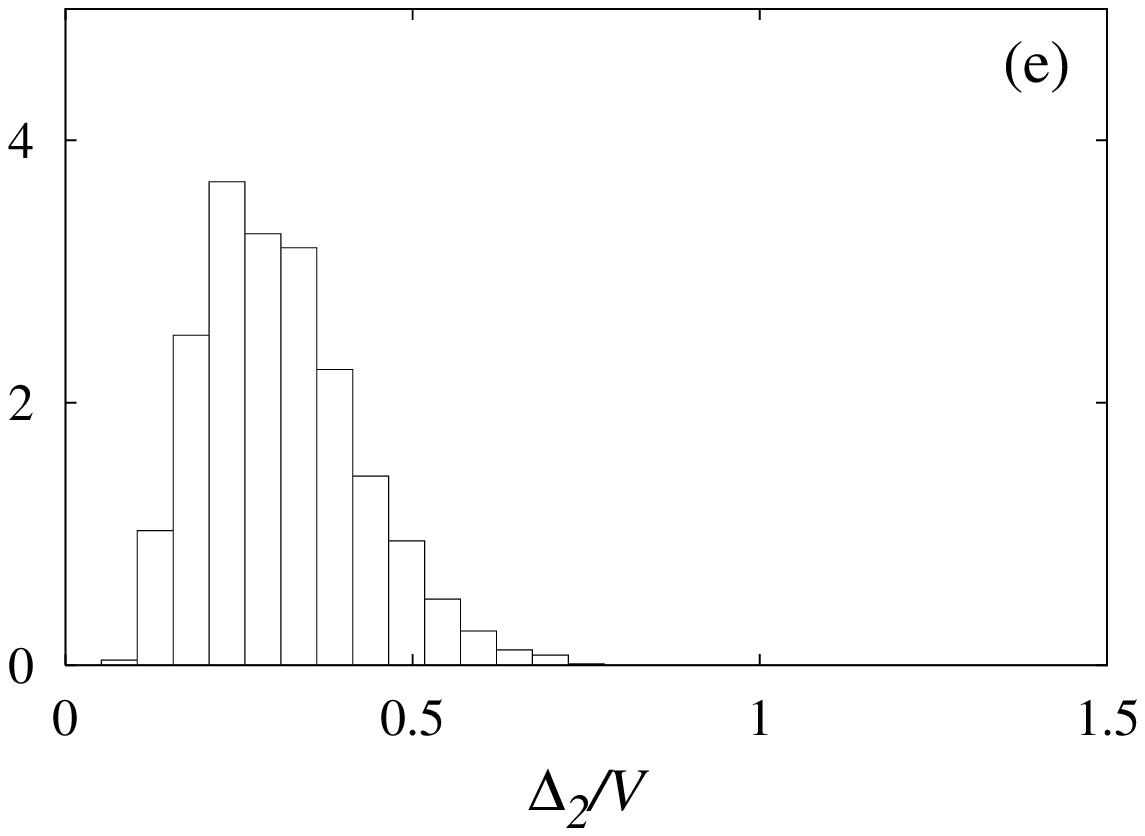}}
  }
  \parbox{0.32\textwidth}{
    \centerline{\epsfig{width=0.33\textwidth,file=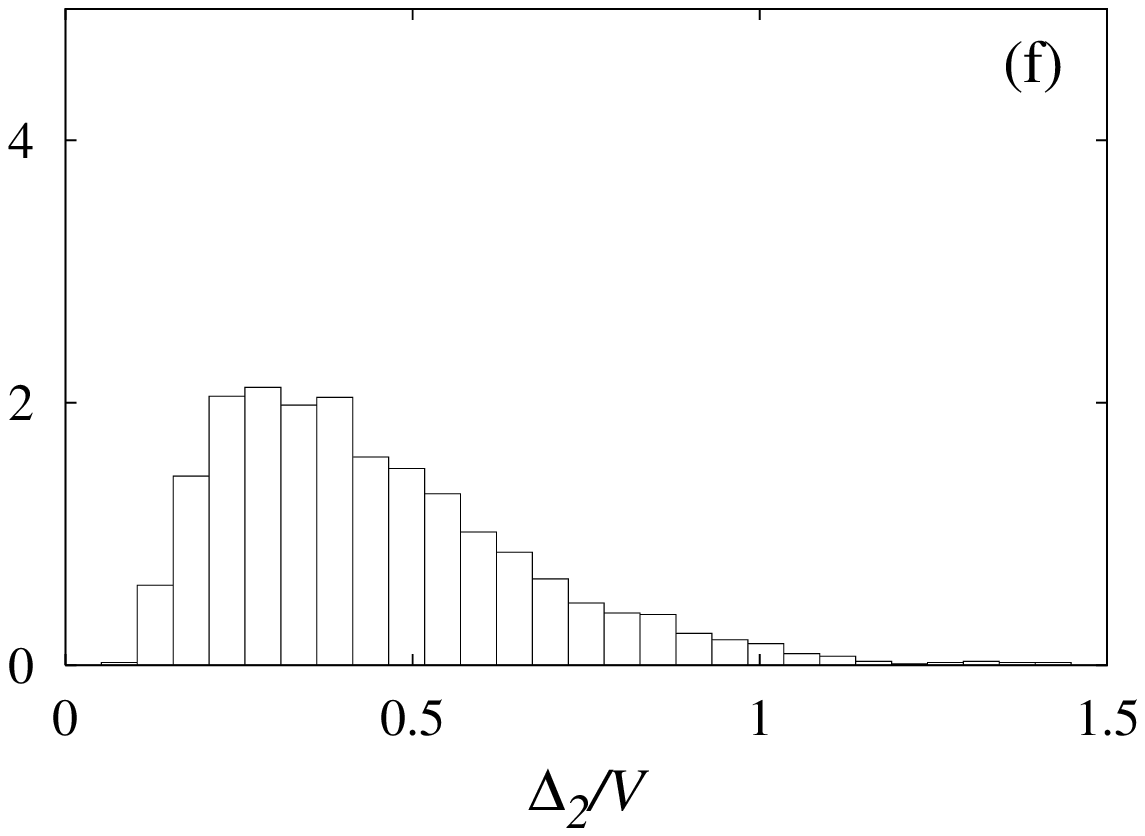}}
  }
  \caption{Distribution of $\ic/V$ in the classical system of size $L=8$ for disorder
    strength $W/V=\,$(a) 0.01, (b) 0.1, (c) 0.2, (d) 0.3, (e) 1, (f) 5.}
  \label{fig:cicdis}
\end{figure*}
\begin{figure}
  \centerline{\epsfig{width=8cm,file=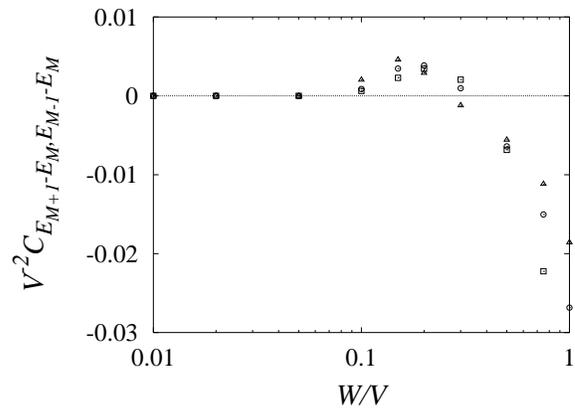}}
  \caption{Correlations between $E_{M{+}1}{-}E_M$ and $E_{M{-}1}{-}E_M$ in units of
    $V^2$ in the classical system of size $L=4(\square)$, $6(\bigcirc)$, and $8(\triangle)$.}
  \label{fig:cfdcor}
\end{figure}

In both extreme limits considered above, one can see that the contributions of the random
potential to the fluctuations of the inverse compressibility are dominant over those of the
interaction.  In the intermediate region, however, the behavior looks rather complicated
and the interaction is expected to play an important role.  For detailed investigation, we
divide the intermediate region into three subregions: (I) $0.02<W/V<0.2$, (II)
$0.2<W/V<0.5$, and (III) $0.5<W/V<\varO(1)$.  In Region I the half-filled state $\ket{M}$
strongly depends on the disorder configuration: Some states are still close to the WC
state while others are rather disordered.
In the latter states, the portion of which increases with $W/V$, the distribution of electrons
is inhomogeneous, reducing the Coulomb energy cost of the addition or removal of one electron.
On the other hand, the former states, which resemble the WC state, rather easily rearrange
the distribution of electrons by the addition or removal of an electron, unlike in the
limit of strong interactions.  Both behaviors generally result in the reduction of $\ic$
although the amount of the reduction depends upon the disorder configuration; this explains
the change of the $\ic/V$ distribution from the $\delta$-function-like one [see
\figref{fig:cicdis}(a)] to the asymmetric broad distribution with a long left tail [see
\figref{fig:cicdis}(b)]. It is observed that the minimum of $\ic/V$ as well as the peak
position of the distribution decreases with $W/V$.  Indeed in a finite system the minimum
of $\ic/V$ hits its lower bound given by the smallest possible Coulomb interaction between
successively inserted electrons,\cite{Koulakov97}
\begin{equation}
  \label{eq:cicmin}
  \frac{\ic|_{\mathrm{min}}}{V} = \frac{1}{r_{\mathrm{max}}} = \frac{1}{(L-1)\sqrt{2}}
\end{equation}
at $W/V\approx0.2$. In addition, as $W/V$ is raised, the emergence of rather disordered
states in turn leads to the increase of the fluctuations of $\ic/V$, which is reflected as
well in the correlations between $E_{M{+}1}-E_M$ and $E_{M{-}1}-E_M$ shown in
\figref{fig:cfdcor}.  In this regime the correlations are positive and increase with $W/V$
until $W/V\approx0.2$, keeping parallel with the broader distribution of $\ic/V$.  [Recall
that $\ic = (E_{M{+}1}{-}E_M) +(E_{M{-}1}{-}E_M)$.]  Such positive correlations may be
understood via the argument that the presence of a dilute region in state $\ket{M}$, which
lowers $E_{M+1}-E_M$, accompanies a dense region, allowing a low cost of the electron
extraction.  It should be noted that large fluctuations in this regime arise from the
interplay between the Coulomb interaction and the random potential.

The main feature of Region II is that essentially all the states are disordered.  In this
regime the disorder contributions become important in determining the electron
configuration of $\ket{M}$, alleviating the need for rearrangement due to the insertion or
extraction of an electron.  The fluctuations $\delta[E_{M\pm1}-E_M]$, which increase or
decrease very slightly depending on the system size, are not enough to explain the
noticeable decrease of $\icstd/V$ shown in \figref{fig:cicmndv}(b), suggesting that the
correlations between them play a crucial role.  \Figref{fig:cfdcor} shows that the
correlations decrease to negative values.  This behavior, providing an evidence for the
mounting role of the random potential, can be understood as follows: In the noninteracting
case $E_{M{+}1}{-}E_M$ and $E_{M{-}1}{-}E_M$ are simply given by $w_{M{+}1}$ and $-w_M$,
respectively, where $w_n$ is the $n$th smallest strength of the on-site potential in a
disordered sample.  The closeness of $w_{M+1}$ and $w_M$ gives positive correlations
between them, thus leading to negative correlations between $E_{M\pm1}-E_M$.  This
argument is expected to remain valid in the presence of sufficiently weak interactions.
It is remarkable that such correlations indeed begin to show up in the regime of rather
strong interactions.  Furthermore, as can be seen from \figref{fig:cicdis}(c) and (d), the
distribution of $\ic/V$ is bounded below by the minimum of $\ic/V$ given by
\eqnref{eq:cicmin}, while the maximum of $\ic/V$ decreases with $W/V$, accounting for the
corresponding decrease of the fluctuations $\icstd/V$.

Finally, one can observe that Region III is characterized by weak dependence of
$\icavg/V$, $\icstd/V$, and $\icres$ upon the disorder strength.  Although fluctuations of
$E_{M{+}1}{-}E_M$ and of $E_{M{-}1}{-}E_M$ increase with $W/V$, they are canceled out due
to their negative correlations as disclosed in \figref{fig:cfdcor}.  It should also be
noted that the inverse-compressibility distribution follows the WD form throughout this
regime [see \figref{fig:cicdis}(e)] while the universal relative fluctuations reported in
Refs.~\onlinecite{Koulakov97} and \onlinecite{Jeon99} emerge only for $W/V=1$.

Before closing this section, we add a remark on the shape of the distribution of the inverse
compressibility, which is displayed in \figref{fig:cicdis} for several values of $W/V$.
The noninteracting system follows the Pauli distribution, which changes to the WD
distribution for $0.5<W/V<\varO(1)$ as pointed out above.  Further decrease to
$0.05\lesssim W/V \lesssim 0.2$ generates quite a peculiar distribution:
the asymmetric one with a peak biased to the right and a long tail in the left,
which resembles neither the WD nor the Gaussian distribution.
As $W/V$ becomes even smaller, this distribution eventually reduces
to the $\delta$-function distribution, manifesting the WC state.

\section{Quantum Coulomb Glass\label{sec:qs}}

\begin{figure*}
  \parbox{0.32\textwidth}{
    \centerline{\epsfig{width=0.33\textwidth, file=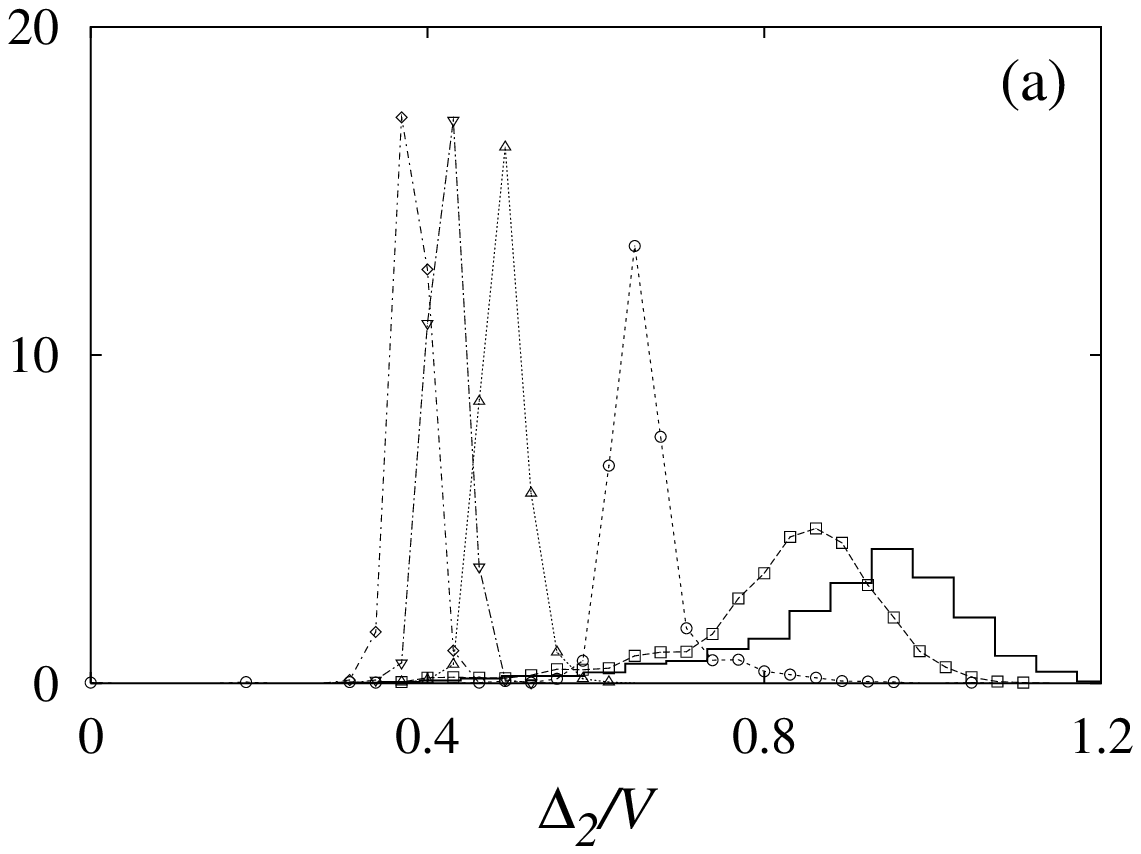}}
  }
  \parbox{0.32\textwidth}{
    \centerline{\epsfig{width=0.33\textwidth, file=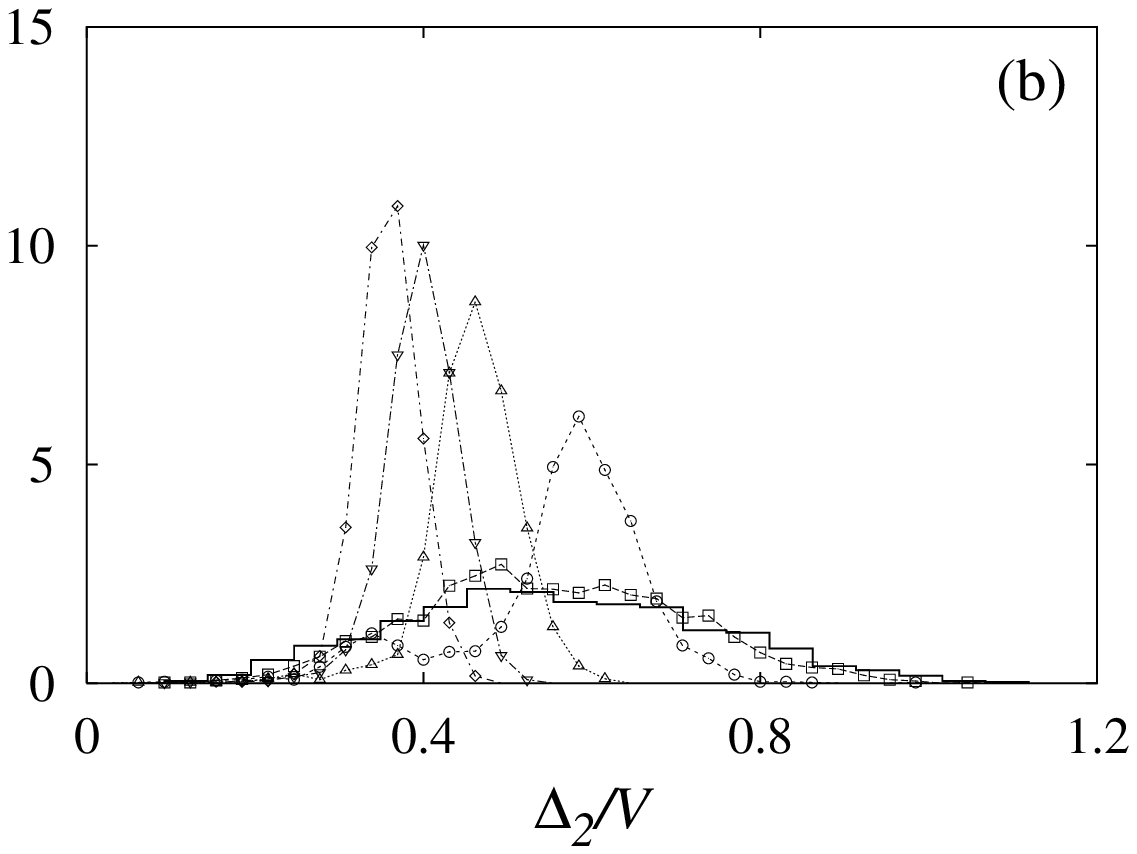}}
  }
  \parbox{0.32\textwidth}{
    \centerline{\epsfig{width=0.33\textwidth, file=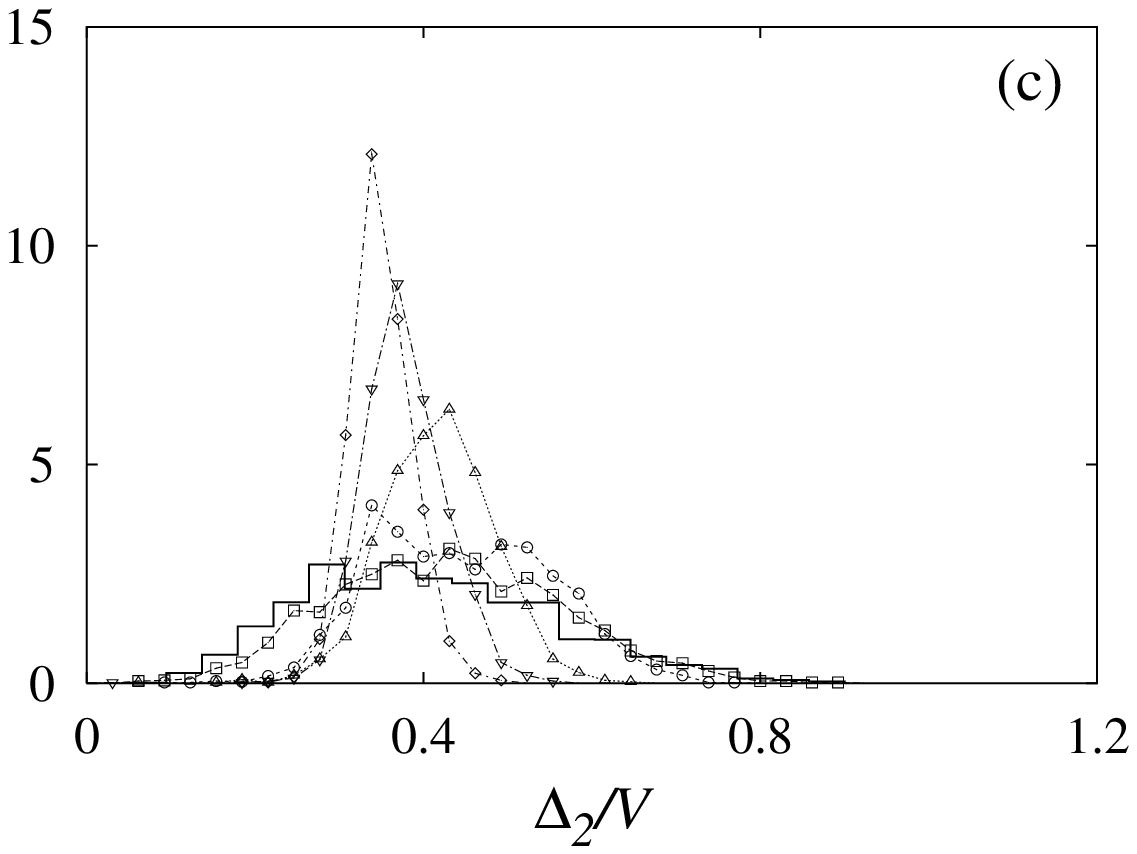}}
  }\\
  \parbox{0.32\textwidth}{
    \centerline{\epsfig{width=0.33\textwidth, file=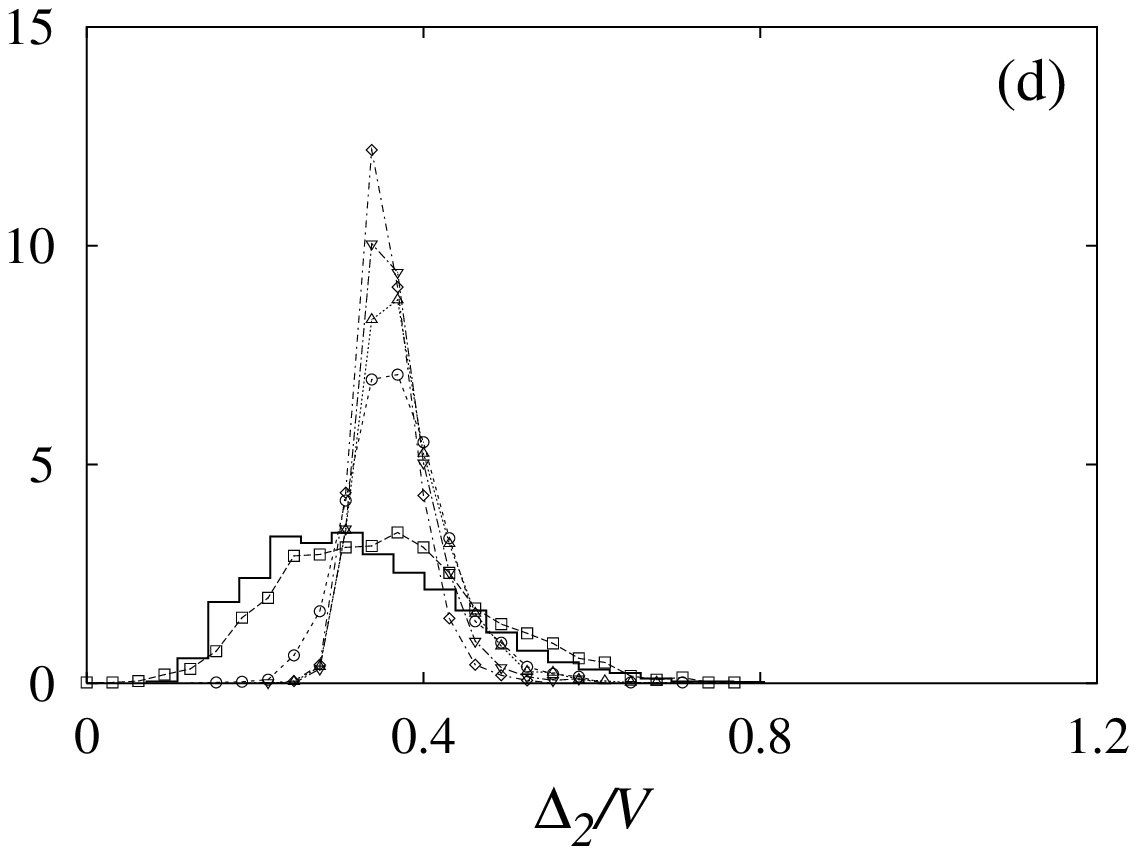}}
  }
  \parbox{0.32\textwidth}{
    \centerline{\epsfig{width=0.33\textwidth, file=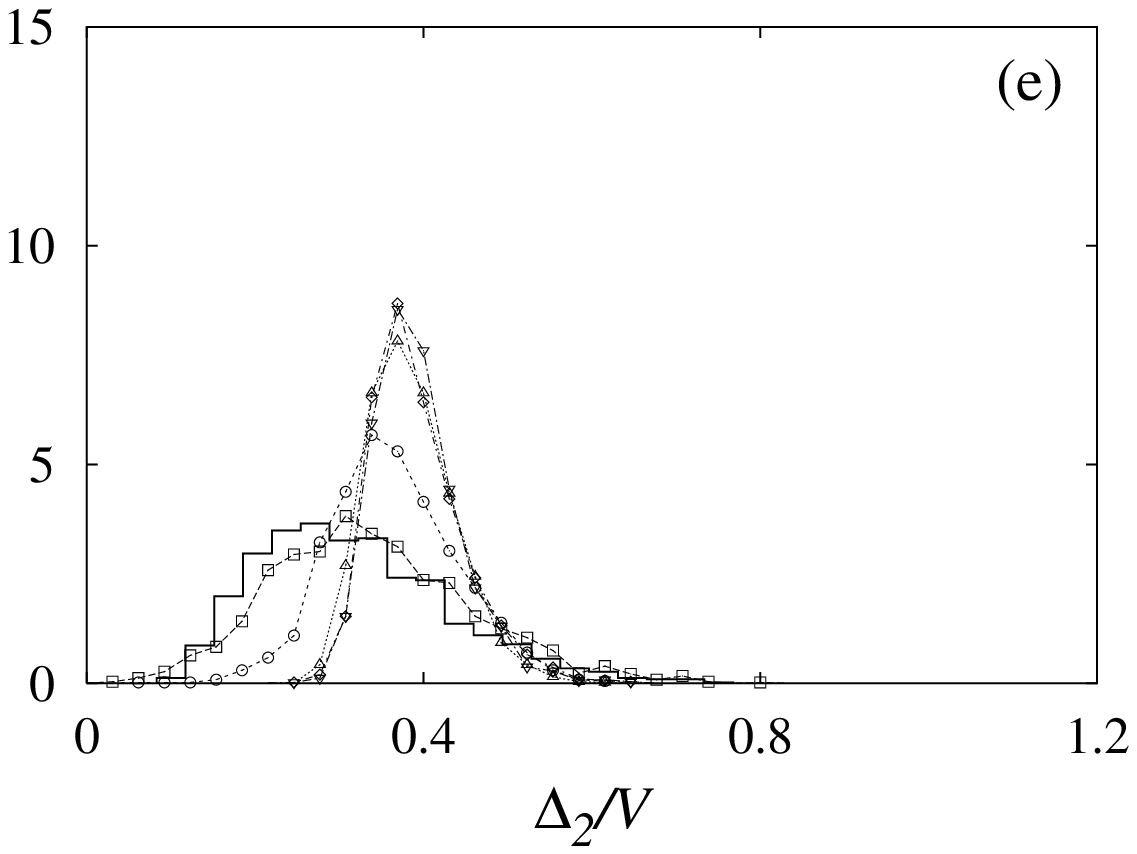}}
  }
  \parbox{0.32\textwidth}{
    \centerline{\epsfig{width=0.33\textwidth, file=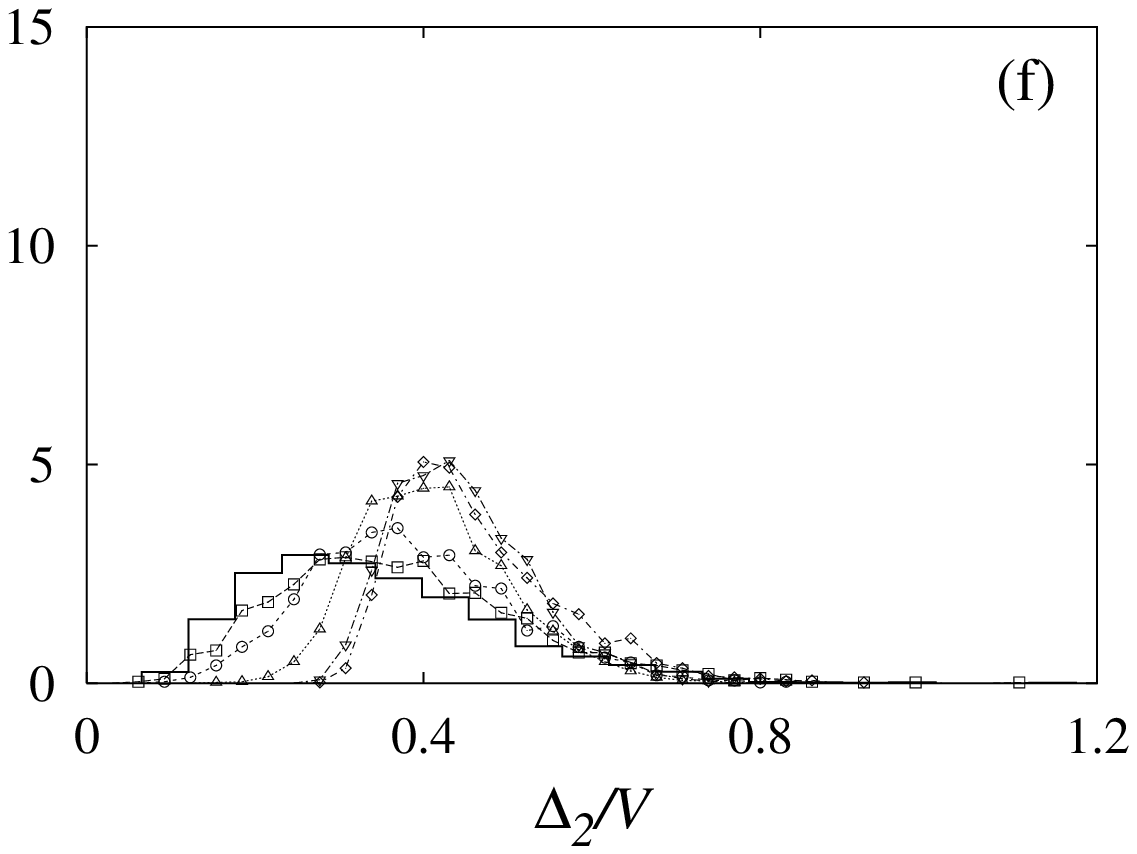}}
  }
  \caption{Distributions of $\ic/V$ in the quantum system of $L=8$ for $W/V=$ (a) $0.1$,
    (b) $0.2$, (c) $0.3$, (d) $0.5$, (e) $1$, and (f) $2$.  For comparison, the
    distribution in the classical system ($t/V=0$) is plotted by a thick solid line in
    each figure.  The data marked by squares, circles, triangles, inverted triangles, and
    diamonds correspond to the hopping strength $t/V = 0.1, 0.5, 1, 2$, and $10$,
    respectively.}
  \label{fig:ichdis}
\end{figure*}
In this section we allow hopping of electrons between sites and investigate the corresponding
quantum effects.  The Hamiltonian for such a quantum Coulomb glass reads\cite{quantum,jeon2}
\begin{eqnarray}
  \nonumber
  \varH & = & t \sum_{\nn{i,j}}(\hat{c}_i^\dagger \hat{c}_j
  + \hat{c}_j^\dagger \hat{c}_i ) + \sum_i w_i \hat{n}_i \\
  \label{eq:qH}
  & & \mbox{} + \half \sum_{i\ne j} (\hat{n}_i - K) U_{ij} (\hat{n}_j - K),
\end{eqnarray}
where $\hat{c}_i^\dagger$/$\hat{c}_i$ is the creation/annihilation operator at site $i$
and the number operator is given by $\hat{n}_i \equiv \hat{c}_i^\dagger \hat{c}_i$.  The
hopping of an electron is allowed only between nearest neighboring sites as indicated by
the first term.  We employ the HF approximation to treat the Coulomb interaction in
\eqnref{eq:qH}, and obtain
\begin{eqnarray}
  \nonumber
  \varH_{\rm HF} & = & t \sum_{\nn{i,j}}(\hat{c}_i^\dagger \hat{c}_j
  + \hat{c}_j^\dagger \hat{c}_i ) + \sum_i w_i \hat{n}_i \\
  \nonumber
  & & \mbox{}
  + \sum_{i\ne j} (\langle \hat{n}_j\rangle - K) U_{ij} \hat{n}_i
  - \sum_{i\ne j} \langle \hat{c}_j^\dagger \hat{c}_i \rangle U_{ij}
  \hat{c}_i^\dagger \hat{c}_j ,
\end{eqnarray}
apart from constant terms.  The third and the fourth terms represent the direct and
the exchange energies of the electron-electron interactions, respectively, and the angular
brackets denote the expectation values with respect to the HF ground state, which should
be determined in a self-consistent manner.  During the iteration we use the method of
potential mixing, where the averages over a few previous iterating steps are inserted into
the expectation values in the Hamiltonian; this not only ensures convergence of the iteration
but also helps to reduce iteration steps necessary for convergence.

\begin{figure}
  \centerline{\epsfig{width=8cm,file=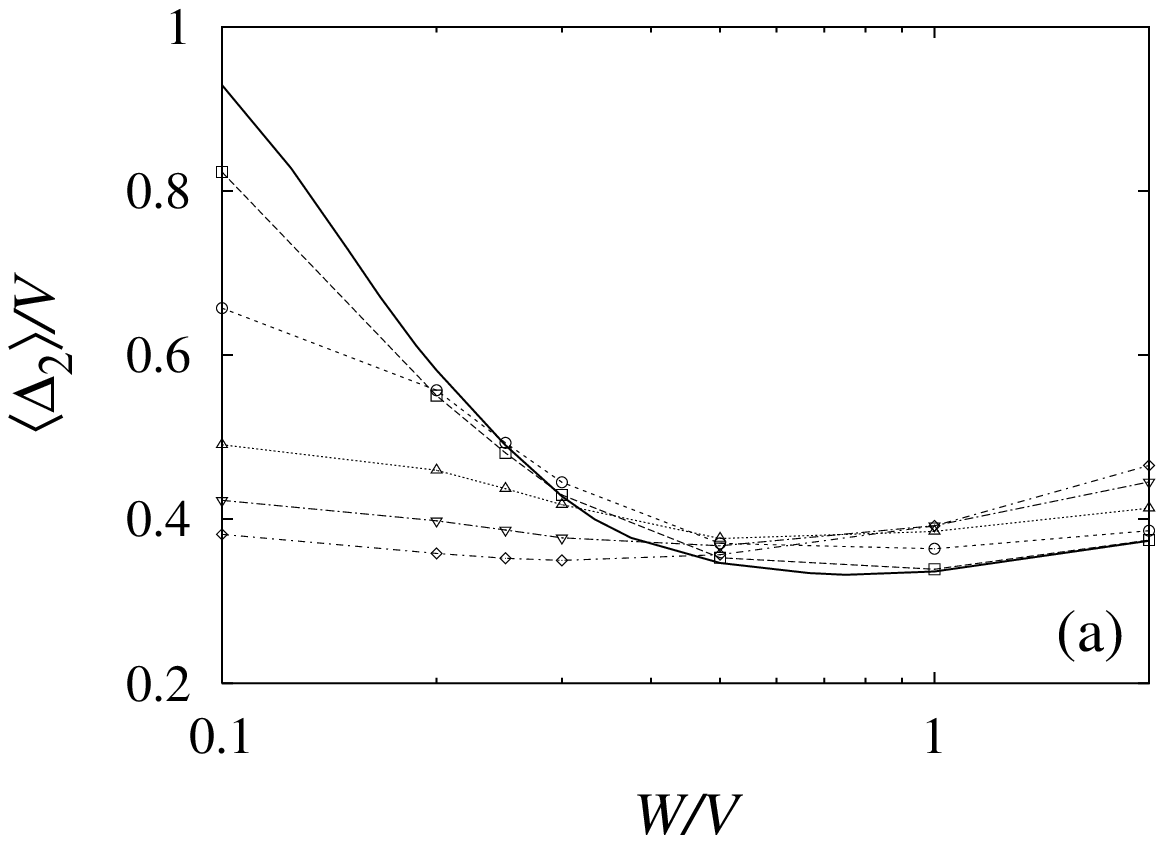}}
  \centerline{\epsfig{width=8cm,file=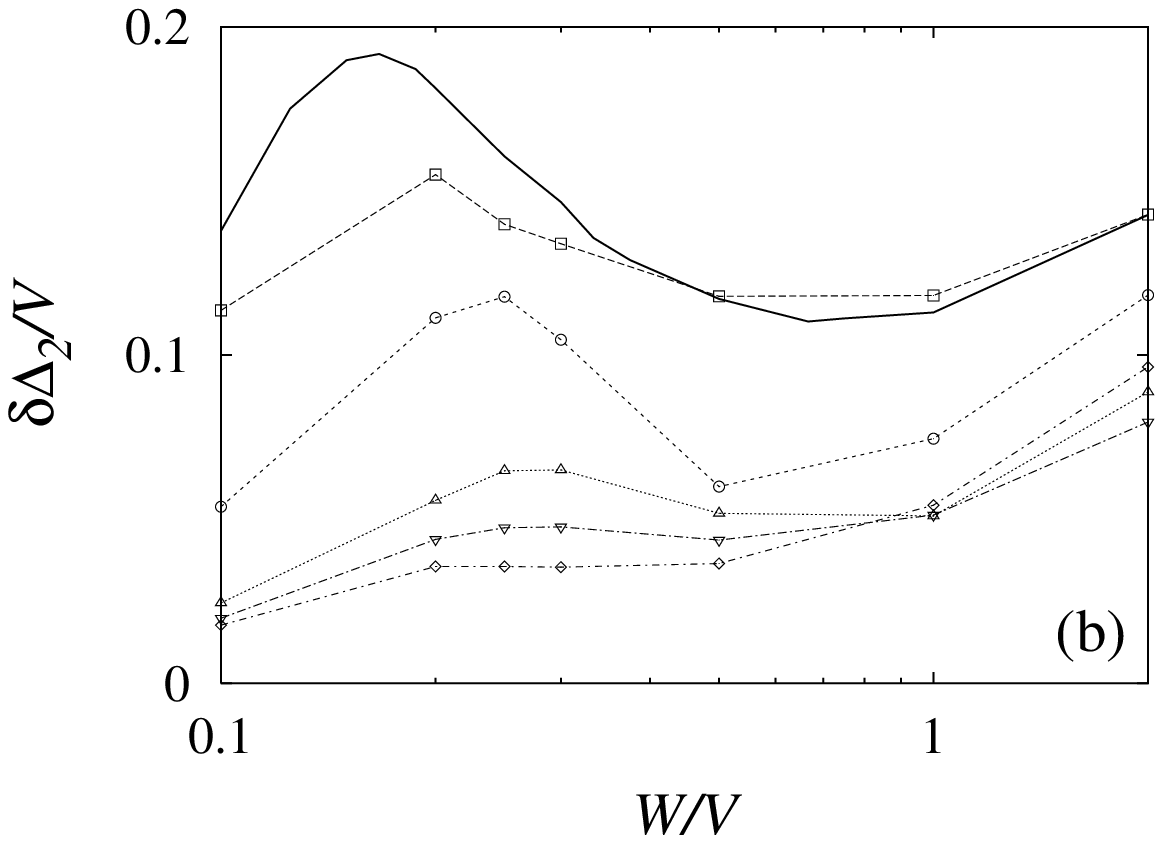}}
  \centerline{\epsfig{width=8cm,file=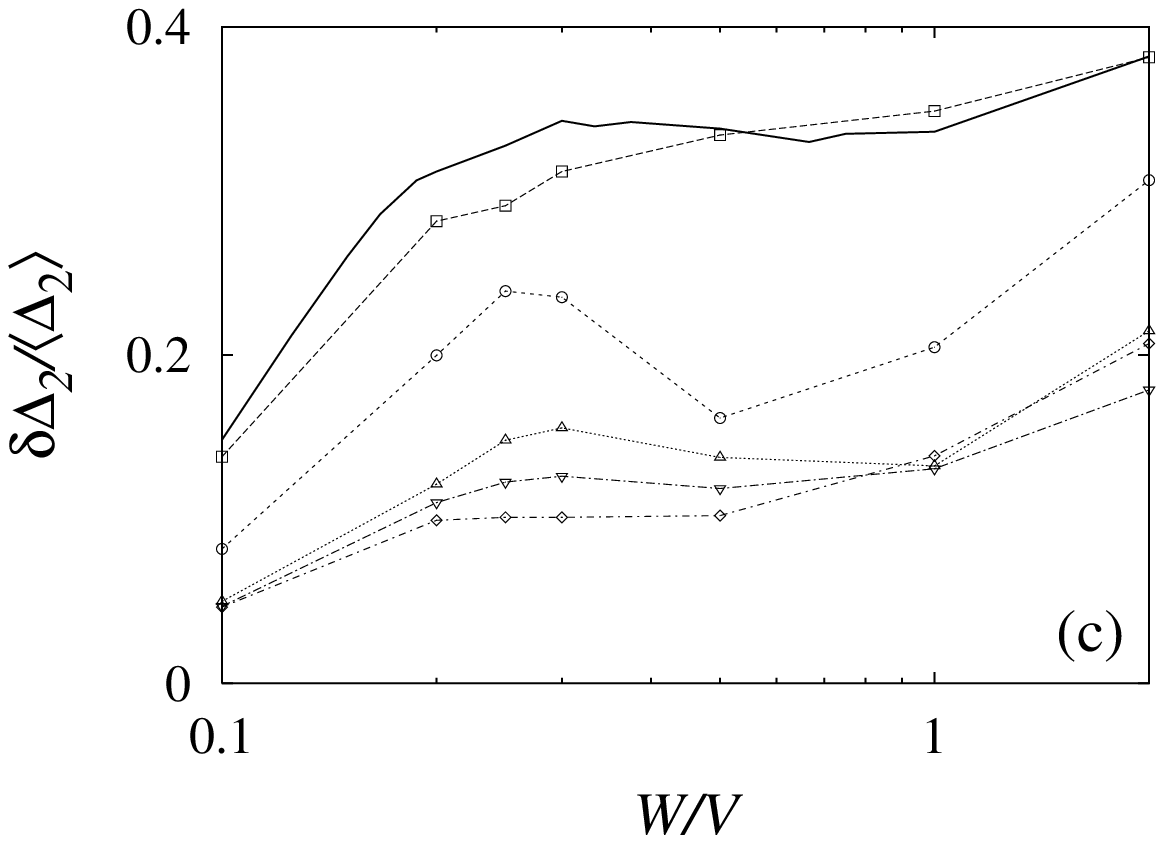}}
  \caption{(a) Average, (b) fluctuations, and (c) relative fluctuations of the
    inverse compressibility as functions of $W/V$ in the quantum system of size $L=8$.
    The data marked by squares, circles, triangles, inverted triangles, and diamonds
    correspond to the hopping strength $t/V=0.1,\, 0.5,\, 1,\, 2$, and $10$, respectively,
    while the thick lines represent the data in the classical limit ($t/V=0$).}
  \label{fig:icv}
\end{figure}
We have computed the inverse compressibility up to 2000 disorder configurations for
several values of the hopping strength $t/V$ and the disorder strength $W/V$.
\Figref{fig:ichdis} displays the distribution of the inverse compressibility for various
values of $t/V$ and $W/V$, together with the distribution in the corresponding classical
system.  For small $W/V$ shown in \figsref{fig:ichdis}(a), (b), and (c), as electron
hopping comes into play, the peak position of the distribution moves toward smaller values
of $\ic$, with the probability for large $\ic$ reduced.  One can see that for $W/V=0.1$
the presence of even very weak hopping produces a noticeable change in the distribution.
\begin{figure*}
  \parbox{0.32\textwidth}{
    \centerline{\epsfig{width=0.33\textwidth,file=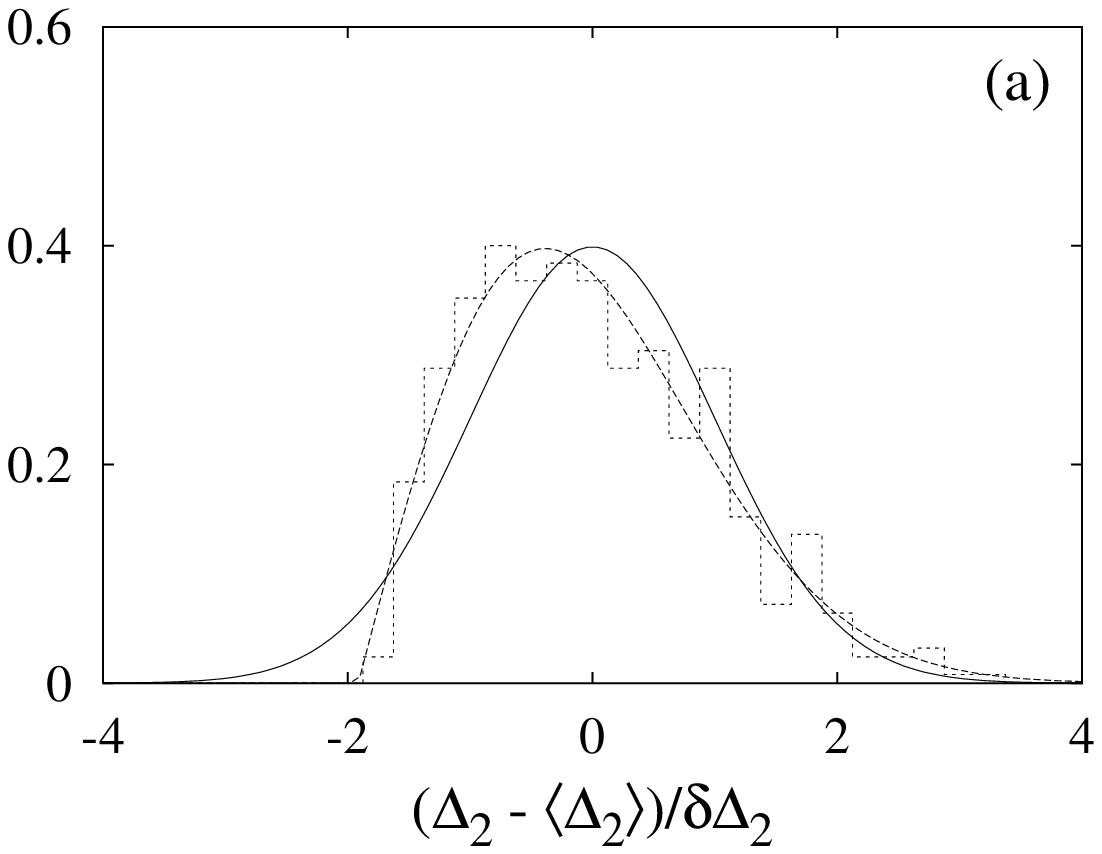}}
  }
  \parbox{0.32\textwidth}{
    \centerline{\epsfig{width=0.33\textwidth,file=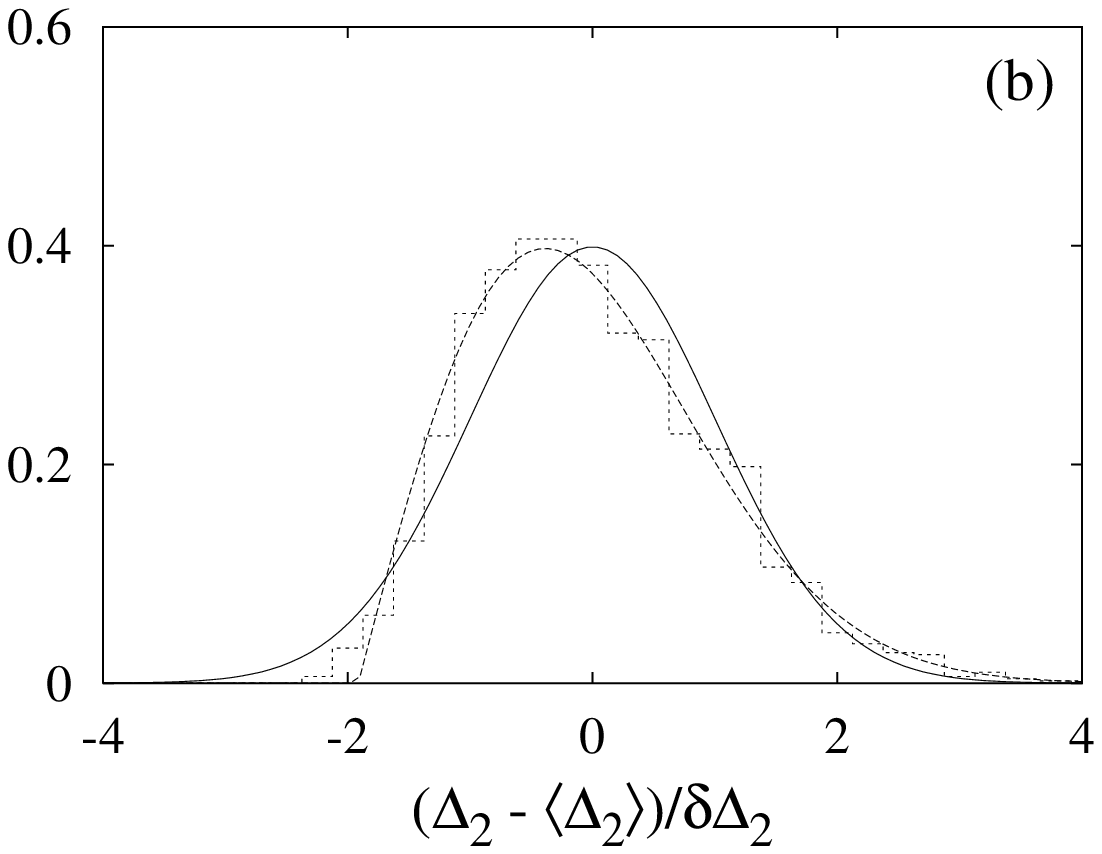}}
  }
  \parbox{0.32\textwidth}{
    \centerline{\epsfig{width=0.33\textwidth,file=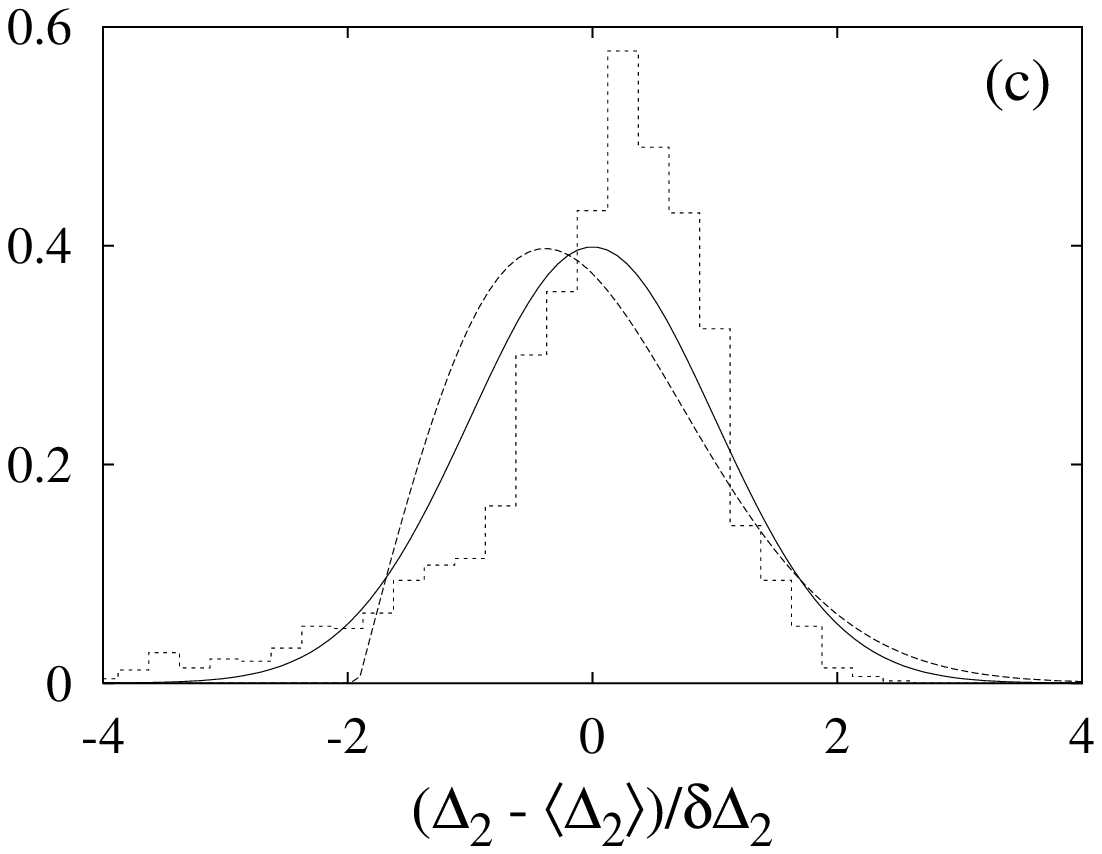}}
  }\\
  \parbox{0.32\textwidth}{
    \centerline{\epsfig{width=0.33\textwidth,file=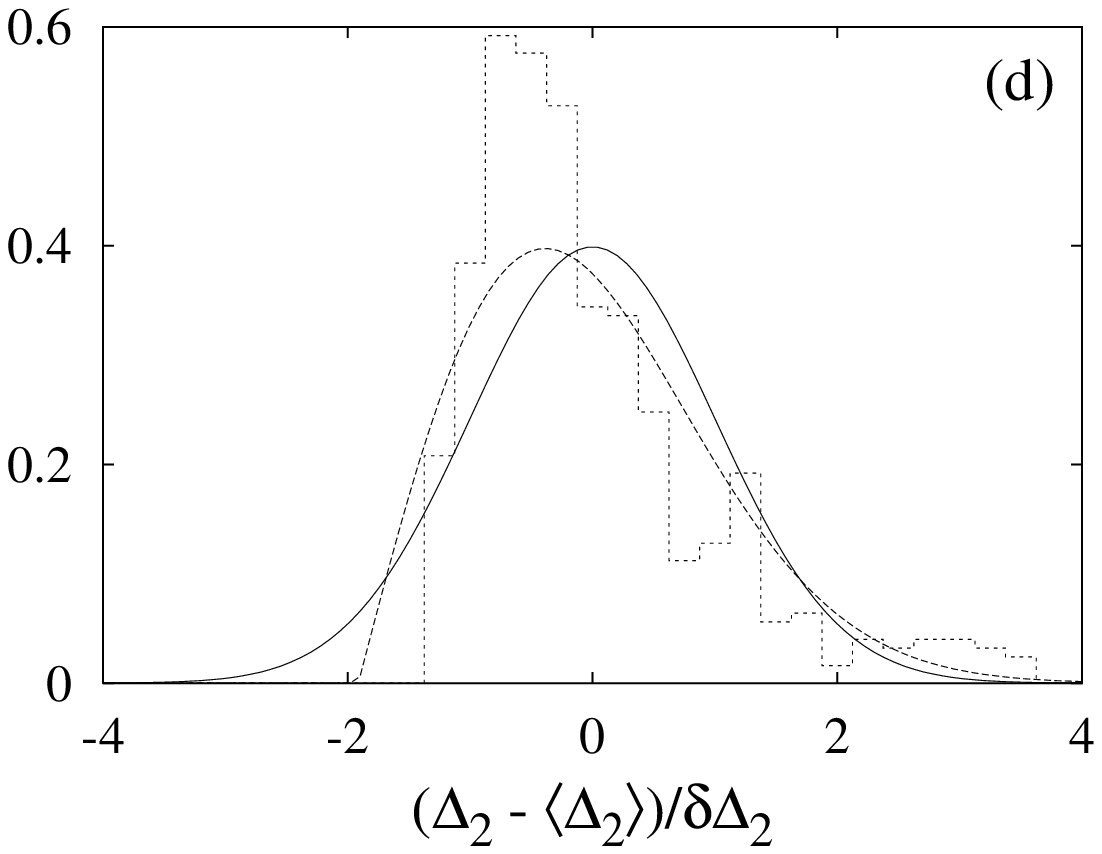}}
  }
  \parbox{0.32\textwidth}{
    \centerline{\epsfig{width=0.33\textwidth,file=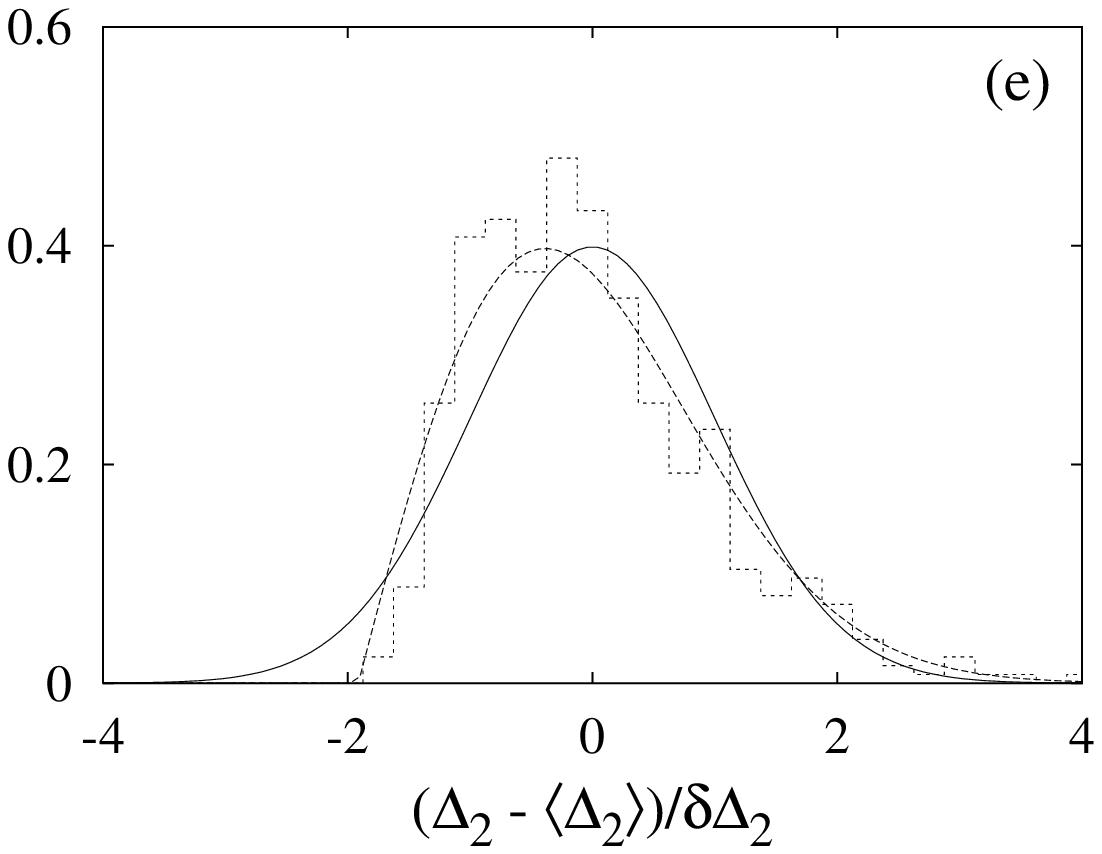}}
  }
  \parbox{0.32\textwidth}{
    \centerline{\epsfig{width=0.33\textwidth,file=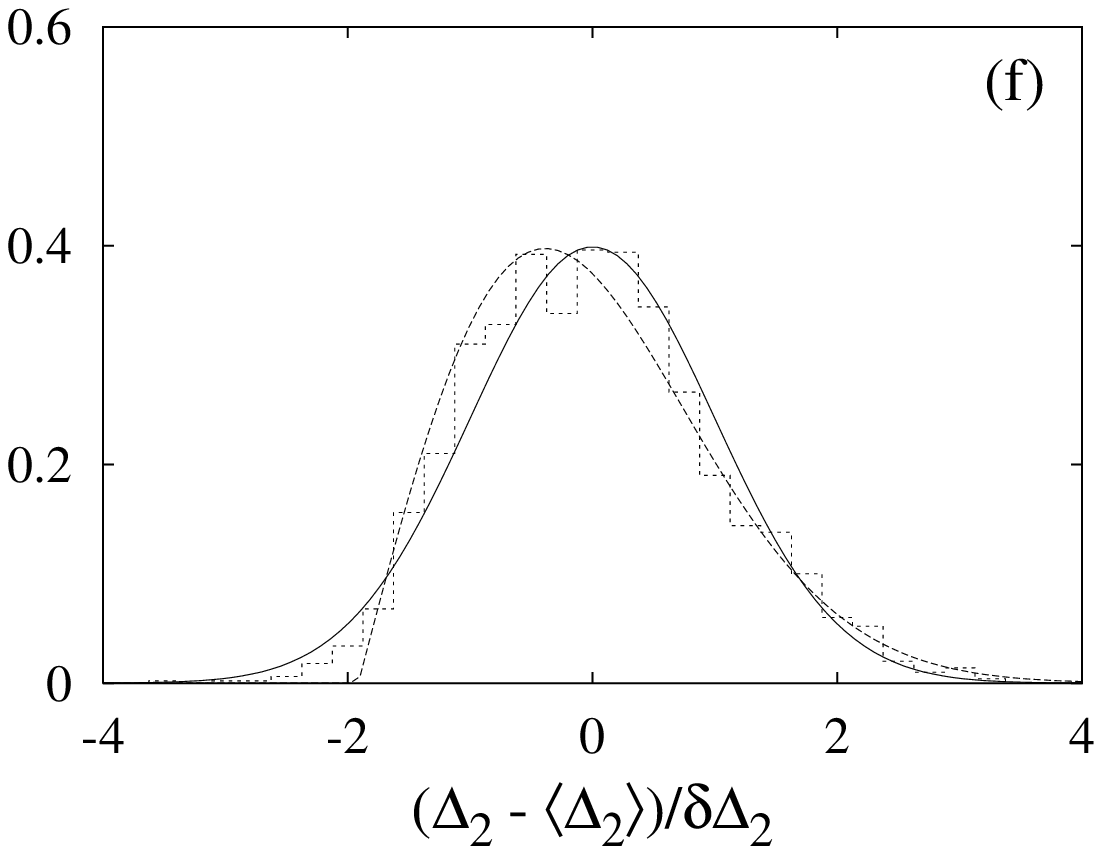}}
  }
  \caption{Distributions of the normalized inverse compressibility in the quantum
    system of $L=8$ for $W/t=1$ and $V/t=$ (a) $0.1$, (b) $1$, (c) $10$; for
    $W/t=5$ and $V/t=$ (d) $0.1$, (e) $1$, (f) $10$.  The solid and the dashed curves
    represent the best fit to the normalized Gaussian and to the WD distributions,
    respectively.}
  \label{fig:icvdis}
\end{figure*}
Indeed the clean system ($W=0$) is observed to show rapid decrease of the inverse
compressibility with the increase of the hopping strength, suggesting that the decrease of
the probability for large $\ic$ originates from the interplay between the Coulomb
interaction and hopping.  \Figref{fig:ichdis}(d), (e), and (f) corresponding to
$W/V\ge0.5$, in contrast, demonstrates that the peak shifts slightly to the right and
appears to saturate eventually as the hopping strength is increased.  In addition, the
probability for small $\ic$ is suppressed and the minimum of $\ic$ rises, which appears to
begin at somewhat weaker disorder $W/V \approx 0.2$.  Recalling that such suppression in
the probability for small $\ic$ is also observed in the noninteracting system, we can
infer that random disorder plays a crucial role here.  The effects of electron hopping are
also manifested by the statistics of the inverse compressibility, which is presented in
\figref{fig:icv}.  For $W/V<0.5$, both $\icavg$ and $\icstd$ decrease with $t/V$ due to
the reduction of the probability for large $\ic$; for $W/V>0.5$, on the other hand, the
suppression of small $\ic$ induces slight increase of $\icavg$ and decrease of $\icstd$,
which results in overall suppression of $\icres$ in the whole region.

\Figref{fig:icvdis} illustrates the interaction effects on the inverse-compressibility
distribution for given hopping and disorder strength.  As the interaction strength is
increased, the peak in the distribution of the normalized inverse compressibility
$(\ic-\icavg)/\icstd$ moves to the right both in the clean case ($W/t=1$) and in the dirty
case ($W/t=5$).  Here it is of interest to notice the growth of a tail in the left side.
The increase of the interaction modifies the WD-like distribution to a symmetric one and
even drives the distribution into a right-biased shape at sufficiently strong interactions
[see \figref{fig:icvdis}(c)].  The distribution returns to a symmetric narrow one upon
further increase of the interaction.  Strong disorder in dirty samples, on the other hand,
tends to suppress such influence of the interaction: Growth of the left tail is retarded
and the distribution remains in the left-biased shape even at strong interactions.

\begin{figure}
  \centerline{\epsfig{width=8cm,file=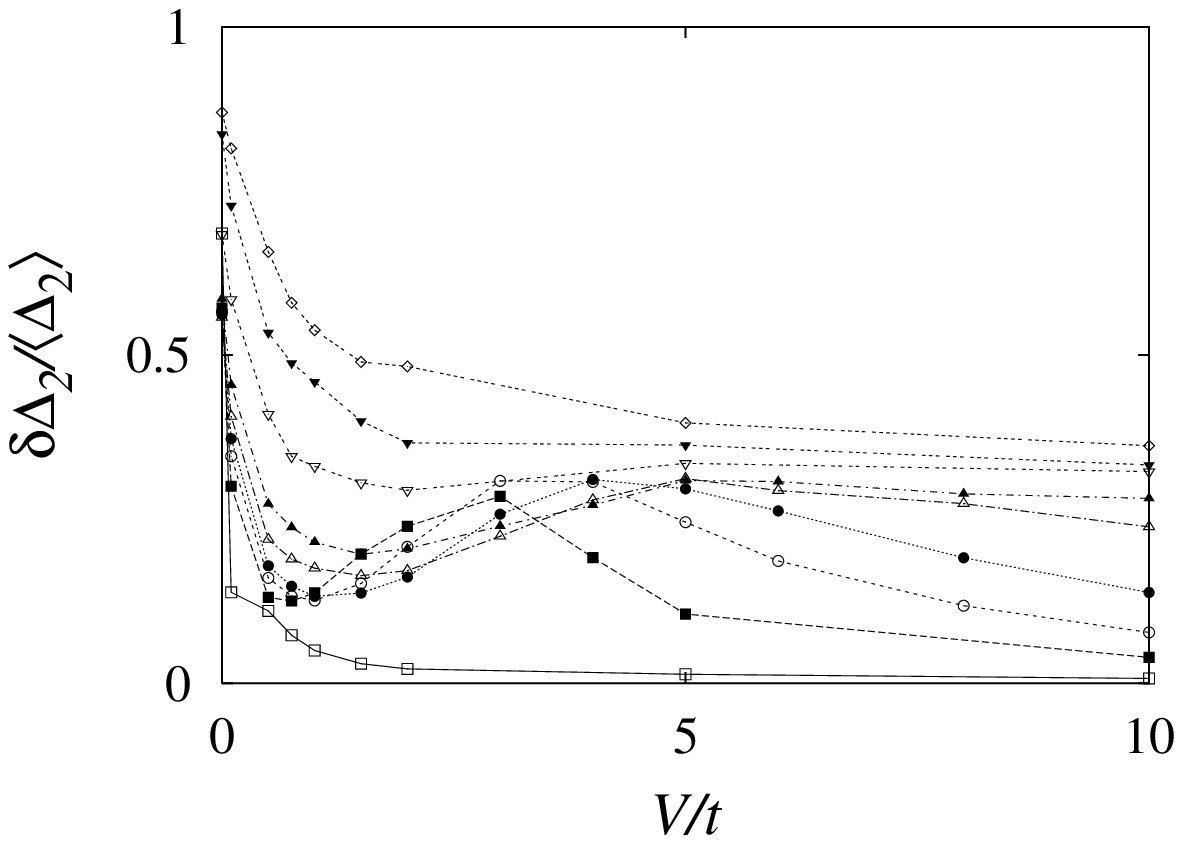}}
  \caption{Relative fluctuations of the inverse compressibility versus $V/t$ in
    the quantum system of $L=8$ for various values of $W/t=$ 0.1($\square$),
    0.5($\blacksquare$), 0.75($\circ$), 1($\bullet$), 1.5($\triangle$),
    2($\blacktriangle$), 4($\triangledown$), 6($\blacktriangledown$), and 10($\diamond$).}
  \label{fig:ichres}
\end{figure}
Several experiments\cite{Sivan96,Simmel99,Patel98} reported that the distribution of the
normalized inverse compressibility follows a roughly symmetric Gaussian form with
non-Gaussian tails, which is very different from the WD distribution predicted by the
random matrix theory.  The interaction strength in those experiments is characterized by
the dimensionless parameter $r_s \approx 1$ to $2$, which, via the relation $r_s =
V/(2t\sqrt{\nu\pi})$ with the filling factor $\nu=1/2$, corresponds to $V/t \approx 2$ to
$5$ in our model.
\begin{figure*}
  \parbox{0.45\textwidth}{
    \centerline{\epsfig{width=8cm,file=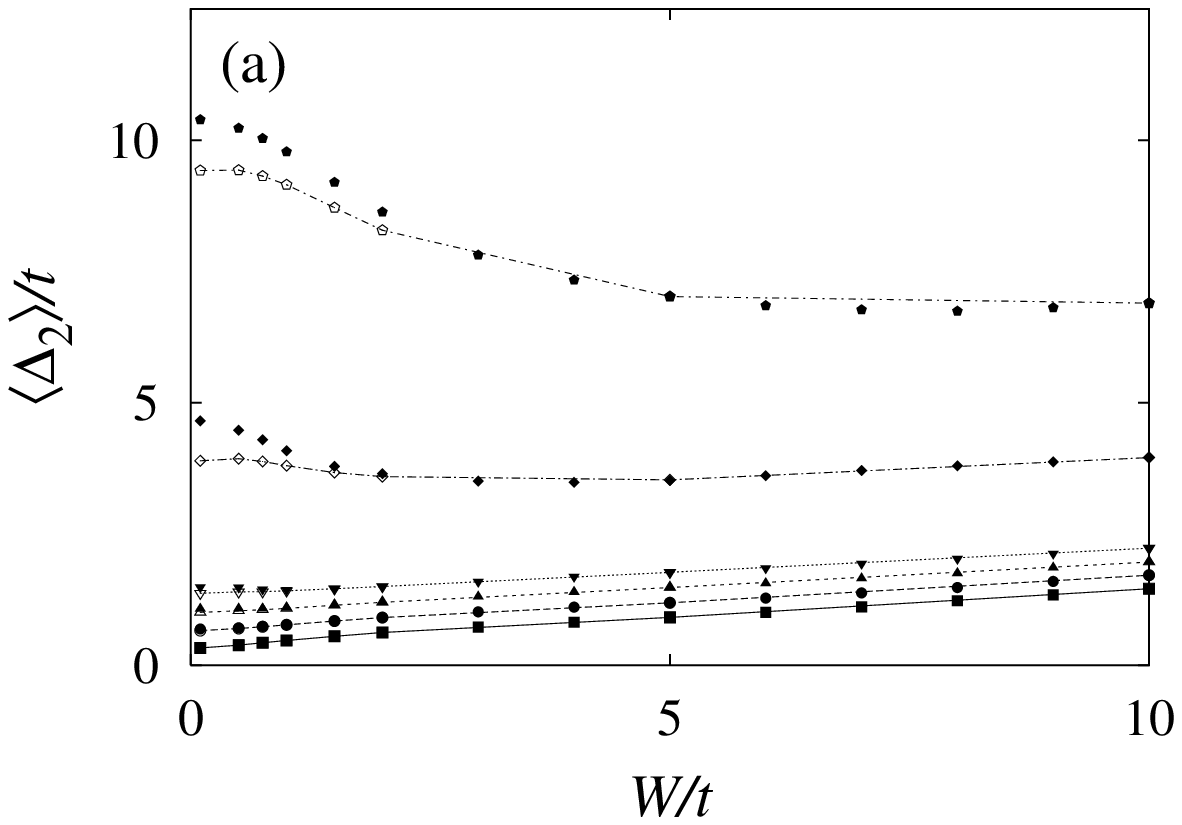}}
    \centerline{\epsfig{width=8cm,file=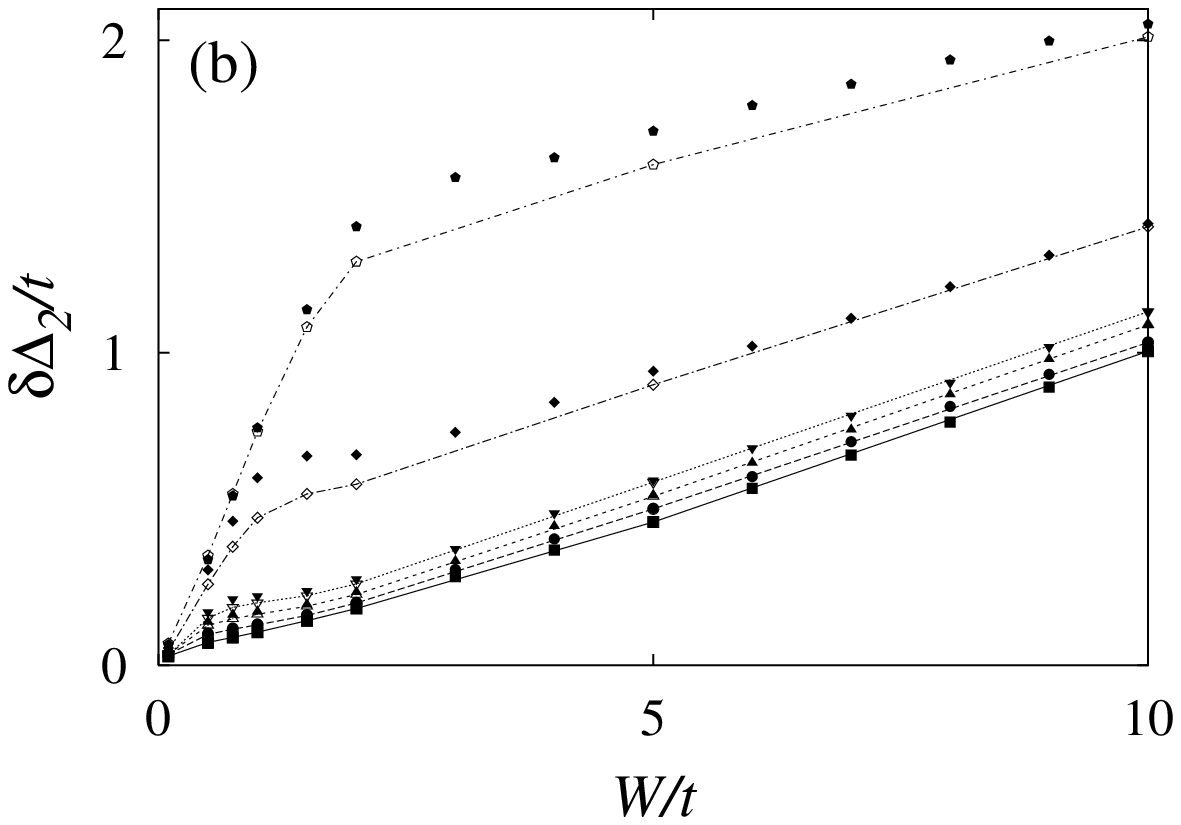}}
  }
  \hspace{0.05\textwidth}
  \parbox{0.45\textwidth}{
    \centerline{\epsfig{width=8cm,file=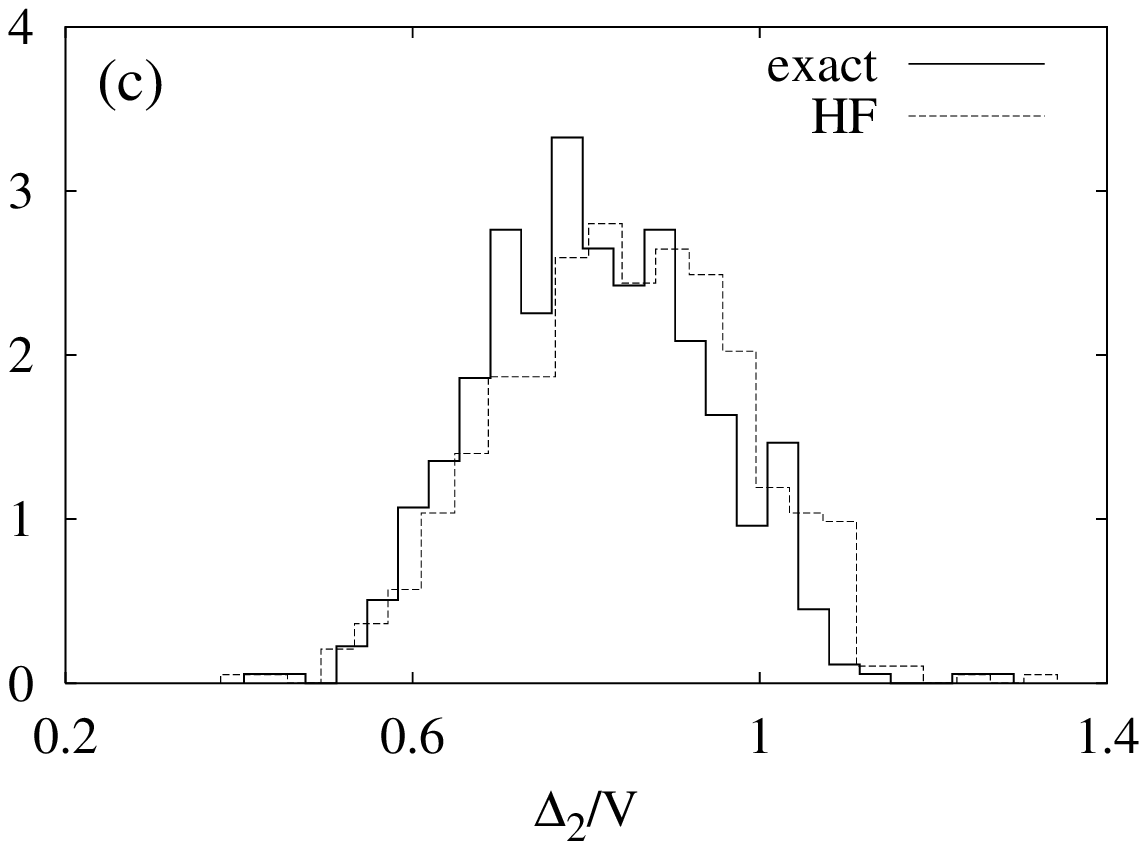}}
    \centerline{\epsfig{width=8cm,file=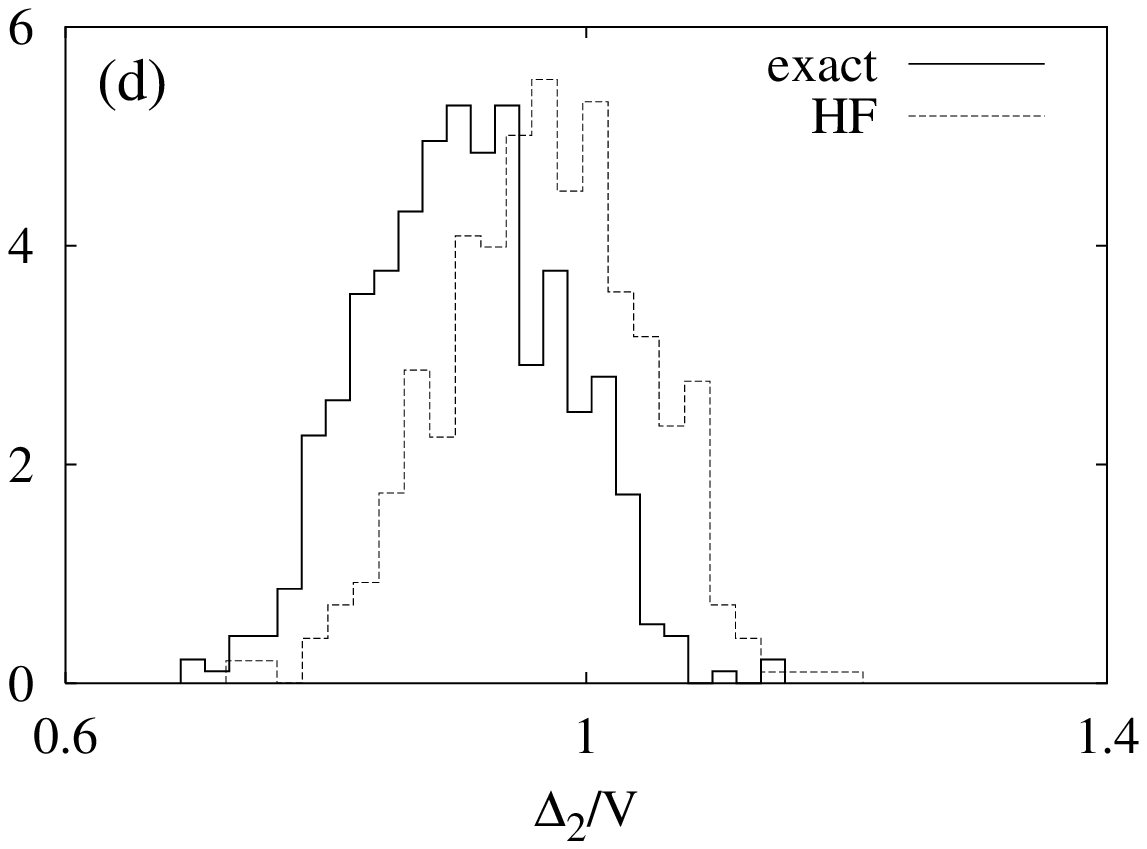}}
  }
  \caption{Comparison of the HF results (filled symbols) and the exact ones
    (empty symbols with lines) for (a) the average and (b) fluctuations of the inverse
    compressibility in the quantum system of $L=4$ for various values of $V/t = 0.5,\, 1,\, 1.5,\,
    2,\, 5$, and $10$ from below; comparison of the inverse-compressibility distribution for
    $V/t=10$ and $W/V= $ (c) $0.2$ and (d) $0.1$. The discrepancy between the two results
    is considerable in (d).}
  \label{fig:ln}
\end{figure*}
In this interaction range and also at weaker interactions we have identified such a
Gaussian-like symmetric distribution as long as disorder is sufficiently weak
($W/V\lesssim0.3$).  In addition to this symmetric distribution, we have also observed an
interesting asymmetric right-biased distribution with long left tails for $0.05<W/V<0.2$
and $t/V \lesssim0.2$ [see Fig.~7(c)].  Indeed, such an asymmetric distribution, which
also appears in Fig.~3(b) for the classical system, has been observed
experimentally\cite{Simmel99} in the samples with strong interactions $r_s\approx 2.1$ and
rather weak disorder. [See Fig.~3 in Ref.~\onlinecite{Simmel99}.]  Furthermore, we have
obtained in this regime relative fluctuations smaller than the values $0.1 {\sim} 0.2$
obtained in previous numerical calculations;\cite{Sivan96,Koulakov97,Jeon99} this allows
one to explain the weak relative fluctuation $\icres\approx0.06$ observed in
experiment\cite{Simmel99} without considering thermal broadening.  All these observations
thus support the conclusion that the peak spacing fluctuations in the Coulomb blockade
conductance originate mainly from charging energy fluctuations.

We also investigate the effects of the interaction on the statistics of the inverse
compressibility by computing the relative fluctuations, which are shown as a function of
$V/t$ for several values of $W/t$ in \figref{fig:ichres}.  It is observed that
particularly for intermediate disorder strengths a broad peak develops at a certain value
of $V/t$, which turns out to increase with $W/t$.  The decrease of $\icres$ at
sufficiently strong interactions has its origin in the fast increase of $\icavg$ as well
as the decrease of $\icstd$ with $V/t$ in this regime, where the half-filling state is
close to the WC state.  Note here that such decrease was not observed in
Ref.~\onlinecite{Walker99b}, which investigated the quarter-filled system and reported
quadratic increase of the relative fluctuations with $V/t$ at strong interactions.  In
general stronger interactions are required for systems with smaller filling factors to
form WC states, and accordingly it is expected that relative fluctuations begin to
decrease at larger values of $V/t$.  We thus believe that further increase of the
interaction strength should yield reduction of the relative fluctuations even in the
quarter-filled case.

Finally, we check the validity of the HF approximation by comparing the HF results with
those obtained via exact diagonalization of the Hamiltonian in \eqnref{eq:qH} for small
systems.  In \figref{fig:ln}(a) and (b), which displays the statistics of $\ic$ obtained
via the two methods in the system of $L=4$, one can see that the HF results are in good
agreement with the exact ones except for some parameter regions: For $V/t >1$, observed in
\figref{fig:ln}(a) is that the HF approximation overestimates $\icavg$ for small $W/V\,
(\lesssim0.3)$.  \Figref{fig:ln}(b) also shows such overestimation of the fluctuations in
the HF approximation for the intermediate disorder strength $0.1\lesssim W/V\lesssim1$ and
$V/t>1$.  The distribution function plotted in \figref{fig:ln}(c) and (d) provides a clue
to the overestimation: For $W/V=0.2$ corresponding to moderate disorder, the HF
approximation generates more frequently large values of $\ic$ and hardly changes the
minimum of $\ic$, resulting in the overestimation of both $\icavg$ and $\icstd$.  On the
other hand, for smaller $W/V=0.1$ the overall distribution merely shifts to the right with
the shape almost unchanged.  Accordingly, whereas the HF approximation tends to give
larger values of $\icavg$, rather accurate values are obtained for $\icstd$ in the case of
small $W/V$, as shown in \figref{fig:icv}.

\section{Effects of Magnetic Fields\label{sec:mf}}

In this section we examine the effects of magnetic fields on the distribution and
fluctuations of the inverse compressibility.  Electrons hopping in applied magnetic
fields acquire additional phases, leading to the kinetic term in the Hamiltonian
\begin{equation}
  H_K = t \sum_{\nn{i,j}}(e^{-iA_{ij}} c_i^\dagger c_j + e^{-iA_{ji}}c_j^\dagger c_i).
\end{equation}
The phase $A_{ij}$ associated with the hopping between sites $i$ and $j$ is given by
\begin{equation}
  \nonumber
  A_{ij}  =  \left\{
    \begin{array}{l l}
      \pm 2 \pi f y_i & \hbox{for } j = i \pm \ux  \\
      0 & \hbox{for } j = i \pm \uy
    \end{array}
  \right.
\end{equation}
in the Landau gauge, where $f \equiv Ba^2/\Phi_0$ is the frustration parameter in the presence
of a uniform transverse magnetic field $\bfB=B \uz$.  We also adopt
the free boundary conditions, which are convenient for describing the system with arbitrary
values of the frustration parameter (i.e., under arbitrary magnetic fields)
without any mismatch on the boundaries.  Periodic boundary conditions in general cannot avoid
such mismatch except for rather low-lying rational values of the frustration
restricted by the system size.

Some experimental groups\cite{Sivan96,Patel98} have measured the conductance peak spacing
in the presence of finite magnetic fields, reporting that the distribution of $\ic$
remains symmetric and Gaussian-like, as in the case of zero magnetic field, with slightly
smaller width: $(\icres)_{B=0}(\icres)_{B\ne0}^{-1} = 1.2\pm0.1$.\cite{Patel98} Note that
those experiments probed the weak-field regime: The magnetic flux per electron in units of
the flux quantum roughly amounts to $0.08$,\cite{Sivan96} or $0.002$ to $0.04$.\cite{Patel98}

Motivated by this, we first concentrate on the weakly frustrated systems, where magnetic
fields are weak enough not to change significantly the characteristics around the Fermi
levels of corresponding noninteracting clean systems.  Such a case is given by the
frustration range $0<f<f_{d1}$, where $f_{d1}$ is the smallest nonzero value of the {\em
  degenerate frustration} defined below.  We then consider the systems with large values
of the frustration, especially the fully frustrated system with $f=1/2$.

\begin{figure}
  \centerline{\epsfig{width=8cm,file=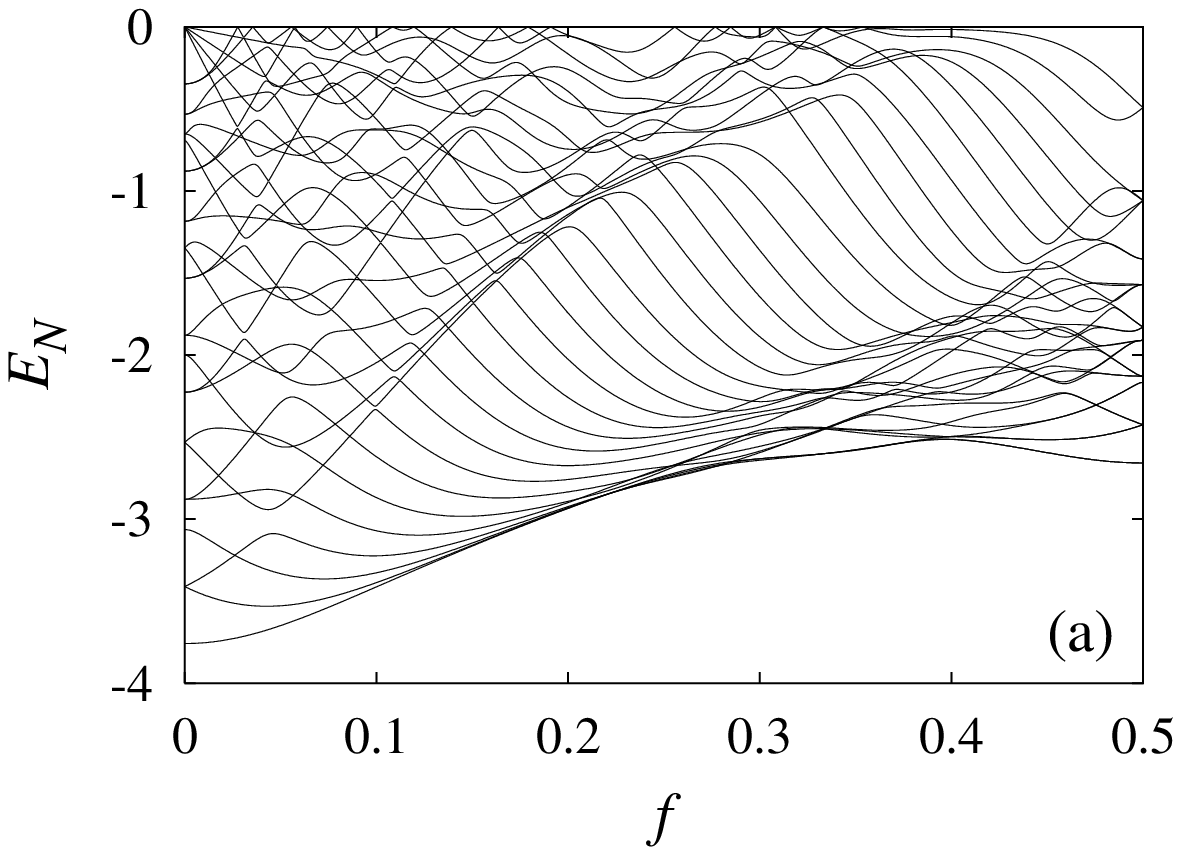}}
  \centerline{\epsfig{width=8cm,file=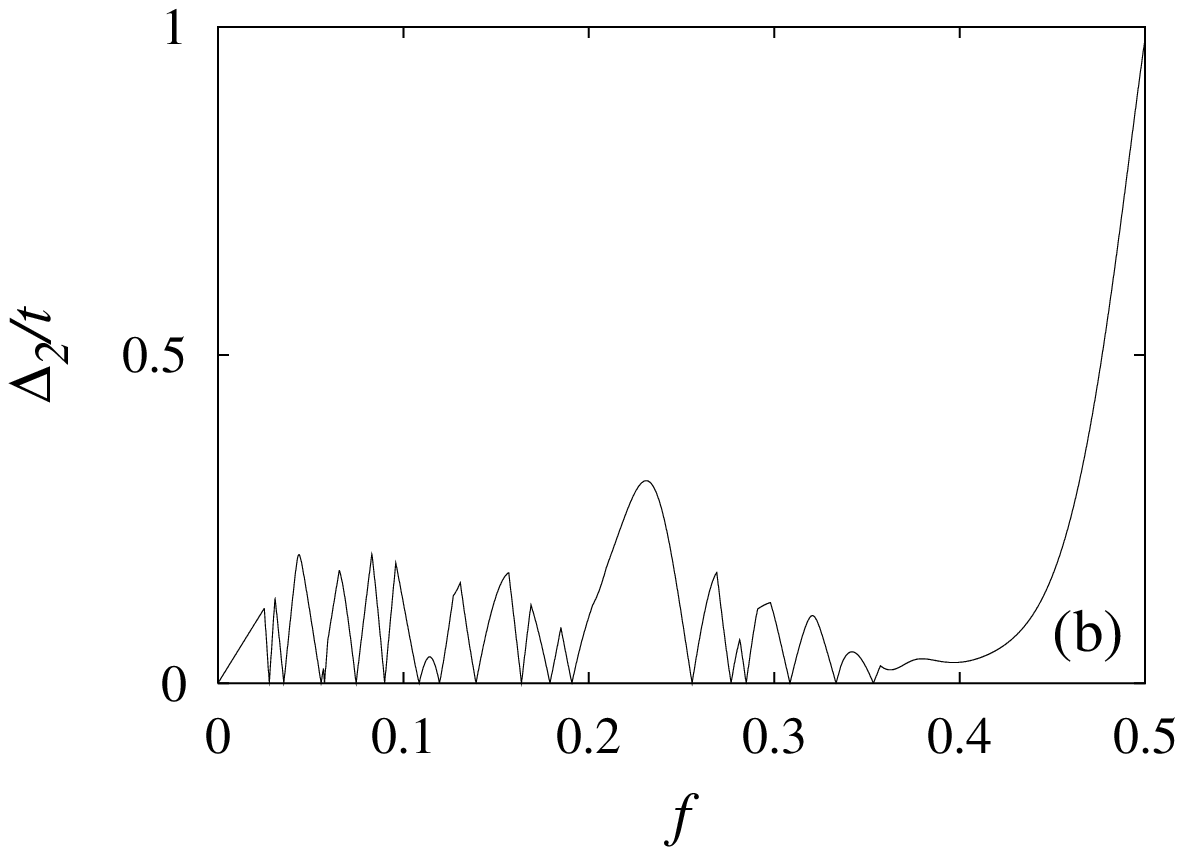}}
  \caption{(a) Single-particle energy spectrum and (b) the inverse compressibility versus
    frustration in the half-filled noninteracting clean system of size $L=8$.
    Since the spectrum possesses reflection symmetry with respect to the $E=0$
    line, only lower half of the spectrum is shown here.}
  \label{fig:free}
\end{figure}
For clear identification of the magnetic-field effects, we first investigate the system of
free electrons on a lattice, in the presence of a magnetic field.  It is well known that
in two dimensions the system is described by Harper's equation, displaying the
characteristic spectrum of a very complex pattern.\cite{Harper} In \figref{fig:free} we
show the single-particle energy spectrum and the corresponding inverse compressibility in
the half-filling case.  It is observed that the Fermi level becomes degenerate at some
values of the frustration, leading the inverse compressibility to vanish.  As the system
size grows, the number of such degenerate frustrations, denoted by $f_d$, increases
rapidly and they spread out over the whole interval $[0,1/2)$ except near the full
frustration ($f=1/2$).

A recent study\cite{Jeon99} has revealed that in the absence of the magnetic field the
degeneracy at the Fermi level in a free-electron system affects the inverse compressibility
and its fluctuations.  The single-particle energies in the system are given by
\begin{equation}
  \epsilon_{nm} = -2t \left( \cos \frac{\pi n}{L+1} + \cos\frac{\pi m}{L+1} \right)
\end{equation}
with $n$ and $m$ being integers from $1$ to $L$.  Accordingly, there are $L$ degenerate
states with $(n,m) = (1,L), (2,L-1),\ldots, (L,1)$ at the Fermi level.  This degeneracy
yields finite average inverse compressibility together with nonvanishing relative
fluctuations in the free-electron limit of the interacting disordered system.\cite{Jeon99}
Such observation leads to a natural question whether the same phenomena -- relatively
small inverse compressibility and non-vanishing relative fluctuations -- remain at nonzero
degenerate values of the frustration and how the distribution of the inverse
compressibility changes with the magnetic field.  An answer to such questions are given
below.  In general, for large systems where the set of the degenerate frustration becomes
very dense, the statistics of the inverse compressibility is expected to remain
qualitatively unchanged irrespectively of the frustration, except around $f=1/2$.

\subsection{Weakly Frustrated System}

We now examine the effects of weak magnetic fields on the statistics of the inverse
compressibility.  Recalling that the influence of the magnetic field enters only in
the hopping term as an additional phase, we can infer that the magnetic-field effects are
not prominent in the system with small hopping strength.  Indeed, for $W/t>2$ or $V/t>2$,
no distinctive change has been found numerically, at least
in the presence of weak magnetic fields.

\begin{figure}
  \centerline{\epsfig{width=8cm,file=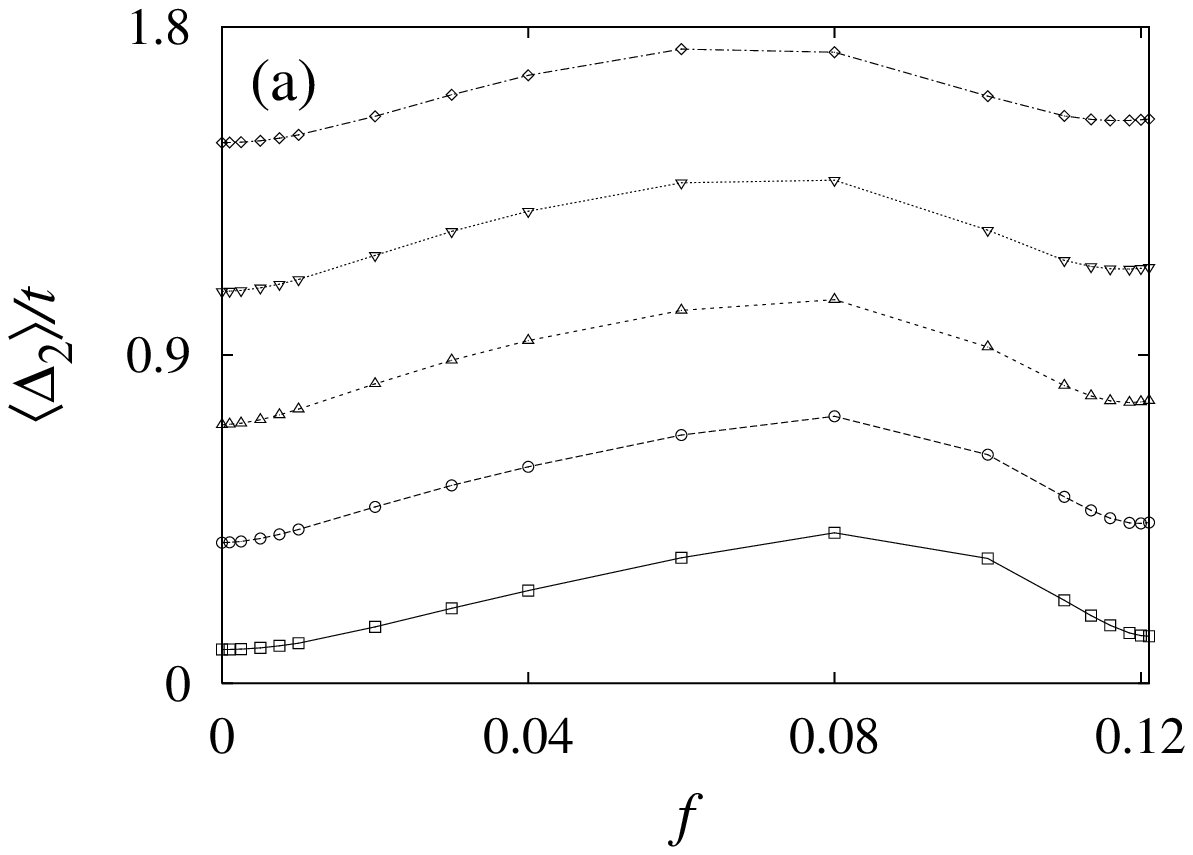}}
  \centerline{\epsfig{width=8cm,file=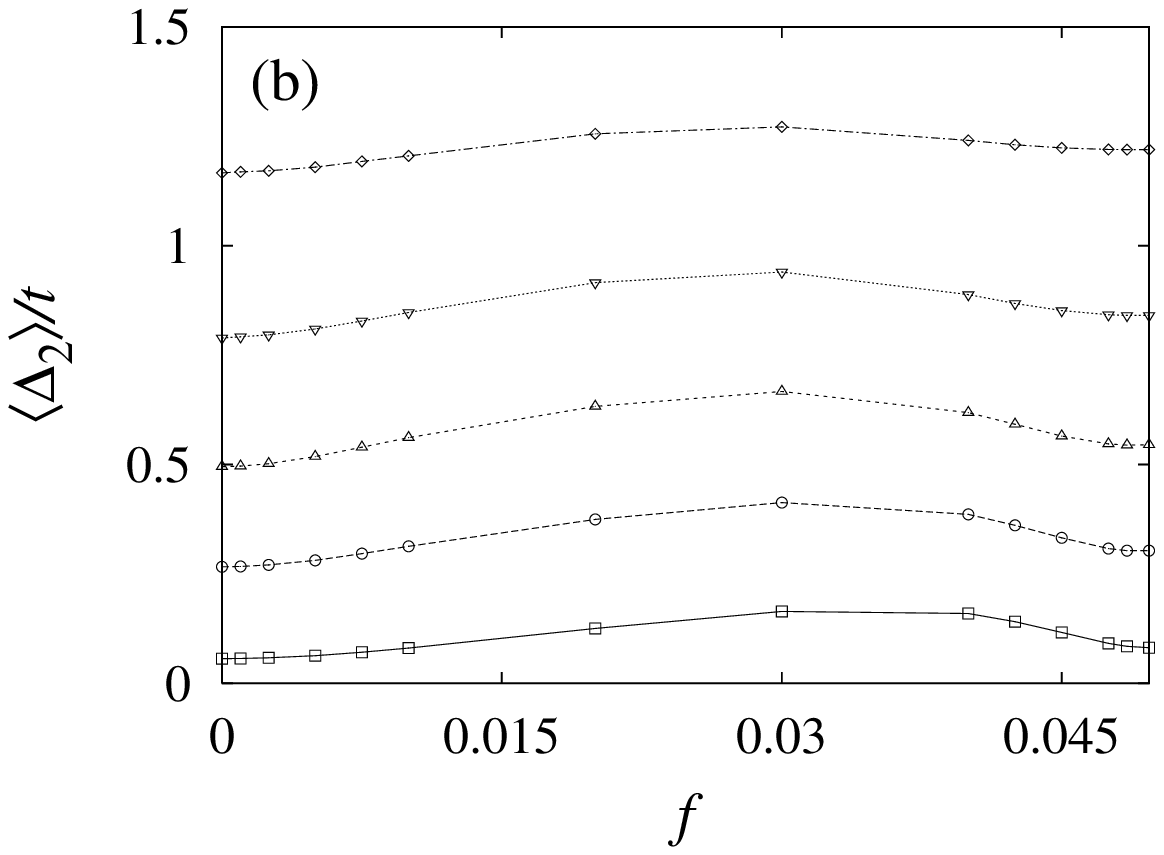}}
  \centerline{\epsfig{width=8cm,file=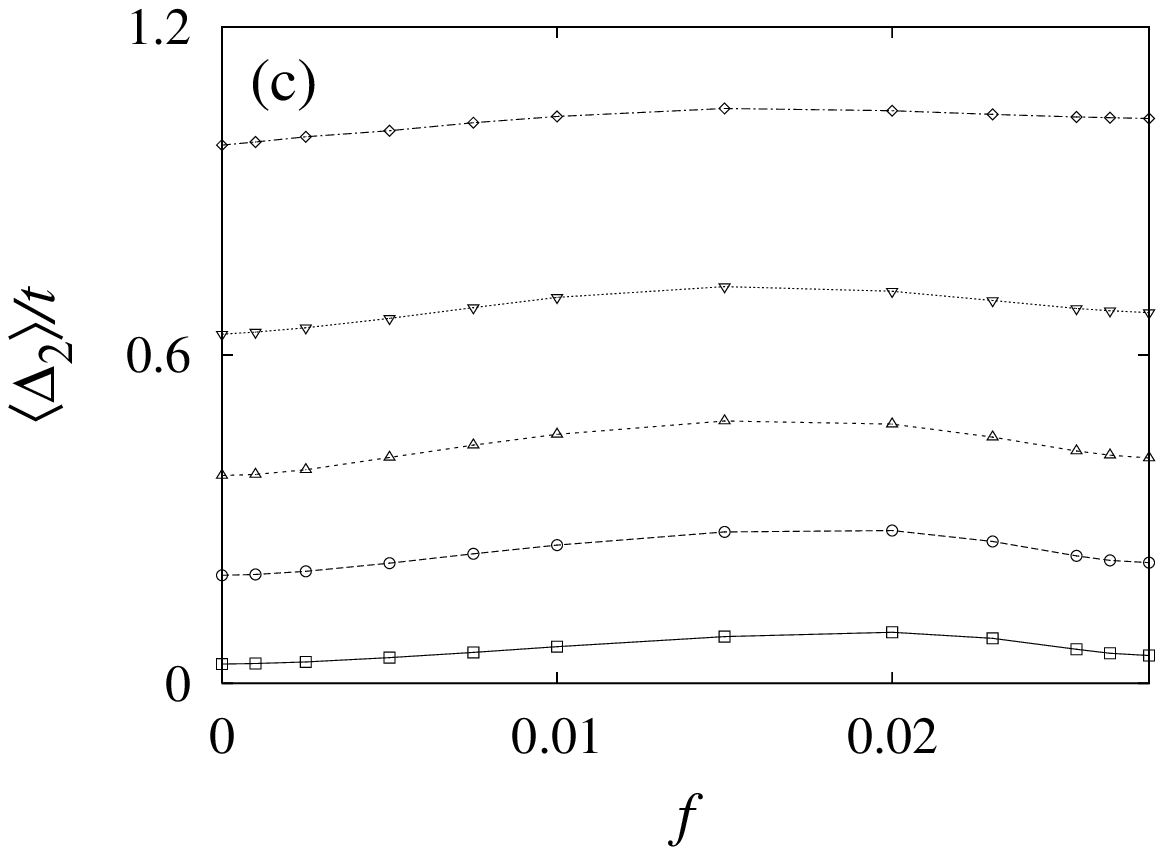}}
  \caption{Average inverse compressibility versus frustration $f$ in the range $[0,f_{d1}]$
    for various values of $V/t =$0($\square$), 0.5($\circ$), 1($\triangle$),
    1.5($\triangledown$), and 2($\diamond$) in the system of size $L=$ (a) $4$, (b) $6$,
    and (c) $8$.  The disorder strength $W/t$ is set to be 0.5.}
  \label{fig:icmnwf}
\end{figure}
\begin{figure*}
  \parbox{0.45\textwidth}{
    \centerline{\epsfig{width=8cm,file=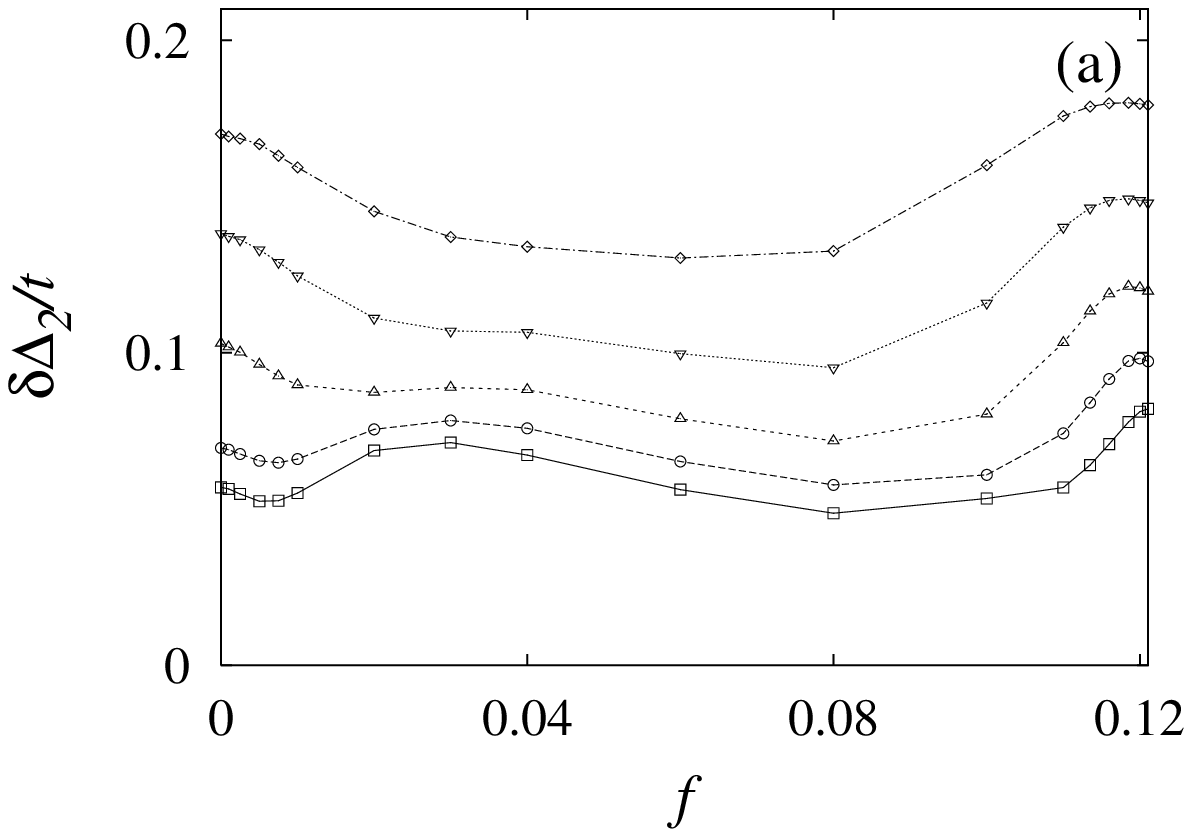}}
    \centerline{\epsfig{width=8cm,file=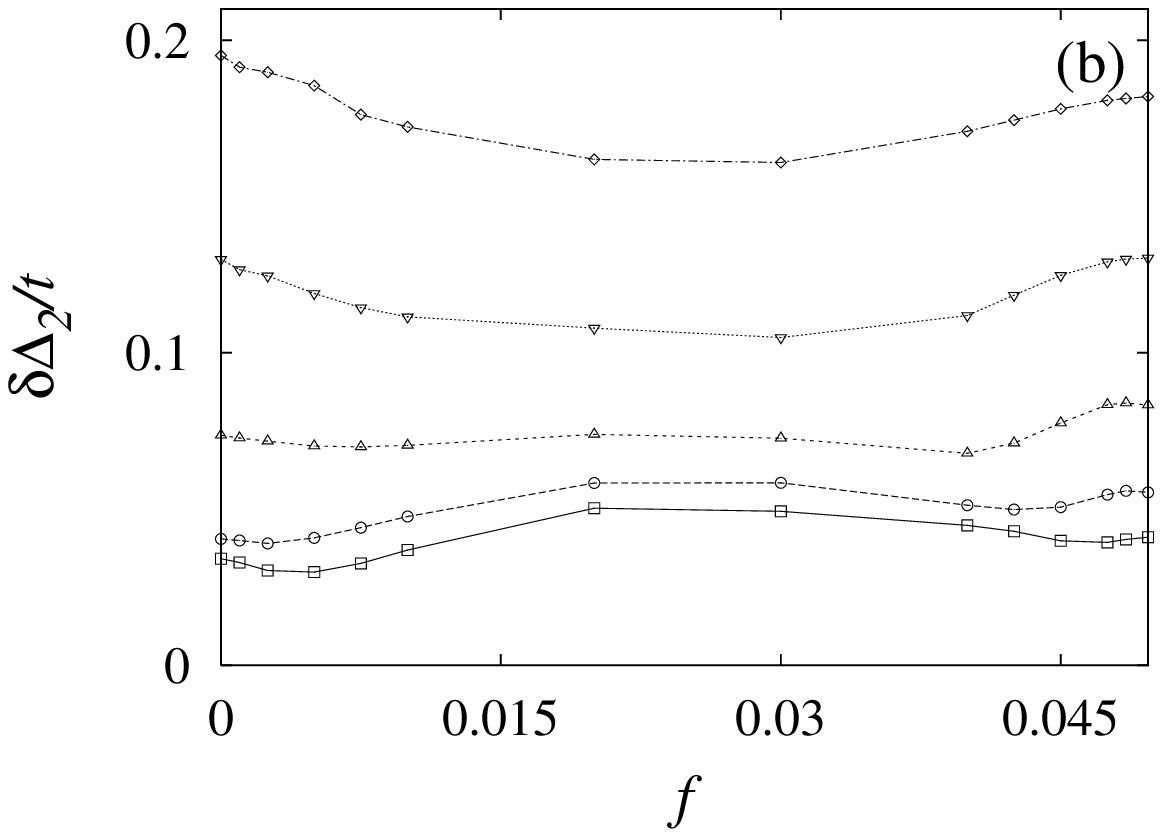}}
    \centerline{\epsfig{width=8cm,file=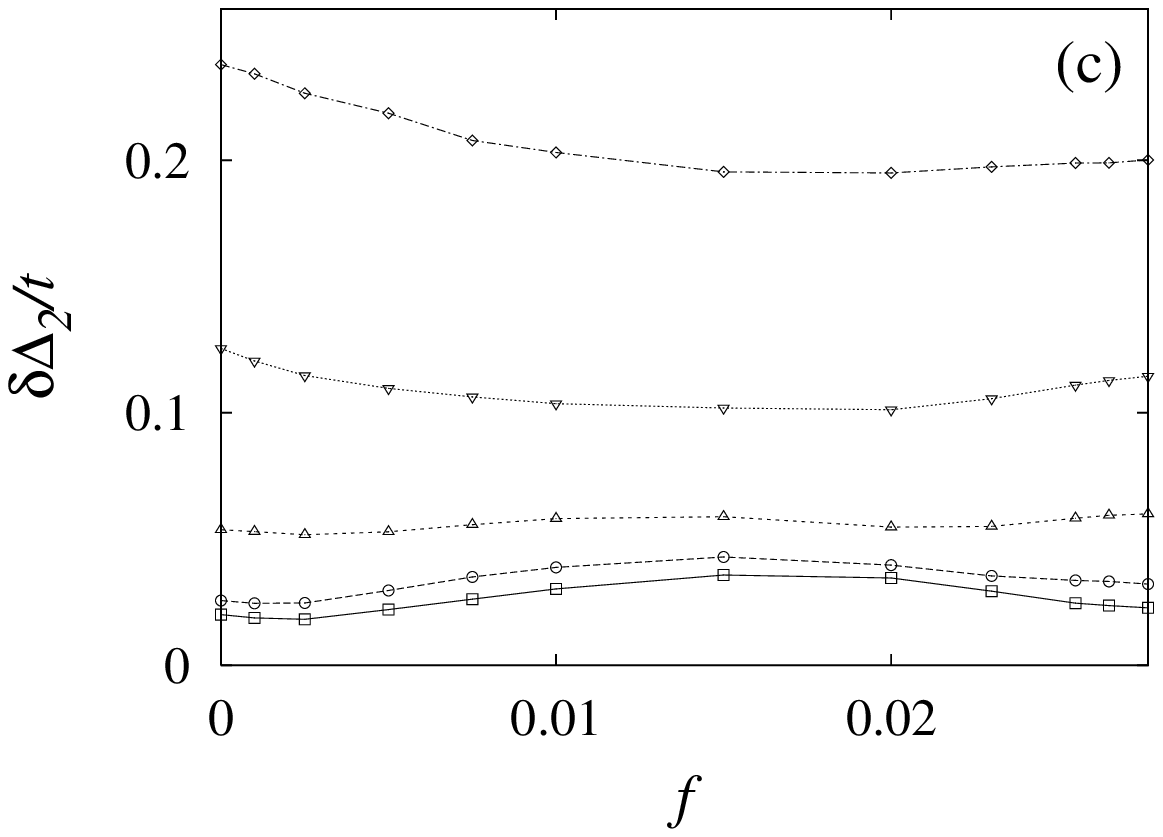}}
  }
  \hspace{0.05\textwidth}
  \parbox{0.45\textwidth}{
    \centerline{\epsfig{width=8cm,file=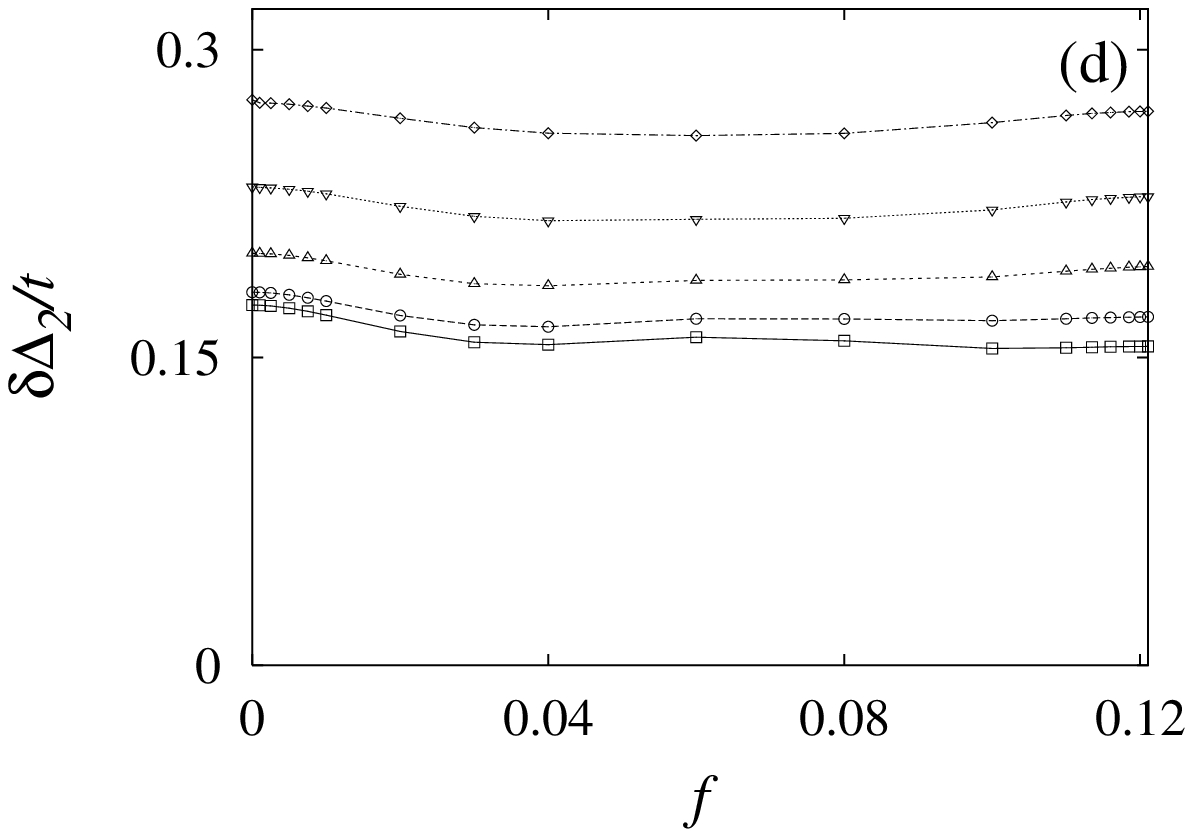}}
    \centerline{\epsfig{width=8cm,file=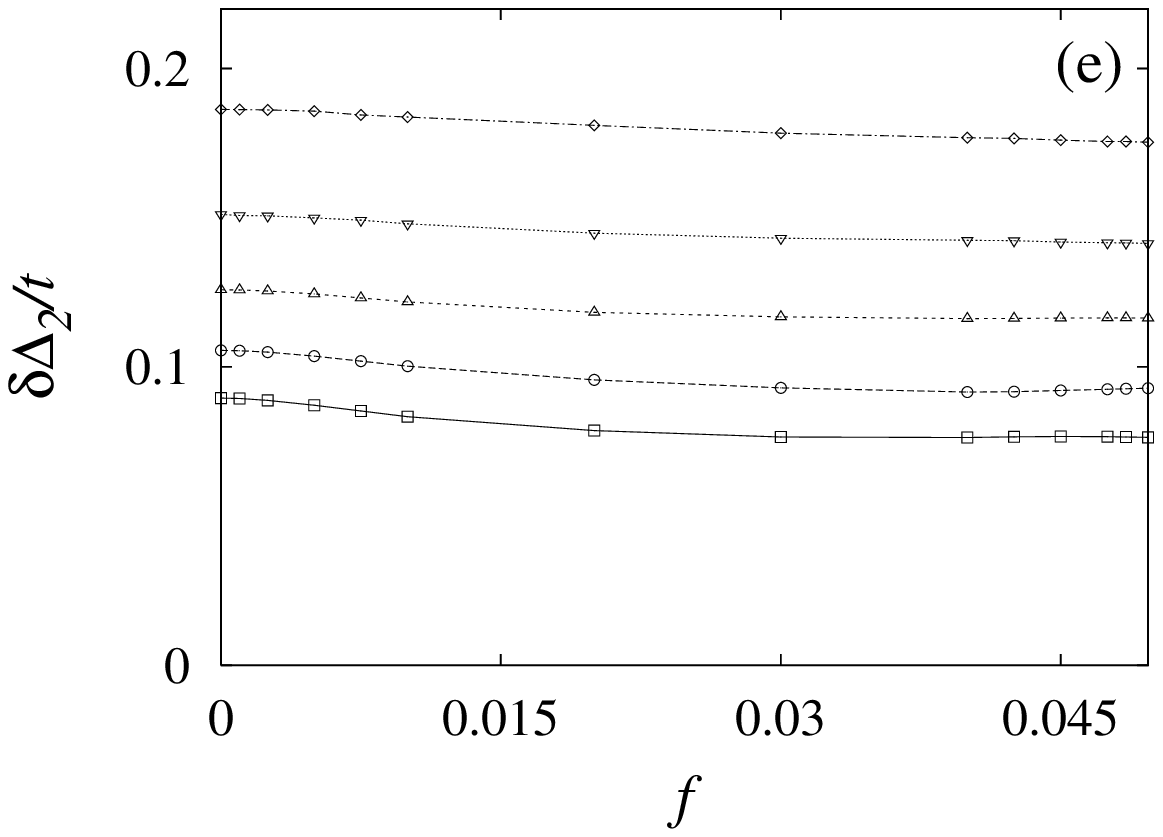}}
    \centerline{\epsfig{width=8cm,file=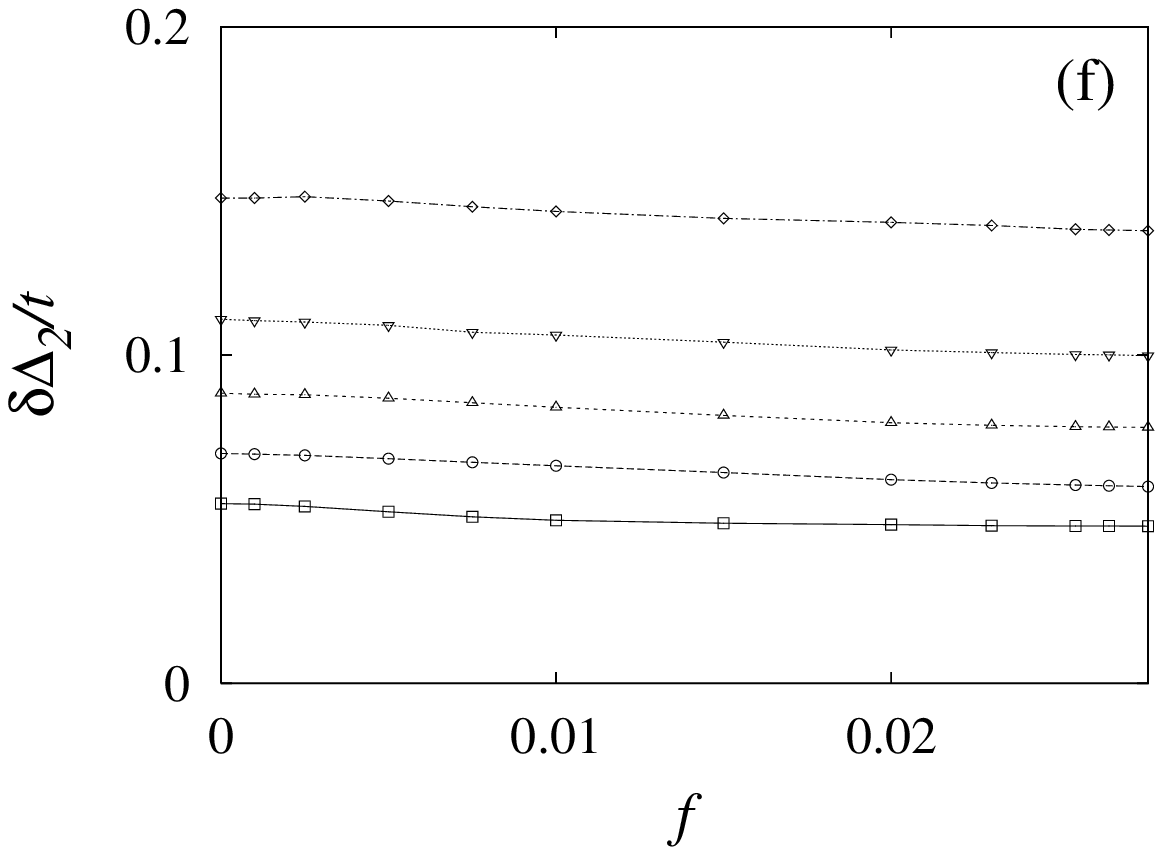}}
  }
  \caption{Fluctuations of the inverse compressibility versus frustration for
    various values of $V/t = $0($\square$), 0.5($\circ$), 1($\triangle$),
    1.5($\triangledown$), and 2($\diamond$) in the weakly disordered system ($W/t=0.5$) of
    size $L= $ (a) $4$, (b) $6$, (c) $8$ and in the strongly disordered system ($W/t=2$) of
    size $L= $ (d) $4$, (e) $6$, (f) $8$.}
  \label{fig:icdvwf}
\end{figure*}
\Figref{fig:icmnwf} shows the dependence of $\avg{\ic}/t$ on the frustration:
The magnetic field initially enhances $\avg{\ic}/t$, which is expected from the increase
of $\ic$ with $f$ in the free-electron system.  As $f$ approaches $f_{d1}$,
the first nonzero degenerate frustration in the corresponding noninteracting clean system,
however, $\icavg/t$ decreases to a value which is close to the zero-frustration value.
Level repulsion due to random disorder smoothes out the dependence on $f$,
especially at the degenerate frustration.

On the other hand, fluctuations of the inverse compressibility show different behaviors
with the frustration, depending on the strength of the interaction and disorder.  In the
system with weak disorder [see \figref{fig:icdvwf}(a)--(c) for $W/t=0.5$], $\icstd/t$
versus $f$ exhibits two minima for weak interactions $V/t(\le2)$, which turn into a single
minimum as the interaction is increased; these features, however, become less pronounced
with the increase of the system size.  For strong disorder ($W/t=2$) shown in
\figref{fig:icdvwf}(d), (e), and (f), the peculiar structure observed in the weak-disorder
regime fades away and $\icstd/t$ tends to decrease with the frustration, particularly in
large systems.

\begin{figure}
  \centerline{\epsfig{width=8cm,file=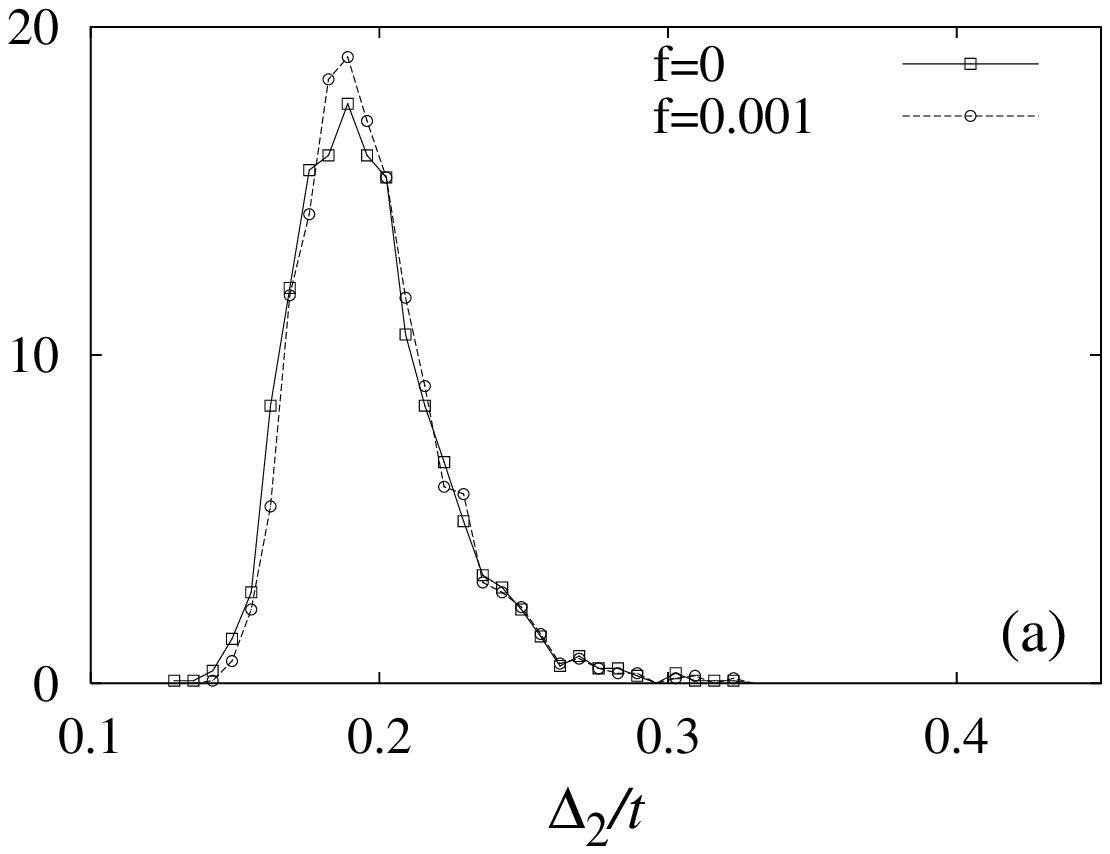}}
  \centerline{\epsfig{width=8cm,file=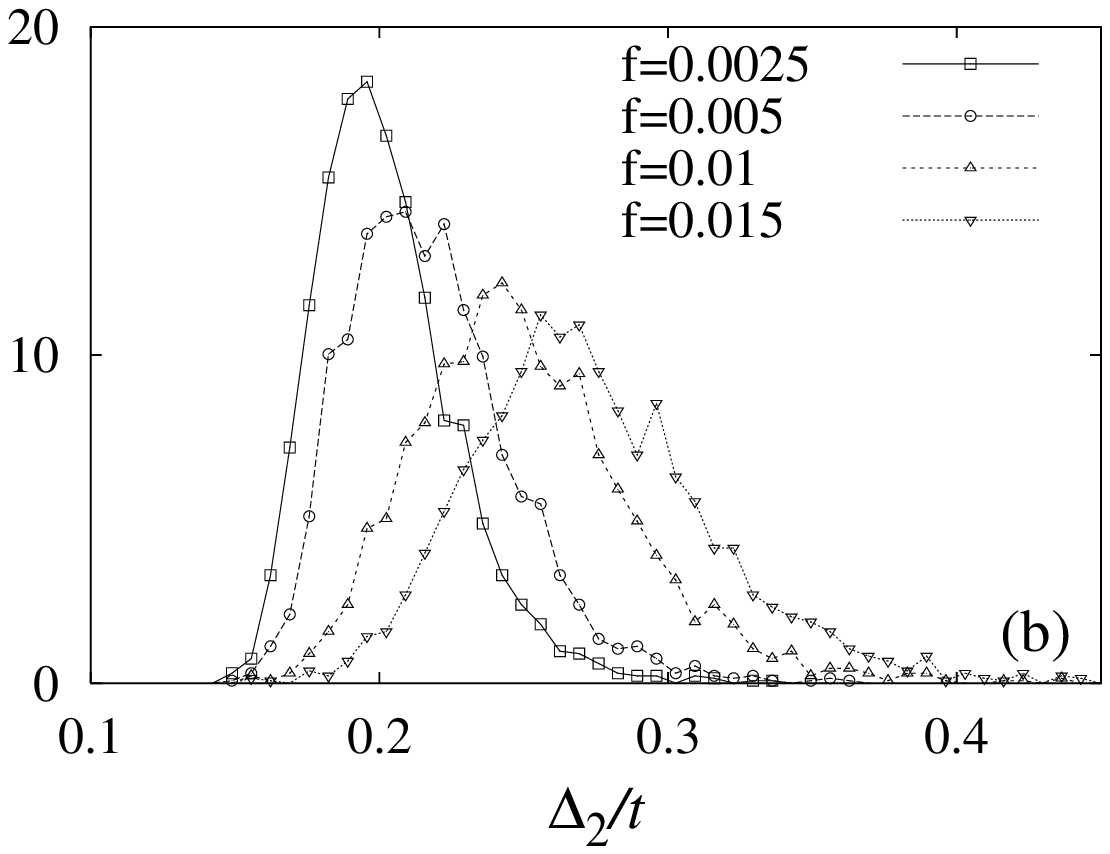}}
  \centerline{\epsfig{width=8cm,file=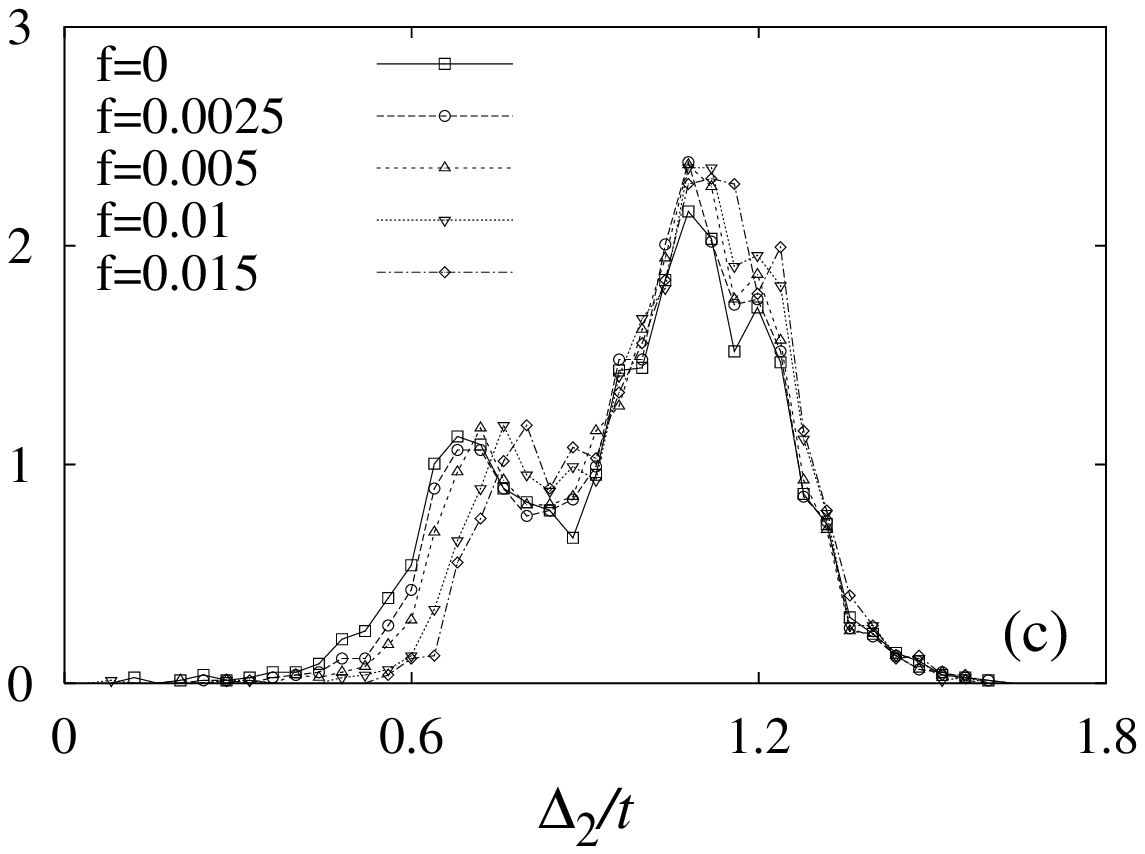}}
  \caption{Distribution of the inverse compressibility in the system of size $L=8$
    for $W/t=0.5$.  The interaction strength is given by $V/t=0.5$ in (a) and (b) and $V/t=2$
    in (c); the frustration $f$ is varied from 0 to $f_{d1}$.}
  \label{fig:icdiswf}
\end{figure}
From the distribution of $\ic$ presented in \figref{fig:icdiswf}, one can get a hint as to
the origin of the behavior of fluctuations in the weakly disordered system.  Turning on
the magnetic fields initially suppresses the appearance of small $\ic$ and thus reduces
fluctuations regardless of the interaction strength, as can be observed in
\figref{fig:icdiswf}(a) and (c).  Further increase of $f$, however, raises significantly
the portion of large $\ic$ in the presence of weak interactions, inducing large
fluctuations [see \figref{fig:icdiswf}(b)].  When $f$ exceeds the value at which
$\icavg/t$ reaches its maximum, the reverse process takes place, producing another minimum
of $\icstd/t$.  In contrast, strong interactions seem to suppress such extension of the
distribution to the large-$\ic$ region, leading to a curve with one minimum for
$\icstd/t$.

To examine the relation between different responses to the magnetic field and localization
properties, we introduce the participation ratio $\beta$ and the inverse
participation number (IPN) $P^{-1}(\nu)$:
\begin{eqnarray}
  \beta & \equiv & \frac{N_e^2}{L^2 \sum_i \rho_i^2} \\
  P^{-1}(\nu) & \equiv & \sum_i |\psi_i(\nu)|^4,
\end{eqnarray}
where $N_e$ is the total number of electrons, $\rho_i$ is the expectation value of the
electron density at site $i$, and $\psi_i(\nu)$ is the probability amplitude for the
single-particle state $\ket{\nu}$ at site $i$.  The participation ratio $\beta$ is equal
to unity in the state with a uniform density profile and decreases as the electron
configuration becomes modulated or localized, eventually reaching the minimal value of the
filling factor $N_e/L^2$ in the limit that all the electrons are located at the $N_e$
sites.\cite{Walker99a} The IPN, on the other hand, assumes unity for the level localized
at one site and decreases as the state becomes delocalized, approaching its minimal value
$L^{-2}$ in the opposite limit of the uniformly extended state.\cite{IPN}

We use these measures to characterize localization or modulation of the half-filled state
$\ket{M}$ and search for the possible relation with the inverse compressibility,
which reveals that for given frustration the participation ratio in general has negative
correlations with the inverse compressibility: The smaller the inverse compressibility,
the larger the participation ratio and vice versa.  This observation indicates that the
electron profiles in the samples with smaller values of $\ic$ tend to be uniform and thus
explains the increase of $\ic$ in the samples displaying relatively small $\ic$ since such
uniform profiles should be sensitive to the change in frustration.  Localized states, on
the other hand, are rather insensitive to frustration, and accordingly, the inverse
compressibility in a sample with large $\ic$ remains unchanged.  To be more definite in
the electron state, we have also calculated the IPN at the Fermi level and observed
positive correlations between the IPN and the inverse compressibility, thus confirming the
above argument.  In the case that hopping is dominant over the interaction and disorder,
the samples mostly have large values of the participation ratio and low values of the IPN
at the Fermi level, implying sensitivity to frustration.  Variations of the sensitivity
over the samples lead to the increase of width of the distribution.

As $f$ approaches $f_{d1}$, the distribution reverses its behavior, restoring that at
$f\approx0$.  However, it should be noted that the distributions at $f=0$ and at
$f=f_{d1}$ are not exactly the same.  Rather, the distribution at $f=f_{d1}$ is similar to
the one at $f=0.005$.  Such difference between the distributions at different degenerate
frustrations may be attributed to the difference of the degree of degeneracy: At $f=0$ the
system has the $L$-fold degeneracy at half-filling\cite{Jeon99} but at $f=f_{d1}$ two-fold
degeneracy is present, as can be seen in \figref{fig:free}(a).  Since large degeneracy
favors a smaller value of $\ic$ even in the presence of the interaction and disorder, the
probability for small $\ic$ at $f=0$ is higher than that at $f=f_{d1}$, which is verified
in numerical calculations.

\begin{figure}
  \centerline{\epsfig{width=8cm,file=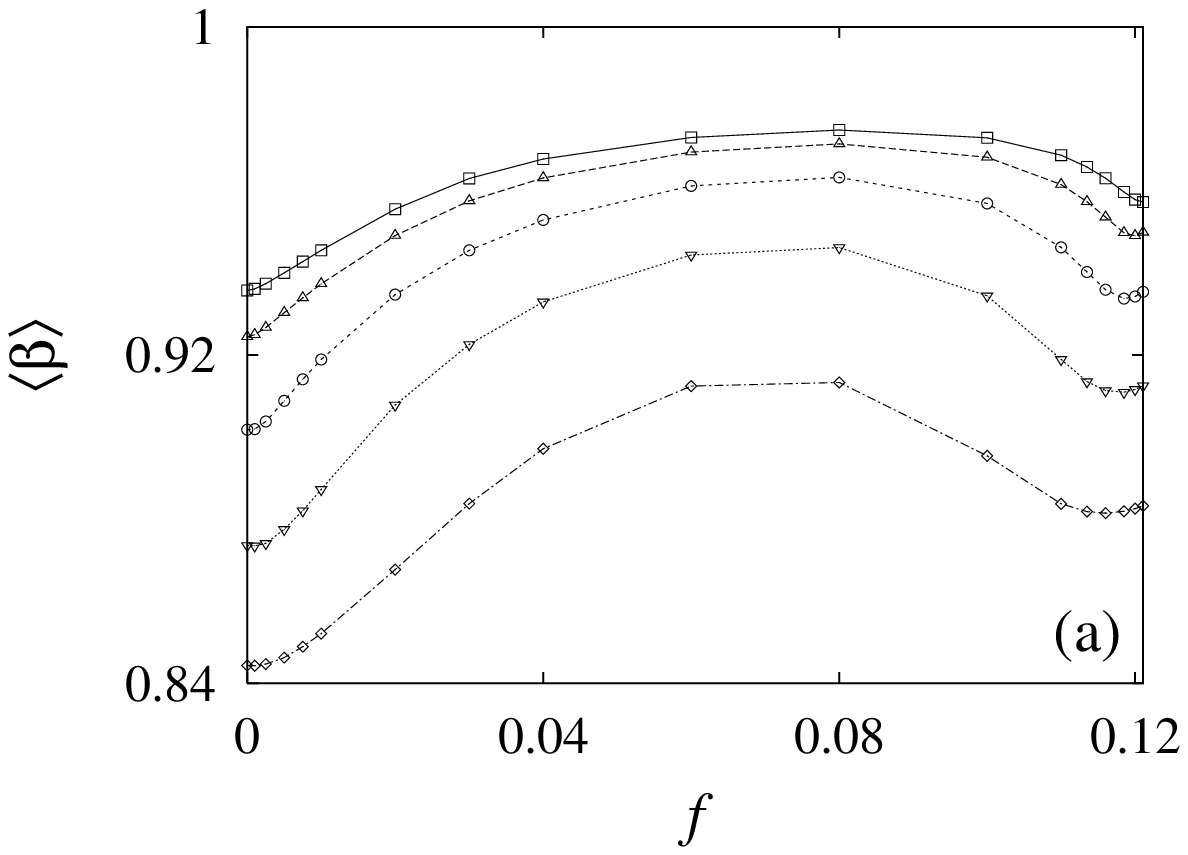}}
  \centerline{\epsfig{width=8cm, file=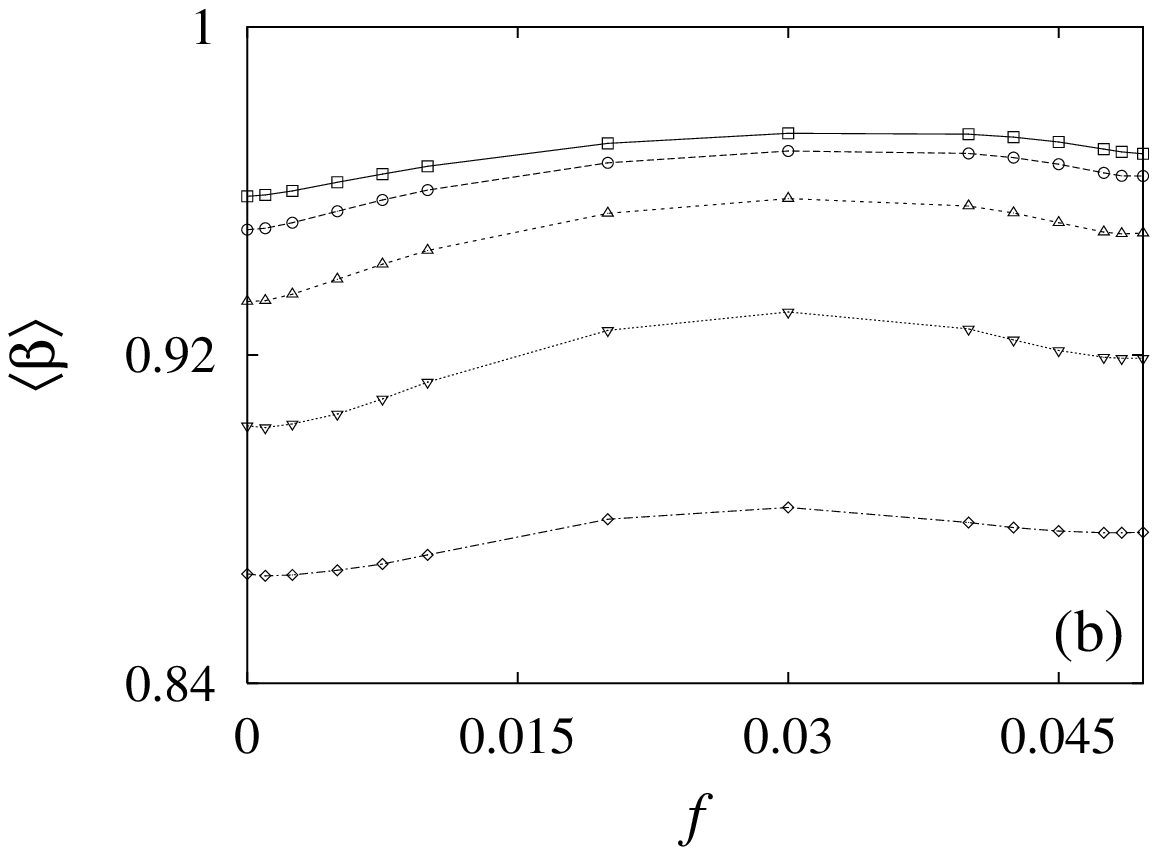}}
  \centerline{\epsfig{width=8cm, file=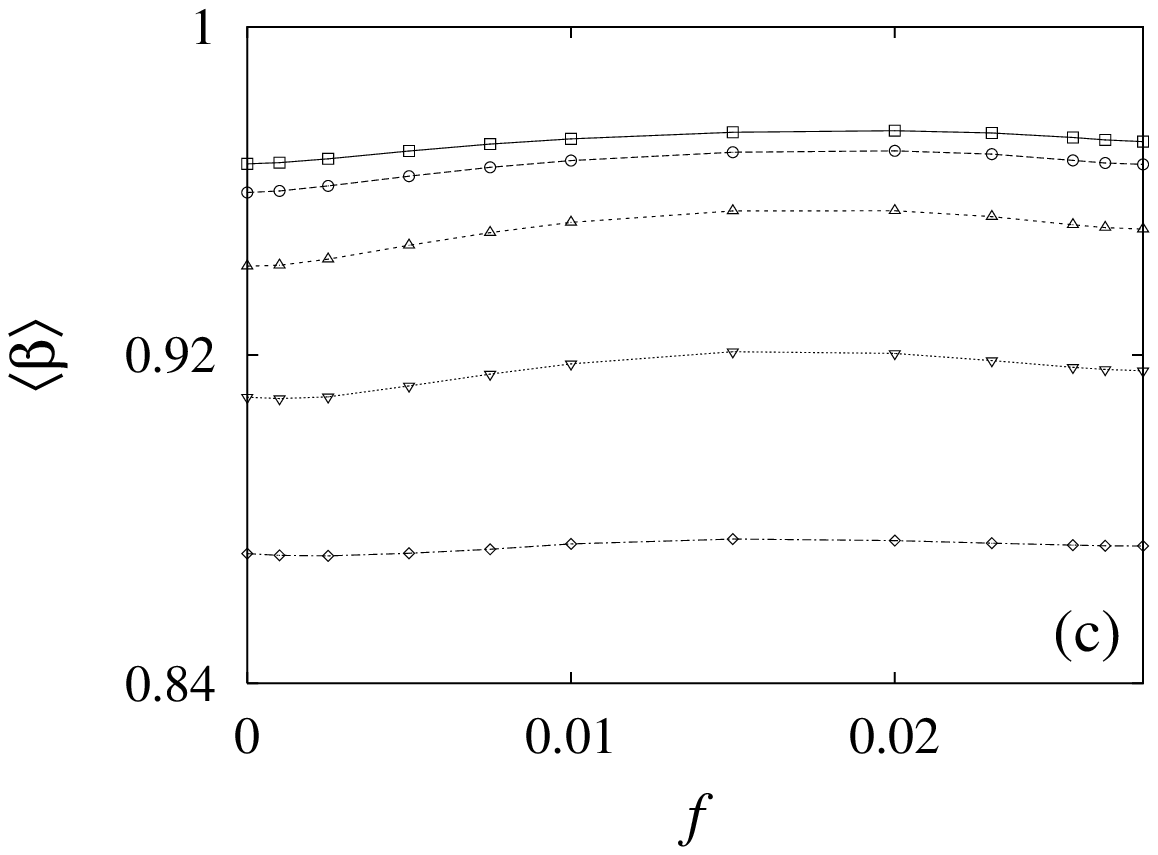}}
  \caption{Average participation ratio versus frustration in the range $[0,f_{d1}]$.
    The system has the size $L= $ (a) $4$, (b) $6$, (c) $8$ together with the interaction
    $V/t = $ 0($\square$), 0.5($\circ$), 1($\triangle$), 1.5($\triangledown$), and
    2($\diamond$).  The disorder strength is given by $W/t=0.5$.}
  \label{fig:prwf}
\end{figure}
To get more insight for large fluctuations at the degenerate frustration in relation to
the localization property, we have computed the average participation ratio $\avg{\beta}$,
which is plotted as a function of $f$ for various interaction strengths in
\figref{fig:prwf}.  The conspicuous decrease in $\avg{\beta}$ near the degenerate
frustration signals an increase in spatial modulation or localization of the total
electron density.  Indeed in a perfectly clean system the participation ratio is exactly
equal to unity, irrespective of the frustration.  As disorder is introduced, $\avg{\beta}$
diminishes faster at the degenerate frustration.  Such spatial localization at the
degenerate frustration or delocalization at non-degenerate frustration are believed to be
related with the pattern of interference of electrons scattered at impurities: Moving
around the lattice under the magnetic field, the electrons acquire additional phases.  At
the degenerate frustration such phases are expected to enhance the constructive
interference.  In connection with the degeneracy we can imagine linear combinations of
degenerate states at the Fermi level; any combination will be an eigenstate of the clean
system, and among them we can choose more modulated or localized one which leads to less
energy in the presence of disorder.  At the non-degenerate frustration such choice is not
possible, yielding rather extended states.

The sensitivity of electron states to disorder at the degenerate frustration in turn
generates larger fluctuations of the inverse compressibility over the samples.  In
Ref.~\onlinecite{Levit99} the enhanced fluctuations of the inverse compressibility were
also attributed to localization of the wavefunctions around the edge of the dot.  Here we
stress that our argument is not restricted merely to the half-filling case which gives
fairly high degeneracy at $f=0$. \Figref{fig:free}(a) shows that there also exist many
degenerate states at $f=0$ at other fillings.

We conclude this section with some remarks on the comparison with the existing experiment.
For the interaction and disorder which are sufficiently strong but not so strong to
overwhelm quantum effects, the fluctuations are observed to diminish with the frustration
$f$, provided that $f\ll f_{d1}$.  This leads to similar decrease in relative
fluctuations, even for weak disorder and interaction, owing to the increase in $\icavg$.
This result agrees quite well with the experimental results mentioned in the beginning of
this section.  Furthermore, the structure of the inverse-compressibility distribution has
been found not to change appreciably, which also coincides with the experimental
observation.

\subsection{Fully Frustrated System}

\begin{figure}
  \centerline{\epsfig{width=8cm, file=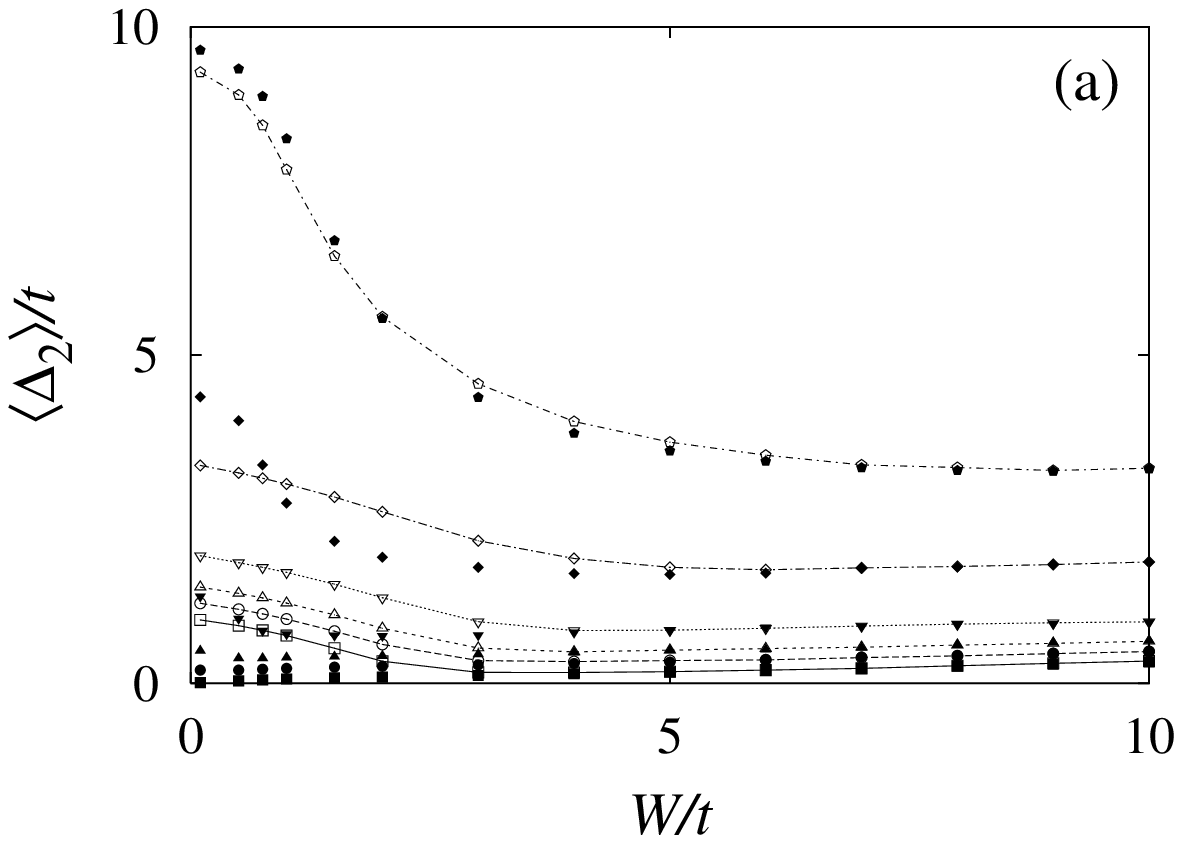}}
  \centerline{\epsfig{width=8cm, file=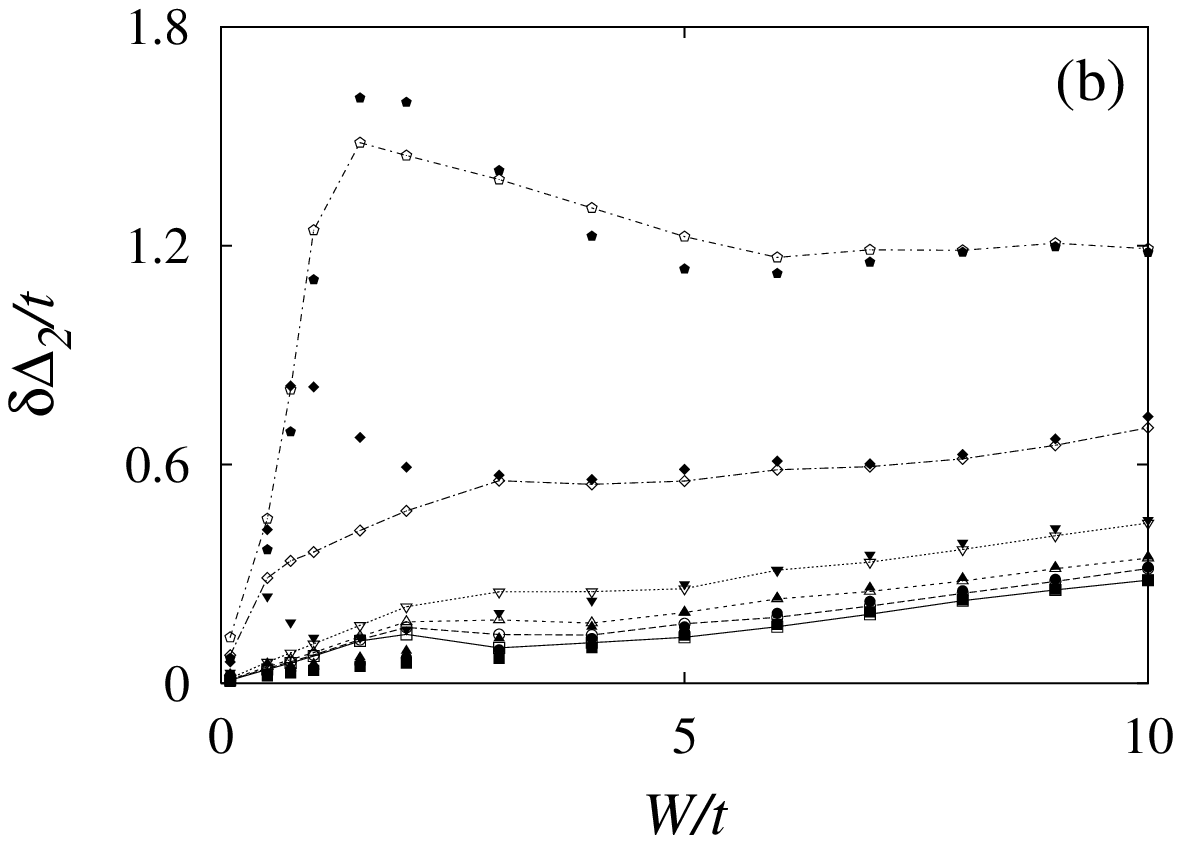}}
  \caption{(a) Average and (b) fluctuations of the inverse compressibility versus $W/t$ at
    $f=1/2$ (empty symbols with lines) and at $f=0$ (filled symbols) in the system of size
    $L=8$ for various values of $V/t$.  The data marked by squares, circles, triangles,
    inverted triangles, diamonds, and pentagons correspond to $V/t = 0, \,0.5,\, 1,\, 2,\,
    5$, and $10$, respectively.}
  \label{fig:icff}
\end{figure}
In the previous section, we have examined the effects of weak magnetic fields, up to the
first degenerate frustration $f_{d1}$, on fluctuations of the inverse compressibility.
The arguments in the previous section as to the fluctuations at degenerate and
non-degenerate frustrations seem to be applicable beyond $f_{d1}$ as well.  However, as
the system size grows, the degenerate frustrations proliferate in number to spread over
the entire region of the frustration and the energy splitting by the magnetic field at the
Fermi level also shrinks.  Accordingly, no distinctive difference is expected to appear in
the distribution of $\ic$ at degenerate and non-degenerate frustrations.  A remarkable
exception happens at the full frustration ($f=1/2$), where quite large energy spacing is
induced at the Fermi level in the noninteracting clean system (see \figref{fig:free}).
For even $L$, the energy spacing reads
\begin{equation} \label{sp}
  4\sqrt{2}t \cos\left[\frac{\pi L}{2(L+1)}\right],
\end{equation}
which decreases to zero in the thermodynamic limit $(L\rightarrow\infty)$.
Nevertheless, for moderate $L$ the finite spacing given by Eq. (\ref{sp}) discerns the full
frustration from other frustration values, yielding qualitative difference in the distribution
even for strong interactions.

\begin{figure}
  \centerline{\epsfig{width=8cm,file=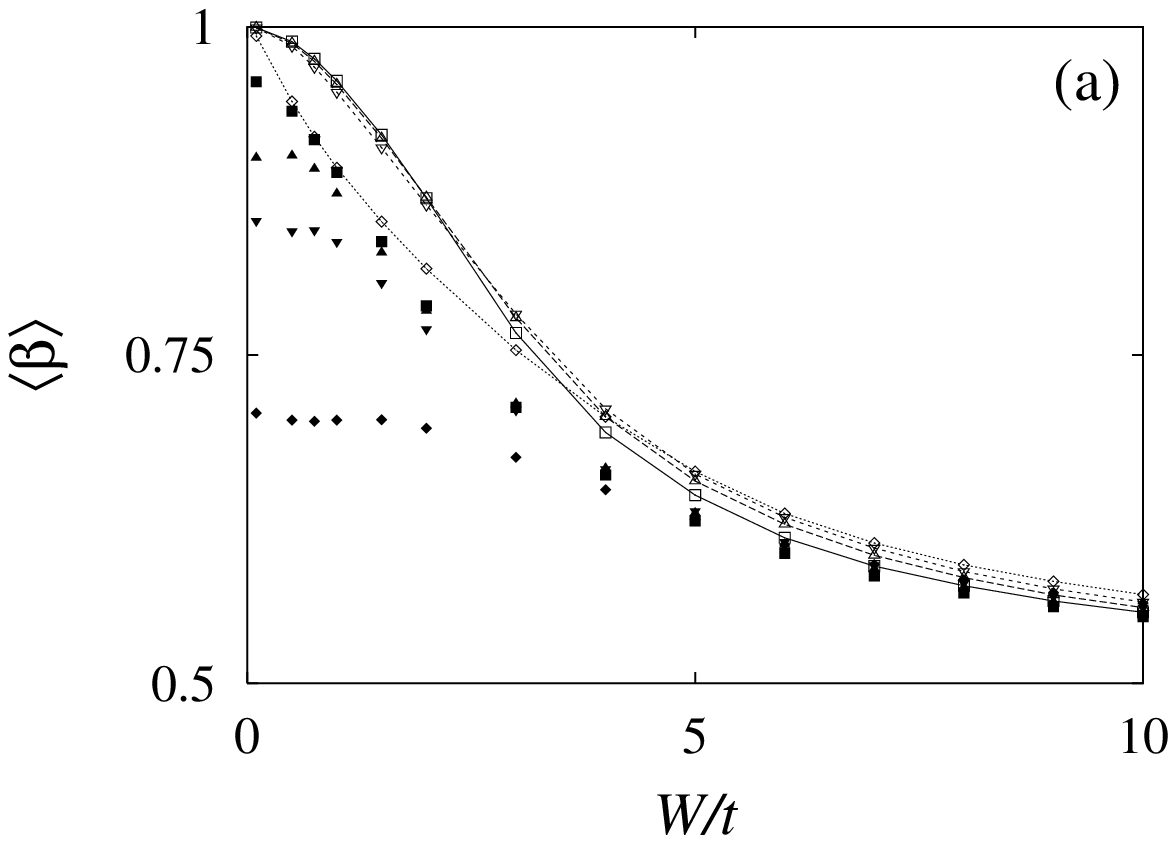}}
  \centerline{\epsfig{width=8cm,file=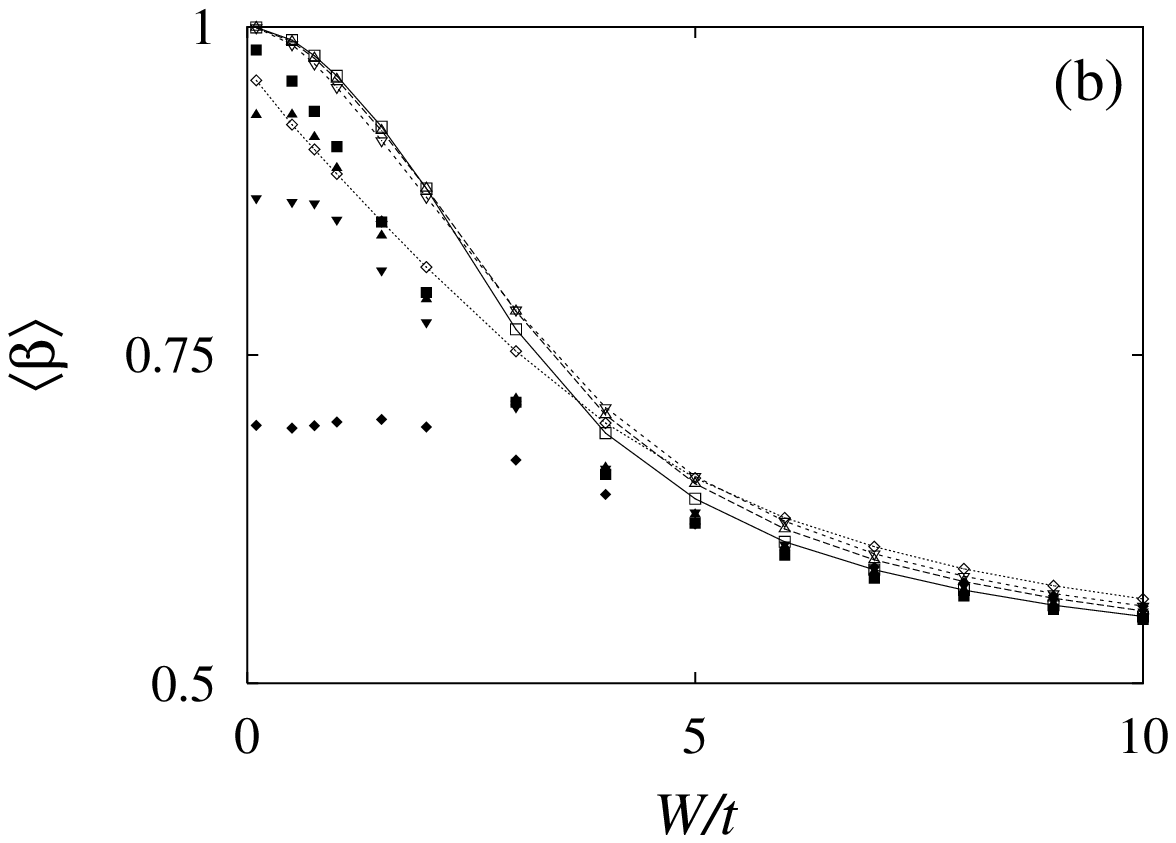}}
  \centerline{\epsfig{width=8cm,file=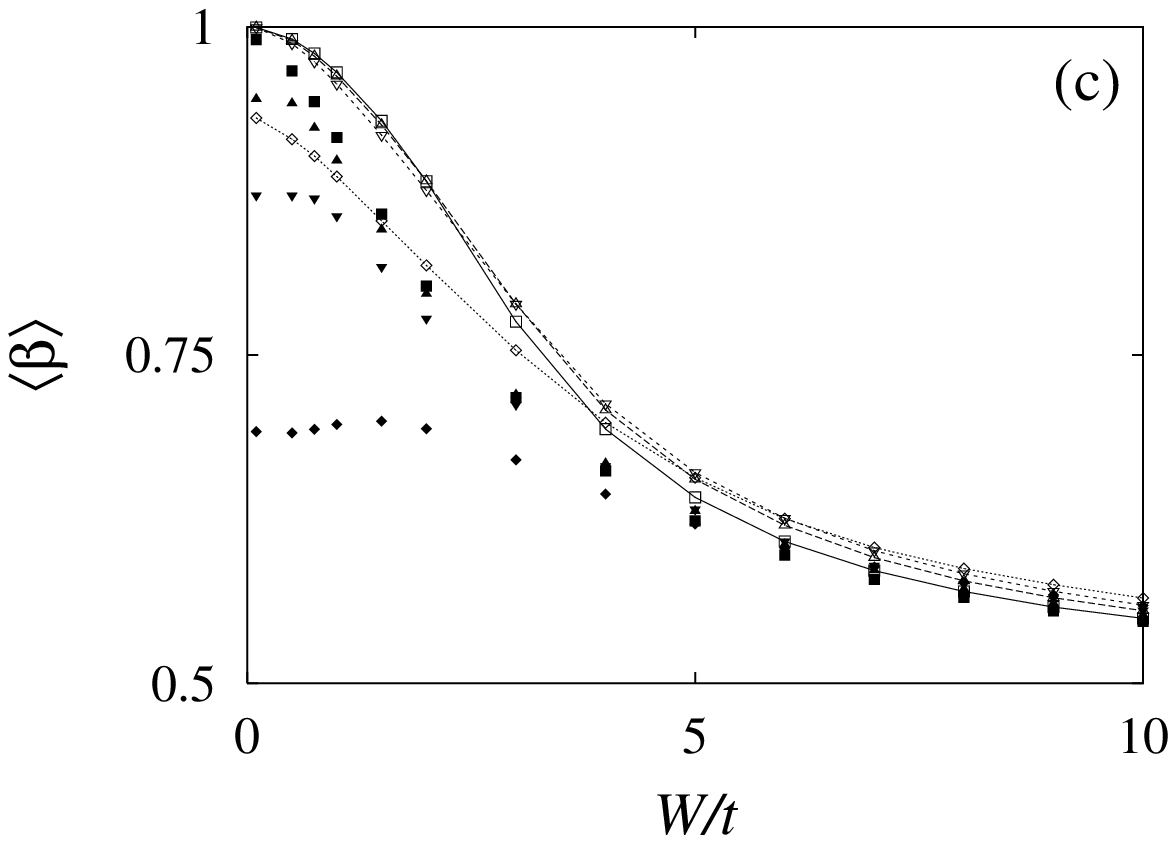}}
  \caption{Average participation ratio versus $W/t$ at $f=1/2$ (empty symbols with lines)
    and $f=0$ (filled symbols) in the system of size $L=$ (a) $4$, (b) $6$, (c) $8$ for
    various values of $V/t$.  The symbols are the same as those in \figref{fig:icff}.}
  \label{fig:prff}
\end{figure}
The average inverse compressibility and its fluctuations are displayed in
\figref{fig:icff}.  For $V/t=2$ and $5$, one can notice the disappearance of both the
lumps in $\icavg/t$ and the overshoots in $\icstd/t$, which develop at weak disorder and
strong interactions in the unfrustrated system ($f=0$).  In \secsref{sec:cs} and
\ref{sec:qs}, both the lumps in $\icavg$ and the overshoots in $\icstd$ have been found to
have their origin in the WC states due to the Coulomb interaction.  Accordingly, the
disappearance of such lumps and overshoots implies that strong magnetic fields suppress
the interaction effects on the electron distribution.  To look into the distribution of
the electrons, we have compared the average participation ratio at $f=0$ and that at
$f=1/2$ in \figref{fig:prff}.  It is observed that the latter, $\avg{\beta}_{f=1/2}$, is
always larger than the former, $\avg{\beta}_{f=0}$, with the difference significant for
$2\le V/t\le5$.  Furthermore, $\avg{\beta}_{f=1/2}$ does not vary much with the
interaction, up to $V/t \approx 2$; even for $V/t=5$, its values are not so different from
the values for $V/t=0$.  We thus conclude that the full frustration restrains the
modulation of electrons induced by the repulsive Coulomb interaction between electrons and
suppresses its influence on the fluctuations of the inverse compressibility.

\begin{figure}
  \centerline{\epsfig{width=8cm,file=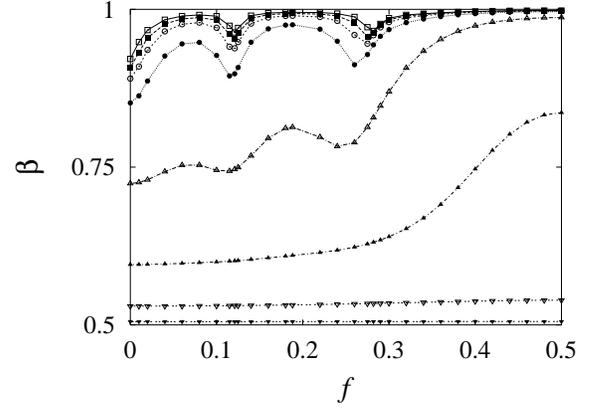}}
  \caption{Average participation ratio versus the frustration $f$ in the system of size $L=4$
    with a single impurity of strength $W/t=1$, located at a corner of the lattice,
    for various values of $V/t$ = 0.1($\square$), 0.5($\blacksquare$),
    1($\circ$), 2($\bullet$), 5($\triangle$), 10($\blacktriangle$), 20($\triangledown$),
    and 50($\blacktriangledown$).}
  \label{fig:prcl}
\end{figure}
To verify the suppression of the Coulomb interaction beyond the HF approximation, we
compute the participation ratio as a function of the frustration parameter via the exact
diagonalization method in the clean system with a single impurity of strength $W/t=1$,
located at a corner of the $4\times4$ lattice, and plot the results for various values of
$V/t$ in \figref{fig:prcl}.  Note that small values of the participation ratio indicate
the formation of the WC-like states due to interactions.  We have intentionally inserted
an impurity in the system in order to eliminate the degeneracy, which allows the two
(degenerate) WC states to superpose and may keep the participation ratio still larger even
when the electrons form a Wigner crystal.  At $f=0$ and other degenerate frustration
values, the average participation ratio is clearly shown to decrease with the interaction;
even rather weak interactions lead to appreciable reduction in the participation ratio.
At the full frustration, in contrast, the participation ratio does not reduce much even in
the presence of substantial interactions, indicating that the fully frustrated system is
reluctant to be in a WC state and thus confirming the above HF results.

\begin{figure}
  \centerline{\epsfig{width=8cm,file=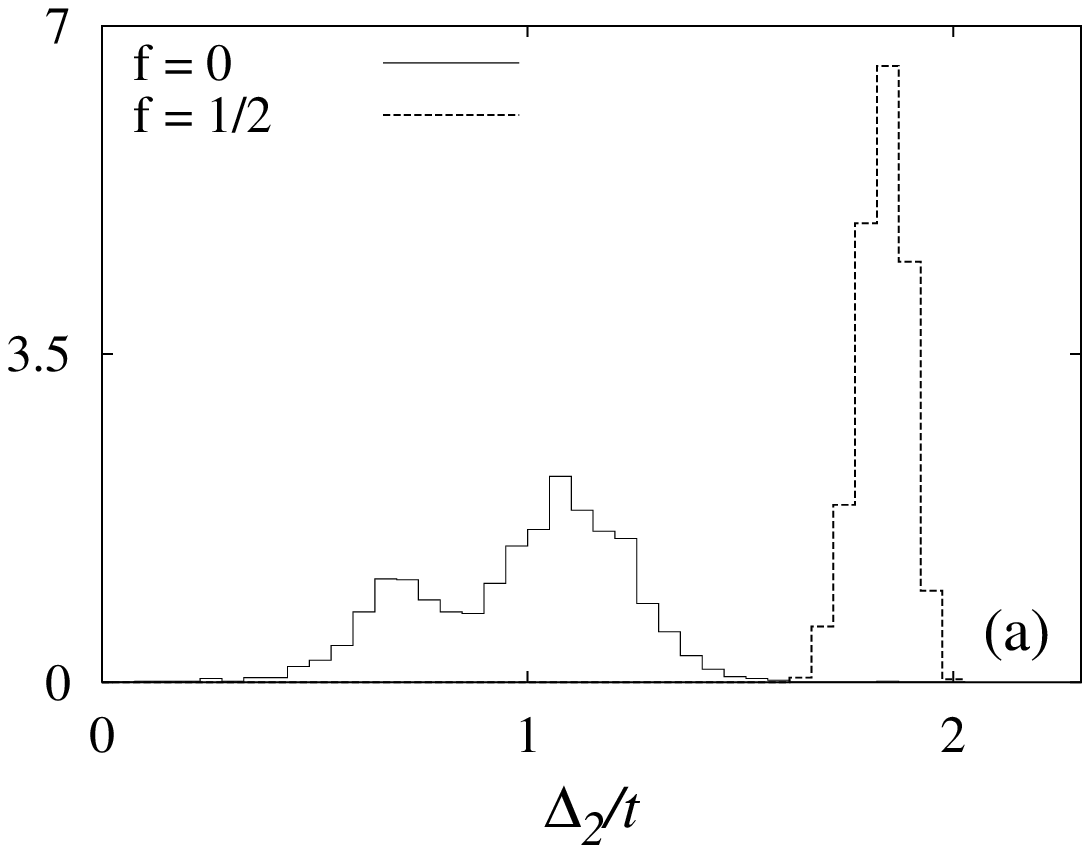}}
  \centerline{\epsfig{width=8cm,file=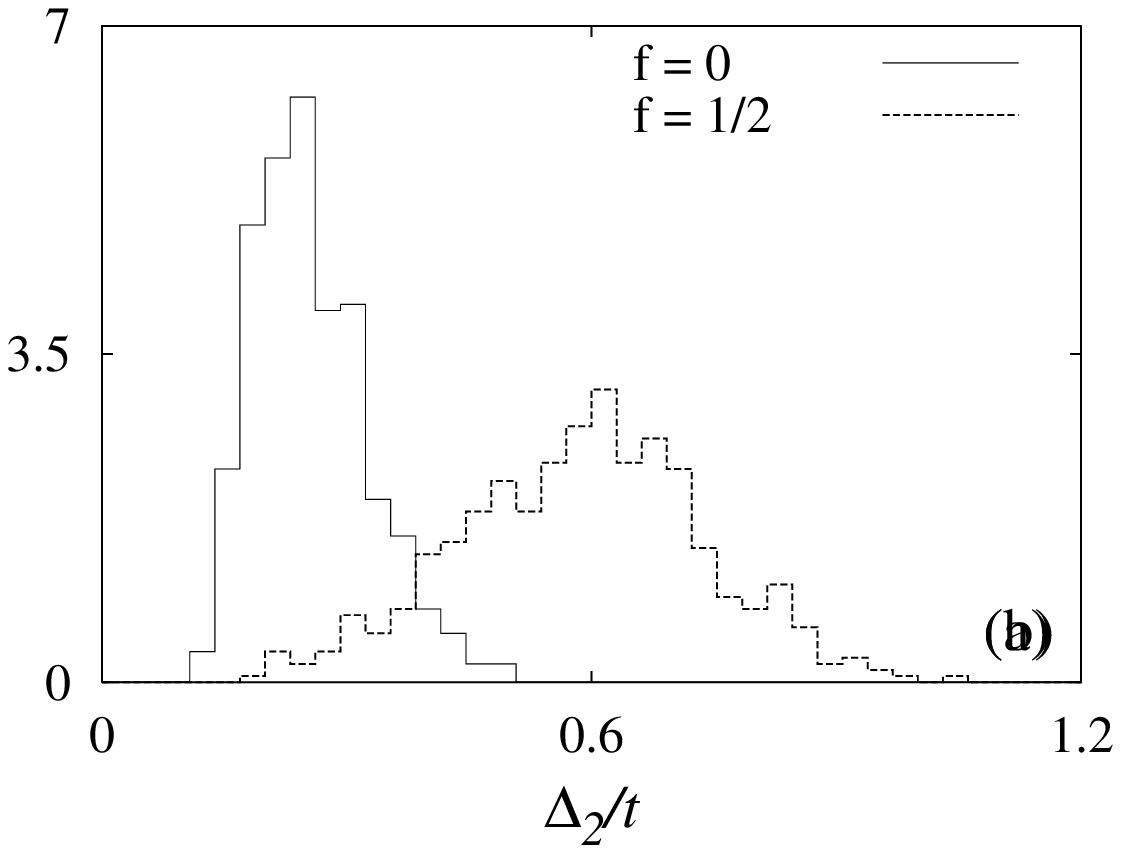}}
  \caption{Distribution of the inverse compressibility at $f=0$ and $f=1/2$ for (a)
    $V/t=2$ and $W/t=0.5$; (b) $V/t=0.5$ and $W/t=2$.}
  \label{fig:icdisff}
\end{figure}
\Figref{fig:icdisff} compares the distribution of $\ic$ at $f=0$ and that at $f=1/2$.  For
$V/t=2$ and $W/t=0.5$, shown in \figref{fig:icdisff}(a), the width of the distribution at
$f=1/2$ is much reduced compared with that at $f=0$.  Here the ubiquity of large values of
$\ic$ at $f=1/2$ comes not from the Coulomb charging energy but from the large
single-particle energy splitting due to the full frustration.  On the other hand, for
$V/t=0.5$ and $W/t=2$, \figref{fig:icdisff}(b) shows a wider distribution at $f=1/2$,
which may be attributed to the fluctuations of the degenerate states just above and below
the Fermi level: The energies of the degenerate states fluctuate severely according to the
disorder configurations, resulting in large fluctuations of the inverse compressibility.

We conclude this section with a comment on the validity of the HF approximation at
$f=1/2$.  As observed in \figref{fig:prcl}, the effects of the full frustration survive
even in the presence of rather strong interactions, with the strength up to $V/t=10$.
Such effects are, however, not disclosed in the average and fluctuations of the inverse
compressibility calculated via the HF method (see \figref{fig:icff} for $V/t=10$).
Further, the average participation ratio obtained within the HF approximation in the same
system as in \figref{fig:prcl} turns out to display considerable discrepancy at $V/t=10$,
which strongly suggests that the HF approximation breaks down for strong interactions $V/t
\gtrsim 10$.

\section{Summary\label{sec:s}}

We have studied numerically the inverse compressibility and its fluctuations in
two-dimensional Coulomb glasses, with emphasis on the quantum effects associated with
electron hopping as well as the frustration effects due to applied magnetic fields.  For a
systematic study of this problem, we have begun with the corresponding classical system,
revealing the detailed dependence of the inverse-compressibility fluctuations on the
interaction and random disorder.  Then the effects of electron hopping on the distribution
of the inverse compressibility have been examined, particularly with regard to the
interplay with the interaction and disorder.  The results, obtained mostly via the
Hartree-Fock approximation, have been compared with those of existing experiments as well
as of previous numerical studies: In addition to the symmetric Gaussian distribution with
non-Gaussian tails, we have observed, in a sufficiently clean sample with strong
interactions, a peculiar right-biased distribution of the inverse compressibility, which
was indeed reported in experiment under similar situations.  Realization of even cleaner
samples, where such an unusual distribution may be attainable with rather weak
interactions, is thus expected to confirm the validity of our results and to clarify the
role of interactions.  Remarkably, the relative fluctuations have been found to decrease
eventually as the interaction strength is increased.  This is contrary to the quadratic
interaction dependence of the fluctuations, suggested in a previous numerical study.

We have next investigated the effects of magnetic fields on the distribution and
fluctuations of the inverse compressibility.  Weak frustration generally suppresses the
fluctuations; in a clean sample with weak interactions, however, fluctuations are enhanced in
some range of the frustration.  The difference in responses to the magnetic fields has been
discussed in relation to localization properties, which provides a good explanation for the
effects of the magnetic fields on the distribution of the inverse compressibility.
Finally, strong frustration has been observed to suppress the effects of the interaction.


\acknowledgments

This work was supported in part by the Ministry of Education through the BK21 Project
(ML, MYC) and by the Korea Science and Engineering Foundation through the
Center for Strongly Correlated Materials Research (GSJ).

\end{document}